\documentclass[letter,12pt]{report}

\usepackage{amsmath,amssymb}             
\usepackage[latin1]{inputenc}
\usepackage[OT1]{fontenc}
\usepackage[left=2.5cm,right=2.5cm,top=2cm,bottom=2cm,includefoot,includehead,headheight=13.6pt]{geometry}
\usepackage{setspace}
\usepackage{lineno}
\usepackage{footmisc}
\usepackage{indentfirst}
\usepackage{siunitx}
\usepackage{lmodern}
\usepackage{bm}
\usepackage{float}
\usepackage{tabu}
\usepackage{multirow}

\usepackage{graphicx,type1cm,eso-pic,color}

\usepackage{color}
\definecolor{linkcol}{rgb}{0,0,0.4} 
\definecolor{citecol}{rgb}{0.5,0,0} 

\usepackage{array}
\newcolumntype{L}[1]{>{\raggedright\let\newline\\\arraybackslash\hspace{0pt}}m{#1}}
\newcolumntype{C}[1]{>{\centering\let\newline\\\arraybackslash\hspace{0pt}}m{#1}}
\newcolumntype{R}[1]{>{\raggedleft\let\newline\\\arraybackslash\hspace{0pt}}m{#1}}

\usepackage{rotating}                    
\usepackage{fancyhdr}                    

\let\headruleORIG\headrule
\renewcommand{\headrule}{\color{black} \headruleORIG}

\usepackage{colortbl}
\arrayrulecolor{black}

\fancypagestyle{plain}{
  \fancyhead{}
  \fancyfoot{}
  
}

\makeatletter

\def\cleardoublepage{\clearpage\if@twoside \ifodd\c@page\else%
  \hbox{}%
  \thispagestyle{empty}
  \newpage%
  \if@twocolumn\hbox{}\newpage\fi\fi\fi}

\makeatother
 
{%

\hrulefill
\vspace*{0.5cm}%
\end{minipage}
}

{ \begin{list}%
	{$\bullet$}%
	{\setlength{\labelwidth}{25pt}%
	 \setlength{\leftmargin}{30pt}%
	 \setlength{\itemsep}{\parsep}}}%
{ \end{list} }

\renewcommand{\epsilon}{\varepsilon}

\usepackage{placeins}
\usepackage{epigraph}

\usepackage{etoolbox}

\makeatletter
\patchcmd{\epigraph}{\@epitext{#1}}{\itshape\@epitext{#1}}{}{}
\makeatother

\usepackage{hyperref}
\hypersetup
{
bookmarksopen=true,
pdftitle="Nuclear GPDs - ALERT Run Group Proposal",
pdfauthor="Z.-E. Meziani", 
pdfsubject="Nuclear GPDs - ALERT Run Group Proposal", 
pdfmenubar=true, 
pdfhighlight=/O, 
colorlinks=true, 
pdfpagemode=UseNone, 
pdfpagelayout=SinglePage, 
pdffitwindow=true, 
linkcolor=linkcol, 
citecolor=citecol, 
urlcolor=linkcol 
}

\begin{document}

\begin{titlepage}
     \begin{center}
       \vspace*{-1.8cm}
       \noindent \Huge \textbf{Partonic Structure of Light Nuclei} \\
     \end{center}
   
\renewcommand{\thefootnote}{\fnsymbol{footnote}}
     \begin{center}
       \vspace*{1.0cm}
       \noindent {W.~R.~Armstrong, J.~ Arrington, I.~Clo\"{e}t, K.~Hafidi\footnote[2]{Spokesperson}, M.~Hattawy$^\dagger$, D.~ Potterveld, P.~Reimer, S.~Riordan, Z.~Yi} \\
       \vspace*{0.2cm}
       \noindent \emph{Argonne National Laboratory, Lemont, IL 60439, USA} \\
       \vspace*{0.7cm}
       \noindent {J.~Ball, M.~Defurne, M.~Gar\c{c}on, H.~Moutarde, S.~Procureur, F.~Sabati\'e} \\
       \vspace*{0.2cm}
       \noindent \emph{CEA, Centre de Saclay, Irfu/Service de Physique Nucl\'eaire, 91191 Gif-sur-Yvette, France} \\
       \vspace*{0.7cm}
       \noindent {W.~Cosyn} \\
       \vspace*{0.2cm}
       \noindent \emph{Department of Physics and Astronomy, Proeftuinstraat 86, Ghent University, 9000 Ghent, Belgium} \\
       \vspace*{0.7cm}
       \noindent {M.~Mazouz} \\
       \vspace*{0.2cm}
       \noindent \emph{Facult\'e des Sciences de Monastir, 5000 Tunisia} \\
       \vspace*{0.7cm}
       \noindent {J.~Bettane, R.~Dupr\'{e}$^\dagger$, M.~Guidal, D.~Marchand, 
       C.~Mu\~noz, S.~Niccolai, E.~Voutier} \\
       \vspace*{0.2cm}
       \noindent \emph{Institut de Physique Nucl\'eaire, CNRS-IN2P3, Univ. Paris-Sud, Universit\'e Paris-Saclay, 91406 Orsay Cedex, France} \\
       \vspace*{0.7cm}
       \noindent {K.~P.~Adhikari, J.~A.~Dunne, D.~Dutta,  L.~El~Fassi,  L.~Ye} \\
       \vspace*{0.2cm}
       \noindent \emph{Mississippi State University, Mississippi State, MS 39762, USA} \\
       \vspace*{0.7cm}
       \noindent {M.~Amaryan, G.~Charles, G.~Dodge} \\
       \vspace*{0.2cm}
       \noindent \emph{Old Dominion University, Norfolk, VA 23529, USA} \\
       \vspace*{0.7cm}
       \noindent {V.~Guzey} \\
       \vspace*{0.2cm}
       \noindent \emph{Petersburg Nuclear Physics Institute, National Research Center "Kurchatov Institute", 
                                  Gatchina, 188300, Russia} \\
       \vspace*{0.7cm}
       \noindent {N.~Baltzell$^\dagger$, F.~X.~Girod, S.~Stepanyan} \\
       \vspace*{0.2cm}
       \noindent \emph{Thomas Jefferson National Accelerator Facility, Newport News, VA 23606, USA} \\
       \vspace*{0.7cm}
       \noindent {B.~Duran, S.~Joosten, Z.-E.~Meziani$^\dagger$\footnote[3]{Contact person}, M.~Paolone$^\dagger$, M.~Rehfuss, N.~Sparveris} \\
       \vspace*{0.2cm}
       \noindent \emph{Temple University, Philadelphia, PA 19122, USA} \\
       \vspace*{0.7cm}
       \noindent {F.~Cao, K.~Joo, A.~Kim, N.~Markov} \\
       \vspace*{0.2cm}
       \noindent \emph{University of Connecticut, Storrs, CT 06269, USA} \\
       \noindent {S.~Scopetta} \\
       \vspace*{0.2cm}
       \noindent \emph{Universit\`a di Perugia, INFN, Italy} \\
       \vspace*{0.7cm}
       \noindent {W.~Brooks, A.~El-Alaoui} \\
       \vspace*{0.2cm}
       \noindent \emph{Universidad T\'ecnica Federico Santa Mar\'ia, Valpara\'iso, Chile} \\
       \vspace*{0.7cm}
       \noindent {S.~Liuti} \\
       \vspace*{0.2cm}
       \noindent \emph{University of Virginia, Charlottesville, VA 22903, USA} \\
       \vspace*{1.1cm}
       \noindent {\Large \textbf{a CLAS Collaboration Proposal} } \\
     \end{center}
\renewcommand*{\thefootnote}{\arabic{footnote}}

\date{\today}

\end{titlepage}
\sloppy

\titlepage

\setcounter{page}{3}
      \renewcommand{\thefootnote}{\fnsymbol{footnote}}  
     \begin{center}
       \vspace*{-1.0cm}
      \noindent {\Large \textbf{Jefferson Lab PAC 45}} \\
      \vspace*{0.8cm}
       \noindent \Huge \textbf{Nuclear Exclusive and Semi-inclusive Measurements with a New CLAS12 Low Energy Recoil Tracker} \\
       \vspace*{0.8cm}
       \noindent \Large \textbf{ALERT Run Group\footnote[2]{Contact Person: Kawtar Hafidi (kawtar@anl.gov)} } \\      
       \vspace*{2.0cm}
       {\large\textbf{EXECUTIVE SUMMARY}}
     \end{center}
 
 \vspace*{0.4cm}

In this run group, we propose a comprehensive physics program to investigate 
the fundamental structure of the $^4$He nucleus. 
An important focus of this program is on 
the coherent exclusive Deep Virtual Compton Scattering (DVCS) and Deep 
Virtual Meson Production (DVMP) with emphasis on $\phi$ meson production. These are 
particularly powerful tools enabling model-independent nuclear 3D tomography 
through the access of partons' position in the transverse plane. These 
exclusive measurements will give the chance to compare directly 
the quark and gluon radii of the helium nucleus. 
Another important measurement proposed in this program is the study of the 
partonic structure of bound nucleons. To this end, we propose next generation 
nuclear measurements in which low energy recoil nuclei are detected. The 
tagging of recoil nuclei in deep inelastic reactions is a powerful technique, 
which will provide unique information about the nature of medium modifications 
through the measurement of the EMC ratio and its dependence on the nucleon 
off-shellness. 
Finally, we propose to measure incoherent spectator-tagged DVCS 
on light nuclei (d, $^4$He) where the observables are sensitive to the
Generalized Parton Distributions (GPDs) of a quasi-free neutron for the case of the deuteron, and bound proton and neutron for the case of $^4$He. The objective is to 
study and separate nuclear effects and their manifestation in GPDs.
The fully exclusive kinematics provide a novel approach for studying
final state interactions in the measurements of the 
beam spin asymmetries and the off-forward EMC ratio.\\

At the heart of this program is the Low Energy Recoil Tracker (ALERT) 
combined with the CLAS12 detector. The ALERT detector is composed of a stereo 
drift chamber for track reconstruction and an array of scintillators for 
particle identification. Coupling these two types of fast detectors will allow 
ALERT to be included in the trigger for efficient background rejection, while 
keeping the material budget as low as possible for low energy particle 
detection. ALERT will be installed inside the solenoid magnet instead of the 
CLAS12 Silicon Vertex Tracker and Micromegas tracker. We will use an 11 GeV 
longitudinally polarized electron beam (80\% polarization) of up to 1000~nA on a gas target 
straw filled with deuterium or $^4$He at 3 atm to obtain a luminosity up to
$6\times10^{34}$~nucleon~cm$^{-2}$s$^{-1}$. In addition we will need to run 
hydrogen and $^4$He targets at different beam energies for detector 
calibration. The following table summarizes our beam time request: \\

\newcommand{\minitab}[2][l]{\begin{tabular}{#1}#2\end{tabular}}
\begin{table}[ht!]
\label{tab:beamTimeRequest}
\center
\bgroup
\def\arraystretch{1.2}%
\tabulinesep=1.5mm
\begin{tabu}{C{3.1cm}C{2.8cm}C{1.6cm}C{2.3cm}C{1.6cm}C{2.5cm}}
\tabucline[2pt]{-}
\bf Configurations  & \bf Proposals & \bf Targets   & \bf Beam time request  & \bf Beam current & \bf Luminosity$^*$ \\
                    &                    &               & days    & nA       & n/cm$^{2}$\!/s     \\
\tabucline[1pt]{-}                                                   
{Commissioning}     & All$^\dagger$      & $^1$H, $^4$He & 5       & Various  & Various            \\
A                   & Nuclear GPDs       & $^4$He        & 10      & 1000      & $6\times10^{34}$   \\
B                   & Tagged EMC \& DVCS & $^2$H         & 20      & 500      & $3\times10^{34}$   \\
C                   & All$^\dagger$      & $^4$He        & 20      & 500      & $3\times10^{34}$   \\
\tabucline[1pt]{-}                                                   
{\bf TOTAL}         &                    & \,            & \bf 55  & \,       & \,                 \\
\tabucline[2pt]{-}
\end{tabu}
\egroup
\end{table}

\footnotetext[1]{This luminosity value is 
   based on the effective part of the target. When accounting for the target's 
   windows, which are outside of the ALERT detector, it is increased by 60\%.}
   
\footnotetext[2]{``All'' includes the four proposals of the run group: Nuclear GPDs, Tagged EMC, Tagged DVCS and Extra Topics. Note that the beam time request is only driven by the three first proposals.}
   
\renewcommand*{\thefootnote}{\arabic{footnote}}

\date{\today}

\sloppy

\titlepage

\renewcommand{\baselinestretch}{1.10}

\setcounter{page}{5}
\addcontentsline{toc}{chapter}{Abstract}

     \begin{center}
{\large\textbf{Abstract}}
    \end{center}

\vspace*{0.4cm}

We propose to study the partonic structure of $^4$He by measuring the Beam Spin 
Asymmetry (BSA) in coherent Deeply Virtual Compton Scattering (DVCS) and the 
differential cross-section of the Deeply Virtual Meson 
Production (DVMP) of the $\phi$. Despite its 
simple structure, a light nucleus such as $^4$He has a 
density and a binding energy comparable to that of heavier nuclei.  Therefore, 
by studying $^4$He nucleus, one can learn typical features of the partonic 
structure of atomic nuclei. In addition, due to its spin-0, only one 
chiral-even GPD, $H_{A}$, parameterizes the $^4$He partonic 
structure at twist-2 allowing for a much simpler extraction of 
its tomography from data.\\

A major goal of this proposal is to cover a wide kinematical range and collect 
high statistics leveraging the knowledge obtained during the CLAS experiment
E08-024 (eg6 run), where, 
for the first time, exclusive coherent DVCS off $^4$He was successfully 
measured. The real and imaginary parts of the 
$^4$He Compton form factors (CFFs) will be extracted in a model independent way 
from the experimental asymmetries, allowing us to access the nuclear transverse 
spatial distributions of quarks.\\ 

An equally important focus of this proposal is to study the gluonic structure 
of nuclei through the measurement of exclusive coherent 
$\phi$ meson electroproduction off a $^4$He target. The kinematic regime 
to be explored includes very low $|t|$ up to 
the first diffractive minimum as found in elastic scattering off $^4$He 
($|t^\prime| \simeq 0.6$ GeV$^2$).  
The $\phi$ meson will be detected primarily through the charged $K^+ K^-$ 
channel, with the neutral $K^0_S K^0_L$ channel also available through $K_S 
\rightarrow \pi^+ \pi^-$. Differential cross-sections for $\phi$ 
electroproduction off $^4$He will be measured for the first time.\\

The combination of CLAS12 and the ALERT detector provides a unique opportunity 
to study both the quark and gluon structure of a dense light nucleus.  
Coherent exclusive DVCS off $^4$He will probe the transverse spatial 
distribution of quarks in the nucleus as a function of the quarks' longitudinal 
momentum fraction, $x$. In parallel, the average spatial transverse gluon 
density of the $^4$He nucleus will be extracted within a GPD framework using 
the measured longitudinal cross-section of coherent $\phi$ production in a 
similar range of longitudinal momentum $x$ as that of the quarks. Additionally, 
threshold effects of $\phi$ production can be explored by exploiting the ALERT 
detector's large acceptance for low $|t|$ events.

\newpage

\tableofcontents

\chapter*{Introduction\markboth{\bf Introduction}{}}
\label{chap:intro}
\addcontentsline{toc}{chapter}{Introduction}

\epigraph{The general evidence on nuclei strongly supports the view that 
    the $\bm{\alpha}$ particle is of primary importance as a unit of 
    the structure of nuclei in general and particularly of the heavier 
    elements.  It seems very possible that the greater part of the mass of 
    heavy nuclei is due to $\bm{\alpha}$ particles which have an independent 
existence in the nuclear structure.}{--- \textup{Rutherford, Chadwick, and 
Ellis~(1930)}\\Radiations from Radioactive Substances\footnote{This was the 
first textbook on nuclear physics and notably published two years before the 
discovery of the neutron.}}

Inclusive deep inelastic scattering (DIS) experiments have been instrumental in 
advancing our understanding of the QCD structure of nuclei and the effect of 
nuclear matter on the structure of hadrons. A great example is the observation 
by the European Muon Collaboration (EMC) of a deviation of the deep inelastic 
structure function of a nucleus from the sum of the structure functions of the 
free nucleons, the so-called EMC effect~\cite{Aubert:1983xm}. It became clear 
that even in a DIS process characterized by high locality of the probe-target 
interaction region, a different picture emerges from the nucleus other than a 
collection of quasi-free nucleons. On the theory side, despite decades of 
theoretical efforts~\cite{Miller-PRC2002,Thomas-Annal-2004,Liuti:2005qj,
Rezaeian-Pirner-2006,Zhen-min-He-1998} 
with increased sophistication, a unifying physical picture of the origin of the 
EMC effect is still a matter of intense debate. To reach the next level of our 
understanding of nuclear QCD and unravel the partonic structure of nuclei, 
experiments need to go beyond the inclusive measurements and focus on exclusive 
and semi-inclusive reactions. \\

Hard exclusive experiments such as Deep Virtual Compton Scattering (DVCS) and 
Deep Virtual Meson Production (DVMP) provide an important new probe that will 
allow us to discern among the different interpretations of nuclear effects on 
the structure of embedded nucleons in the nuclear medium. By introducing a new 
framework to describe both the intrinsic motion of partons and their transverse 
spatial structure in nuclei~\cite{Liuti:2005qj,Rezaeian-Pirner-2006,
Zhen-min-He-1998,Accardi:2005jd,Accardi:2005hk,Kirchner:2003wt}.  
valuable information can be obtained from the measurement of the nuclear 
Generalized Parton Distributions (GPDs) representing the soft matrix elements 
for these processes. The GPDs correspond to the coherence between quantum 
states of different (or same) helicity, longitudinal momentum, and transverse 
position. In an impact parameter space, they can be interpreted as a 
distribution in the transverse plane of partons carrying a certain longitudinal 
momentum~\cite{Burkardt-2000,Diehl-2002,Belitsky-2002}. A crucial feature of 
GPDs is the access to the transverse position of partons which, combined with 
their longitudinal momentum, leads to the total angular momentum of 
partons~\cite{Burkardt-2005}. This information is not accessible to inclusive 
DIS which measures probability amplitudes in the longitudinal plane. \\

A high luminosity facility such as Jefferson Lab offers a unique opportunity to 
map out the three-dimensional quark and gluon structure of nucleons and nuclei.  
While most of submitted proposals to JLab Program Advisory Committee (PAC) have 
focused on the studies of the 3D nucleon structure considered as one of the 
main motivations for the JLab 12 GeV upgrade, we propose here to extend the 
measurements to light nuclei. While this proposal focuses on $^4$He nucleus, we 
also plan to measure few deuteron GPDs\footnote{See the 4$^{th}$ proposal of 
the ALERT run group which summarizes additional measurements we plan to perform 
with no additional beam time.}. Pioneering measurements of exclusive coherent 
DVCS off $^4$He have been successfully conducted during the JLab 6 GeV era 
(E08-024) using the CLAS detector enhanced with the radial time projection 
chamber (RTPC) for the detection of low energy recoils and the inner 
calorimeter for the detection of forward high energy photons. However, the 
experiment covered only limited kinematic range and the results were dominated 
by statistical uncertainties ~\cite{eg6_note}. \\

We propose a new measurement of hard exclusive DVCS and deeply virtual $\phi$ 
production off $^4$He nuclei. The focus of this proposal is on the coherent 
DVCS (DVMP) channel where the scattered electron, the produced photon (the 
$\phi$ meson) and the recoil $^4$He are all detected in the final state. We 
propose to use CLAS12 because of its large acceptance. In addition to the 
coherent DVCS and DVMP off $^4$He, the CLAS12-ALERT setup will allow us to mine 
the data collected in this experiment for other final states as well, such as 
the $\pi^0$, $\rho$ and $\omega$ mesons and other reaction channels described 
in the accompanying proposals of the ALERT run group$^1$.  The novelty of the 
proposed measurements is the use of a new low energy recoil tracker (ALERT) in 
addition to CLAS12.  The ALERT detector is composed of two types of fast 
detectors: a stereo drift chamber for track reconstruction and an array of 
scintillators for particle identification. ALERT will be included in the 
trigger for efficient background rejection, while keeping the material budget 
as low as possible to detect low energy particles.  This was not possible with 
the previous GEM based RTPC due to the long drift time.

\chapter{Physics Motivations}

\begin{figure}[h]
\begin{center}
\includegraphics[width=0.99\linewidth]{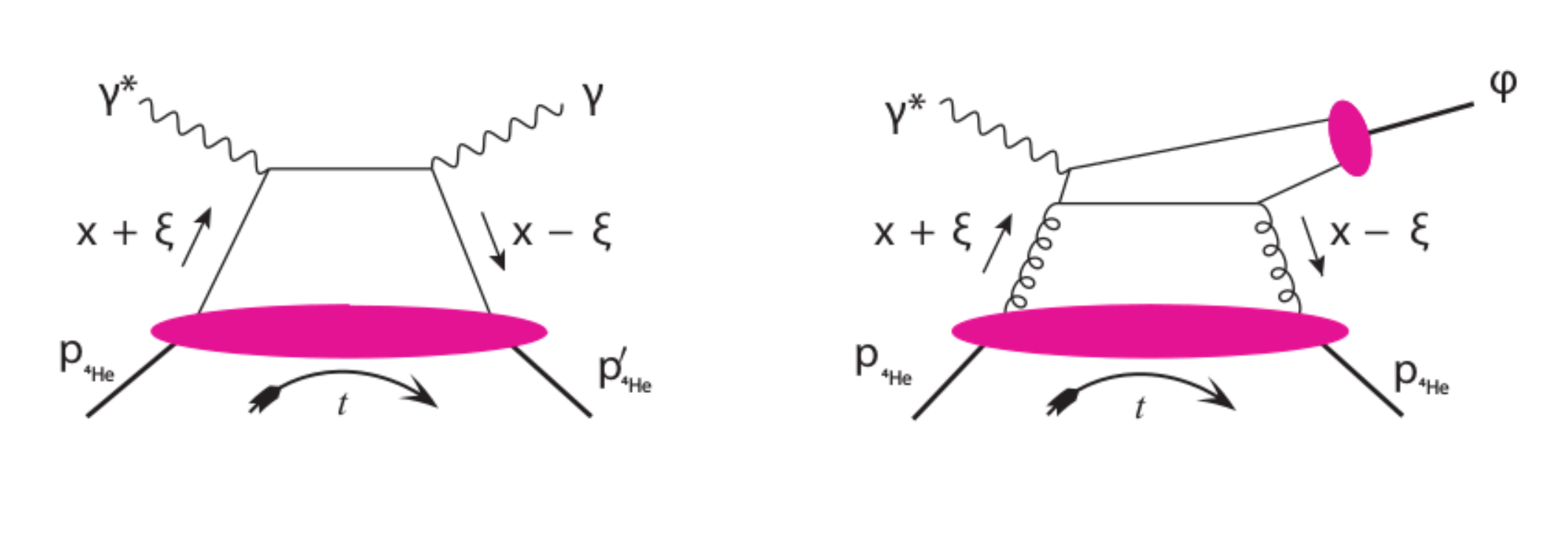}
\caption{Left figure: DVCS process in the handbag approximation. Right figure: 
DVMP diagram at the lowest order, dominated by two-gluon exchange.}
\label{fig:handbag_phi}
\end{center}
\end{figure}

A wealth of information on the QCD structure of hadrons lies in the 
correlations between the momentum and spatial degrees of freedom of the 
constituent partons. Such correlations are accessible via GPDs which, more 
specifically, describe the longitudinal momentum distribution of a parton 
located at any given position in the plane transverse to the longitudinal 
momentum of the fast moving nucleon. Various GPDs extracted from measurements 
of hard exclusive reactions with various probe helicities and target spin 
configurations are necessary to identify this subset of the hadronic 
phase-space distribution, known as the Wigner distribution. The processes 
which are most directly related to GPDs are DVCS and DVMP corresponding to the 
exclusive electroproduction of a real photon or a meson in the final state 
respectively, see Figure~\ref{fig:handbag_phi}.\\

The number of GPDs needed to parametrize the partonic structure of a nucleus 
depends on the different configurations between the spin of the nucleus and the 
helicity direction of the struck quark. For example, for a target of spin $s$, 
the number of chiral-even GPDs is equal to ($2s+1$)$^2$ for each quark flavor.  
DVCS off spin 0 nuclear targets, such as $^4$He, is simpler to study since only 
one chiral-even GPD, $H_{A}$, is present at leading twist.\\

The $^4$He nucleus is a well studied few-body system in standard nuclear 
physics. It is characterized by a strong binding energy and relatively high 
nuclear core density similar to some more complex nuclei. Inclusive scattering 
off $^4$He shows a large EMC effect. By measuring quark and gluon GPDs in nuclei, one also 
accesses transverse spatial degrees of freedom, by which one can infer space 
dependent nuclear modifications directly from data.  

\section{DVCS Measurement}
\label{chap:physics}

The $^4$He nucleus provides a textbook case for DVCS 
measurements since it has only one chiral-even GPD. Therefore, by measuring 
coherent exclusive DVCS, one can, in a model independent way at leading twist, 
access the single Compton form factor and subsequently extract the transverse 
spatial distribution of quarks in the fast moving nucleus. It is also
an ideal target to isolate higher twist effects as is proposed in the
4$^{th}$ proposal of the ALERT run group.\\


The DVCS process off nuclear targets differs from single proton scattering in 
that it can occur via either the coherent or incoherent channels. In this 
proposal, we will consider the coherent channel where the target nucleus 
remains intact and recoils as a whole while emitting a real photon ($eA 
\rightarrow e' A' \gamma$). This process allows one to measure the nuclear 
GPDs, which contain information on the parton correlations and the nuclear 
forces in the target \cite{Liuti:2005qj,Polyakov:2002yz}.
\begin{figure}[h]
\begin{center}
\includegraphics[width=0.60\linewidth]{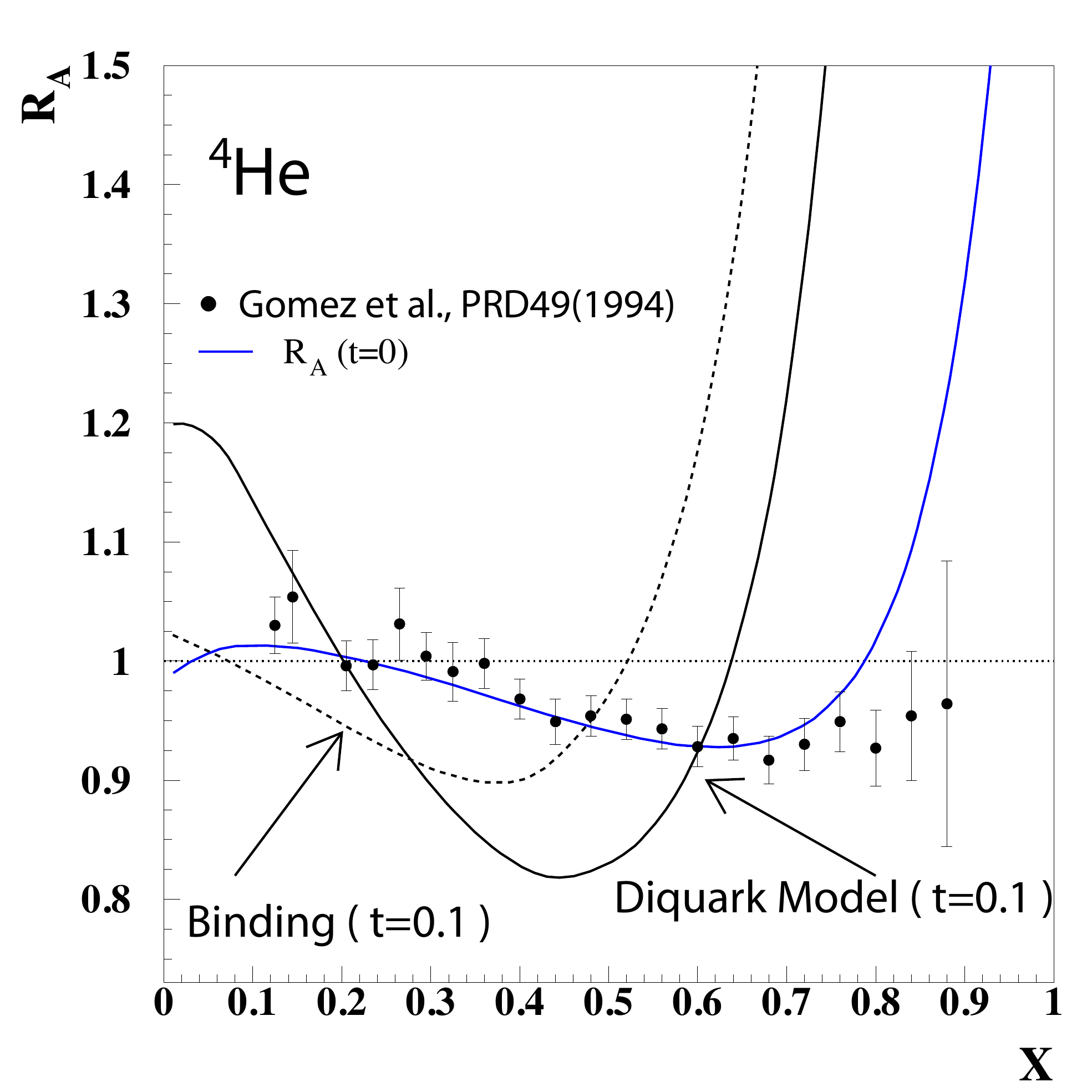}
\caption{Theoretical expectations for Off-forward EMC effect in $^4$He.  
   Predictions at $t=0.1$ GeV$^2$ from both ``conventional'' binding models and 
   within a diquark picture for nuclear modifications are shown. For 
comparison, the effect at $t=0$ is given by experimental data and theory (blue 
curves) on the ratio of inelastic structure functions \cite{Gomez:1993ri} 
(adapted from Ref.~\cite{Liuti:2005qj}).}
\label{Mot_fig0}
\end{center}
\end{figure}
We propose to measure coherent DVCS Beam Spin Asymmetries (BSA) in order to 
extract in a model independent way both the real and imaginary parts of the 
$^4$He nuclear Compton form factor $H_A$. This will lead the way toward the 
determination of the nucleus 3D picture in terms of its basic degrees of 
freedom, namely valence quarks in this case. In addition, the comparison 
between the coherent nuclear BSA and the free proton ones will allow us to 
study a variety of nuclear medium effects, such as the modification of quark 
confinement size in the nuclear medium. In fact, configuration size 
modifications have been advocated as responsible for the behavior of the EMC 
ratio in the intermediate $x_B$ region 
\cite{Close1983,Nachtmann1984,Jaffe1984,Close1988}. The generalized EMC effect 
{\it i.e.} the modification of the nuclear GPDs with respect to the free 
nucleon ones, normalized to their respective form factors was studied in 
Refs.~\cite{Liuti:2005gi,Guzey:2003jh,Guzey:2005ba,Scopetta:2004kj}.  
Measurements in the intermediate $x_B$ range between 0.1 and 0.6, and for an 
appropriate $t$-range are crucial for both establishing the role of partonic 
configuration sizes in nuclei, and for discerning among the several competing 
explanations of the EMC effect. As shown in Ref.~\cite{Liuti:2005gi}, the role 
of partonic transverse degrees of freedom, both in momentum and coordinate 
space, could be important in the generalized EMC effect, thus predicting an 
enhancement of signals of nuclear effects with respect to the forward case 
(Figure~\ref{Mot_fig0}).
\section{Deep Virtual $\phi$ Production Measurement}
As we mentioned in the previous section, over the last two decades, there has 
been an increasing interest in multidimensional imaging of the structure of the 
nucleon. Through measurements of exclusive processes off 
the nucleon, information on Compton form factors and subsequently GPDs has 
been obtained successfully. Through these GPDs, the transverse parton 
density of the nucleon in the infinite momentum frame is obtained by a Fourier 
transform of the momentum transfer dependence leading to a transverse spatial 
parton density description of the nucleon.\\

In contrast with DVCS, which is sensitive to the charge distribution, exclusive 
$\phi$ production provides an access to the gluon GPDs. The leading order
diagrams of these processes, shown in Figure~\ref{fig:handbag_phi},
illustrate nicely this feature. In the DVCS case, the scattering is 
facilitated through quark exchange, while in the exclusive $\phi$ production, 
the mostly strange $\phi$, interacts with the mostly up-down nucleus 
via a two-gluon exchange. \\

A recently approved proposal using the CLAS12 detector, E12-12-007
\cite{Girod:2012PR}, aims to extract the transverse gluon distribution of the 
proton and its gluonic size.  In analogy, this proposal uses a very similar 
framework to the one discussed in E12-12-007 but focuses on the gluon GPD for 
a tightly bound spin zero nucleus, namely $^4$He, thus extending the 
investigation of quark GPDs in a nucleus to the case of gluons.  
An experiment to extract gluon distributions on heavy nuclei through coherent 
$\phi$ electroproduction at the EIC has already been proposed in the most 
recent EIC white paper \cite{Accardi:2012qut}. JLab 12 GeV can start 
such an investigation at large $x$ initiating a full three-dimensional 
partonic structure investigation of a nucleus for the first time.\\

Gluons are the salient partners of the quarks in a nucleon as well as in a 
nucleus. We know they are responsible for the confinement of quarks and for 
their own confinement, and represent a large fraction of the energy or the mass 
of the nucleon.  However, gluons are charge neutral and cannot be probed 
directly using the electromagnetic probe. For example we know the charge 
distribution of $^4$He and how to interpret it through the charge of nucleons.  
For instance, the diffraction minima in the measured charge distribution of 
$^4$He tell us that nucleons are the appropriate degrees of freedom to 
consider when describing the electromagnetic properties of a nucleus. In fact 
we have yet to see unambiguously the elusive signature of quarks in elastic 
scattering off a nucleus even though we know they must be there as the building 
blocks of nucleons. Similarly, we do not know where the gluons are distributed 
in the nucleon and how they participate in the long and short-range 
nucleon-nucleon correlations that are responsible for the structure of the 
nucleus. One can ask whether in a nucleus the gluons are localized in the 
confined volume defined by the nucleons or spread beyond the size of the 
nucleons. Considering only the gluonic matter in a nucleus, a natural question 
arises, is a nucleus the sum of localized gluon density corresponding to that 
of free nucleons, or else? It would be of paramount importance to test our 
naive understanding of the charge neutral gluonic matter when we discuss the 
size of the nucleon and that of the nucleus. Among the interesting questions 
one might ask is whether there is evidence that the gluon transverse spatial 
distribution is homogeneous, or does it appear to be affected by the location of 
the bound nucleons? In the same spirit of the discussion of quarks which was 
carried in the previous section, the discussion of the gluons is at least as 
relevant to our understanding of nuclei from basic principles.\\

Measuring the gluon distributions in the nucleon is an important step and will 
be carried  by the approved experiment E12-12-007 \cite{Girod:2012PR}.  
Understanding how these distributions are modified to provide the binding and 
structure in a nucleus is as fascinating of a question and an integral part of 
our quest of using QCD to explore nuclear matter. The future Electron Ion 
Collider will have the tools to address these questions using heavier mesons 
like $J/\Psi$ and $\Upsilon$. At JLab 12 GeV we can use the lighter vector 
mesons, namely the $\phi$, to initiate this physics program and provide a 
glimpse into the salient features of nuclear matter.

\setlength\parskip{\baselineskip}%
\chapter{Formalism and Experimental Observables}
\label{chap:expobs}

The observables which are sensitive to $^4$He's quark and gluon GPDs are 
noticeably  different, and therefore, so are the techniques for extracting the 
quark and gluon information. In this chapter we first discuss how the GPDs 
relate to the proposed measurements. In \ref{sec:DVCSFormalism} the methods by 
which we extract the quark GPD $H_A$ from the DVCS beam spin asymmetry are 
presented and the current experimental status is discussed.  In 
section~\ref{sec:DVMPFormalism} we show how the gluon GPD $H_g$ is extracted 
from the angular distribution of the $\phi$ meson decay. 


\section{Generalized Partons Distributions}
%
GPDs are universal non-perturbative objects, entering the description of hard 
exclusive electroproduction processes. They are defined for each quark flavor 
$f$ and gluon as matrix elements of light cone operators~\cite{Belitsky:2005qn} 
describing the transition between the initial and final states of a hadron.  
The GPDs depend on two longitudinal momentum fraction variables $(x,\xi)$ and 
on the momentum transfer $t$ to the target. At twist-2 order, $\xi$ 
can be calculated as $x_B/(2 - x_B)$, where $x_{B}$ ($= Q^2/2M_p\nu$) is the Bjorken variable , and -2$\xi$ is the longitudinal fraction of the momentum transfer $\Delta$, with $\Delta^2 =(p-p')^2 = t$. $x$ is the average longitudinal momentum fraction of the parton involved in the process. 

In the limit $\xi \rightarrow 0$, the 2-dimensional Fourier transform yields 
impact parameter GPDs
\begin{equation}
  f(x,\bm{b_{\perp}}) =  \int \frac{d^2 \bm{\Delta}_{\perp}}{4\pi^2} F(x, \xi=0, 
  \Delta)\mathrm{e}^{-i \bm{b}_{\perp}\cdot\bm{\Delta}_{\perp}}
\end{equation}
where $F\in {H,E,\tilde{H},\tilde{E}}$. With azimuthal symmetry in the transverse plane this 
reduces to the Hankel transform
\begin{align}
  f(x,b) &=  \int_{0}^{\infty} J_{0}(b \Delta_{\perp}) 
  F(x,0,\Delta_{\perp}^2) \frac{d\Delta_{\perp}}{\pi} \\ &= \int_{0}^{\infty} 
  J_{0}(b\sqrt{t}) F(x,0,t) \sqrt{t} \frac{dt}{2\pi}\label{eq:HankelTransform}
\end{align}
where $J_0$ is a Bessel function of the first kind.
In this limit the impact parameter GPDs can be interpreted as the probability 
distribution of finding a parton with longitudinal momentum fraction $x$ at a 
transverse position $b$ with respect to the nucleus' center of 
momentum~\cite{Burkardt:2002hr}.

GPDs can also be considered as the off-forward kinematic generalizations of the 
standard Parton Distributions Functions (PDFs) from inclusive DIS.  PDFs can be 
recovered from GPDs in the limit of zero momentum transfer between the initial 
and final protons (the forward limit), with no target spin flip. Similar to 
DIS, higher twist terms describe quark-gluon-quark correlations which are 
suppressed by powers of $1/Q$, we
elaborate in the 4$^{th}$ proposal of the run group on possible methods to 
study these effects. 

The spin zero of the $^4$He target allows for a simple parametrization of its 
partonic structure characterized at leading twist by one chirally-even GPD 
$H_A$. In the forward limit ($t \to 0$), this GPD reduces to the usual parton 
densities of $^4$He measured in DIS. The polynomiality property of GPDs leads 
to interesting consequences: the first Mellin moment provides an explicit link 
with the electromagnetic form factor $F_A$ of the nucleus
\begin{equation}
\sum_f e_{f} \int_{-1}^{1} dx \, H_A^f(x,\xi,t) = F_A(t) \, ,
\end{equation}
and the second moment yields the relationship
\begin{equation}
\int_{-1}^{1} dx \, x \, H_A^f(x,\xi,t) = M_2^{f/A}(t) + \frac{4}{5} \xi^2 d_A^f(t) \,
\label{dterm}
\end{equation}
which constrains the $\xi$-dependence of the GPDs. At $t \to 0$, the first term 
of the right-hand side of Eq.~(\ref{dterm}) is the momentum fraction of the 
target carried by a given quark. The second term of Eq.~(\ref{dterm}) is the 
so-called $D$-term which was shown to encode information about the spatial
distribution of forces experienced by quarks and gluons inside 
hadrons~\cite{Polyakov:2002yz}.

\section{Coherent DVCS}\label{sec:DVCSFormalism}

\subsection{Accessing the Quark GPD}

\begin{figure}[htb]
  \begin{center}
    \includegraphics[width=0.65\linewidth]{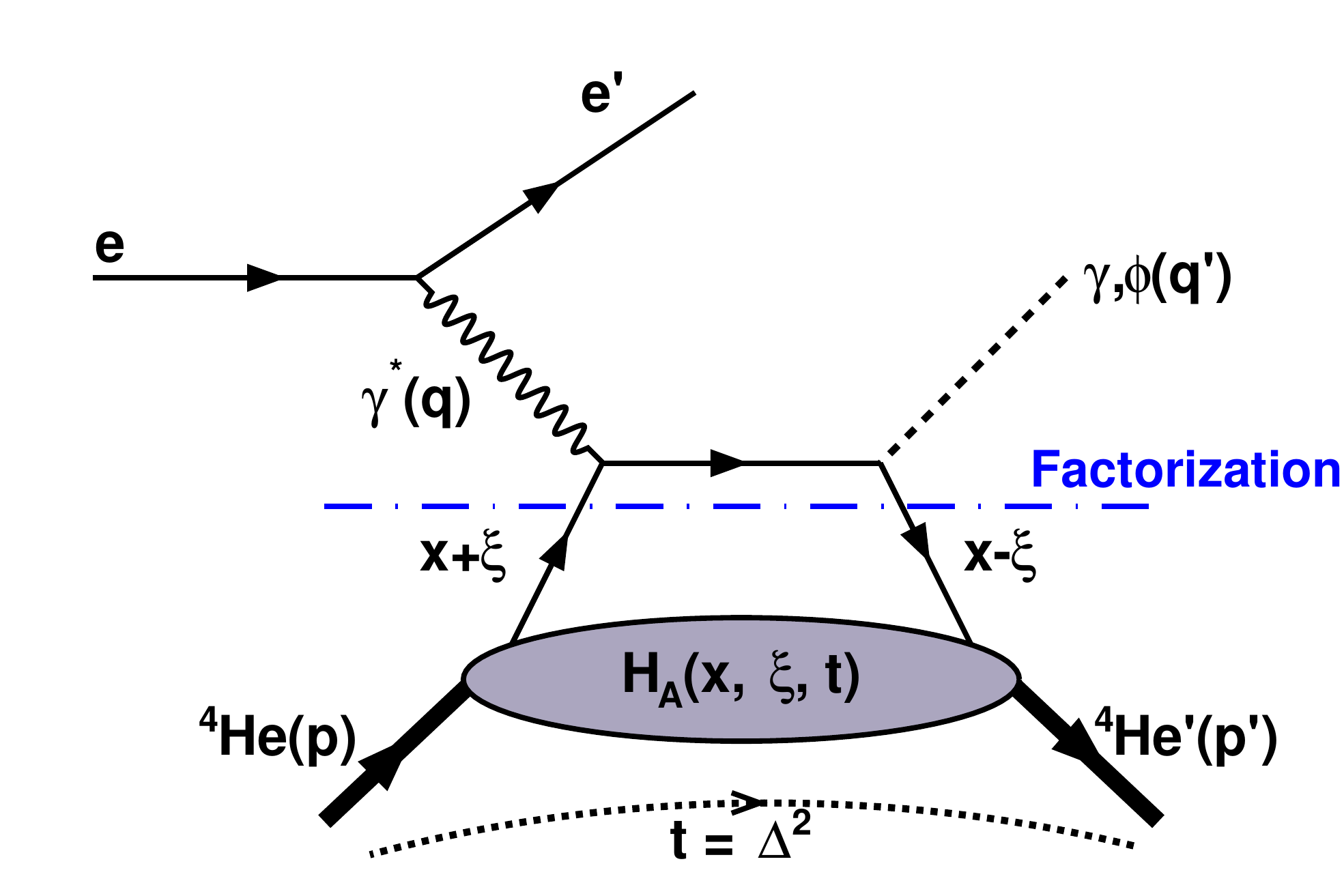}
    \caption{Lowest order (QCD) amplitude for the DVCS and DVMP processes, the 
      so-called handbag diagram. $q$, $q'$ represent the the four-momentum of the 
      virtual and real photons or mesons, and $p$, $p'$ are the initial and final 
    four-momentum of the target nucleus.}
    \label{fig:handbag}
  \end{center}
\end{figure}

The handbag diagram in Figure~\ref{fig:handbag} displays a hard part which is 
calculable in perturbative QCD, and a soft, non-perturbative, part which 
contains the fundamental partonic structure of the nucleus. However, because 
of the loop in the handbag diagram, the $x$ variable is not directly accessible
in the DVCS process, so we access the Compton Form Factor
(CFF), noted $\mathcal{H}_{A}$, and expressed in terms of the GPD as 
\begin{align}
  \begin{split}
    \Re e(&\mathcal{H}_{A}) = \mathcal{P} 
  \int_{0}^{1}dx[H_A(x,\xi,t)-H_A(-x,\xi,t)] \, C^{+}(x,\xi), \end{split} \\
  \Im m(&\mathcal{H}_{A}) = H_A(\xi,\xi,t)-H_A(-\xi,\xi,t),
\end{align}
with $\mathcal{P}$ as the Cauchy principal value integral, and $C^{+}(x,\xi)$ a 
coefficient function ($=  \frac{1}{x-\xi} + \frac{1}{x+\xi}$) 
\cite{Guidal:2013rya}.

\begin{figure}[htb]
  \begin{minipage}[c]{.46\linewidth}
    \hspace{-0.2in}\includegraphics[height=6.0cm]{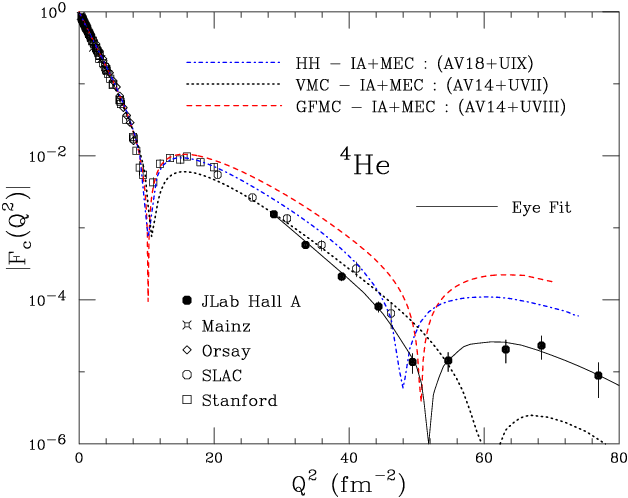}
    \caption{$^4$He charge form factor measurements at Stanford, SLAC, Orsay, Mainz 
      and JLab Hall A compared with theoretical calculations. The figure is from 
    \cite{PhysRevLett.112.132503}. }
    \label{fig:He-4_FFs}
  \end{minipage} \hfill
  \begin{minipage}[c]{.46\linewidth}
    \hspace{-0.3in}\includegraphics[height=7.1cm]{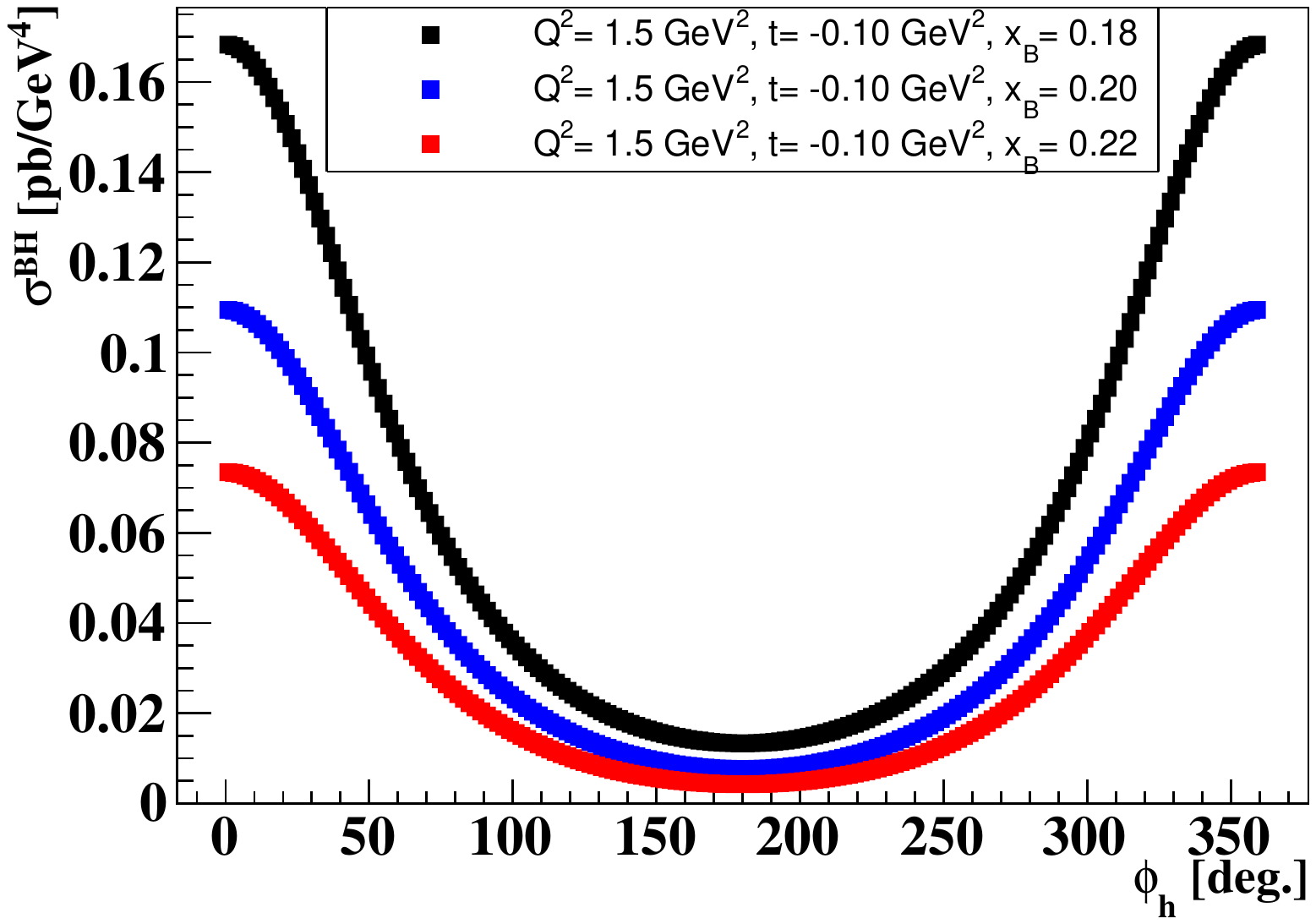}
    \caption{The calculated BH cross section as a function of $\phi$ on a $^4He$ 
      target at three values of $x_{B}$ and fixed values of $Q^{2}$ and $t$.  
      ($t$~=~-~0.1~GeV$^2$/c$^2$ corresponds to $Q^2$~$\approx$~2.57~fm$^{-2}$ on 
    figure \ref{fig:He-4_FFs}).}
    \vspace{+0.3in}
    \label{fig:BH_cross_section_4He}
  \end{minipage}
\end{figure}

Experimentally, the DVCS reaction is indistinguishable from the Bethe-Heitler 
(BH) process, which is the reaction where the final photon is emitted either 
from the incoming or the outgoing leptons. The BH process is not sensitive to 
GPDs and does not carry information about the partonic structure of the 
hadronic target. The BH cross section is calculable from the well-known 
electromagnetic FFs. The DVCS amplitude is enhanced through the interference 
with the BH process. Figure \ref{fig:He-4_FFs} shows the world measurements of 
the $^4$He $F_{A}(t)$ along with theoretical calculations. Following the $^4$He 
$F_{A}(t)$ parametrization by R.  Frosch and his collaborators 
\cite{PhysRev.160.874} (valid at the small values of $-t$ which are of interest 
in this work), Figure \ref{fig:BH_cross_section_4He} shows the calculated BH as 
a function of the azimuthal angle between the leptonic and the hadronic planes 
($\phi$), using 11~GeV electron beam on a $^4$He target.

The differential cross section of leptoproduction of photons for a 
longitudinally-polarized electron beam ($\lambda = \pm 1$) and an unpolarized 
$^4$He target can written as:
\small
\begin{equation}
\frac{d^{5}\sigma^{\lambda}}{dx_{A} dQ^{2} dt d\phi_{e} d\phi} = 
\frac{\alpha^{3}}{16 \pi^{2}} \frac{x_{A} \, y}{Q^{4} \sqrt{1 + \epsilon ^{2}}} 
\frac{
|\mathcal{T}_{BH}|^{2} + |{\mathcal{T}}_{DVCS}^{\lambda}|^{2} + {\mathcal{I}}_{BH*DVCS}^{\lambda}}{e^{6}}
\label{eq:sigdiff}
\end{equation}
\normalsize
where $y = \frac{p \cdot q}{p \cdot k}$, $\epsilon  =  \frac{2 x_{A} 
M_{A}}{Q}$, $x_A  =  \frac{Q^2}{2 p \cdot q}$, ${\mathcal{T}}_{DVCS}$ is the 
pure DVCS scattering amplitude, $ {\mathcal{T}}_{BH}$ is the pure BH amplitude 
and ${\mathcal{I}}^{\lambda}_{BH*DVCS}$ represents the interference amplitude.  
Similarly to a nucleon target one can write out the azimuthal angle, $\phi$, 
dependence for the nuclear BH, DVCS and interference terms in the cross 
section: each modulation in $\phi$ is multiplied by a structure function 
containing the GPDs of interest. The different amplitudes are written as 
\cite{Belitsky:2008bz},
\small
\begin{equation}
 |\mathcal{T}_{BH}|^{2} =  \frac{e^{6} (1 + \epsilon^{2})^{-2}}{x^{2}_{A} y^{2} 
 t \mathcal{P}_{1}(\phi) \mathcal{P}_{2}(\phi)} \left[ c_{0}^{BH} + c_{1}^{BH} 
 \cos(\phi) + c_{2}^{BH} \cos(2\phi)\right] \label{TTBH}
\end{equation}
\begin{equation}
 |\mathcal{T}_{DVCS}|^{2} =  \frac{e^{6}}{y^{2} Q^{2}} \left[ c_{0}^{DVCS} + 
 \sum_{n=1}^{2} \Bigg( c_{n}^{DVCS} \cos(n \phi) + \lambda s_{n}^{DVCS} \sin(n 
 \phi)\Bigg) \right] \label{TTDVCS}
\end{equation}
\begin{equation}
 \mathcal{I}_{BH*DVCS} =  \frac{\pm e^{6}}{x_A y^{3} t \, \mathcal{P}_{1}(\phi) 
 \mathcal{P}_{2}(\phi)} \left[ c_{0}^{I} + \sum_{n=0}^{3} \Bigg( c_{n}^{I} 
 \cos(n \phi) + \lambda s_{n}^{I} \sin(n \phi) \Bigg) \right]
  \label{TTinter}, 
 \end{equation}
The explicit expressions of the coefficients can be found in Appendix \ref{app:Helium_cross_section}.
It is convenient to use the beam-spin asymmetry as DVCS observable because most of the experimental normalization and acceptance issues cancel out in an asymmetry ratio. The beam-spin asymmetry is measured using a longitudinally polarized lepton beam (L) on an unpolarized target (U) and defined as:
\begin{equation}
A_{LU} = \frac{d^{5}\sigma^{+} - d^{5}\sigma^{-} }
                {d^{5}\sigma^{+} + d^{5}\sigma^{-}}.
\label{BSA_equation}
\end{equation}
where $d^{5}\sigma^{+}$($d^{5}\sigma^{-}$) is the DVCS differential cross section for a positive (negative) beam helicity.
 At leading twist, the beam-spin asymmetry ($A_{LU}$) with the two opposite helicities of a  longitudinally-polarized electron beam (L) on a spin-zero target (U) can be written as:        
 \begin{eqnarray}
 \label{eq:coh_BSA}
A_{LU}& =& \frac{x_A(1+\epsilon^2)^2}{y} \, s_1^{INT} \sin(\phi) \, 
\bigg/ \, \bigg[ \, \sum_{n=0}^{n=2}c_n^{BH}\cos{(n\phi)} +  \\
& & \frac{x_A^2 t {(1+\epsilon^2)}^2}{Q^2} P_1(\phi) P_2(\phi) \, c_0^{DVCS} + 
\frac{x_A (1+\epsilon^2)^2}{y} \sum_{n=0}^{n=1} c_n^{INT} \cos{(n\phi)} \bigg].  \nonumber 
\end{eqnarray}
where $\mathcal{P}_1(\phi)$ and $\mathcal {P}_2(\phi)$ are the Bethe-Heitler 
propagators. The factors: $c_{0,1,2}^{BH}$, $c_0^{DVCS}$, $c_{0,1}^{INT}$ and 
$s_1^{INT}$ are the Fourier coefficients of the BH, the DVCS and the 
interference amplitudes for a spin-zero target 
\cite{Kirchner:2003wt,Belitsky:2008bz}.
The beam-spin asymmetry ($A_{LU}$) can be rearranged as
\begin{equation}
A_{LU}(\phi) = \frac{\alpha_{0}(\phi) \, \Im m(\mathcal{H}_{A})}
{\alpha_{1}(\phi) + \alpha_{2}(\phi) \, \Re e(\mathcal{H}_{A}) + \alpha_{3}(\phi) \, 
\big( 
\Re e(\mathcal{H}_{A})^{2} + \Im m(\mathcal{H}_{A})^{2} \big)}
\label{eq:A_LU-coh}
\end{equation}
where $\Im m(\mathcal{H}_{A})$ and $\Re e(\mathcal{H}_{A})$ are the imaginary and real parts of the CFF $\mathcal{H}_{A}$ associated to the GPD $H_A$. The $\alpha_{i}$'s are $\phi$-dependent kinematical factors that depend on the nuclear form factor $F_A$ and the independent variables $Q^2$, $x_{B}$ and $t$. These factors are simplified as:
\small
\begin{eqnarray}
   \alpha_0 (\phi) & = &\frac{x_{A}(1+\epsilon^2)^2}{y} S_{++}(1) \sin(\phi)\\
    \alpha_1 (\phi) & = & c_0^{BH}+c_1^{BH} \cos({\phi})+c_2^{BH} \cos(2\phi)\\ 
   \alpha_2 (\phi) & = & \frac{x_{A}(1+\epsilon^2)^2}{y}  \left( C_{++}(0) +  
C_{++}(1) \cos(\phi) \right)\\
\alpha_3 (\phi) &=& \frac{x^{2}_{A}t(1+\epsilon^2)^2}{y} {\mathcal P}_1(\phi) 
{\mathcal P}_2(\phi) \cdot 2 \frac{2-2y+y^2 + \frac{\epsilon^2}{2}y^2}{1 + 
\epsilon^2}
\end{eqnarray}
\normalsize
Where $S_{++}(1)$, $C_{++}(0)$, and $C_{++}(1)$ are the Fourier harmonics in 
the leptonic tensor. Their explicit expressions can be found in Appendix 
\ref{app:Helium_cross_section}. Using the $\alpha_{i}$ factors, one can obtain 
in a model-independent way $\Im m(\mathcal{H}_{A})$ and $\Re 
e(\mathcal{H}_{A})$ from fitting the experimental $A_{LU}$ as a function of 
$\phi$ for given values of $Q^2$, $x_B$ and $t$.

From a practical point of view, to access the quarks' density distributions 
of the target, we need to extract $H_{A}$ from the CFF $\mathcal{H}_{A}$, that 
appear directly in the cross section expressions. With the assumption that the 
anti-quark contribution is small in our kinematical region, $H_A(\xi,\xi,t) = 
\Im m(\mathcal{H}_{A})$ is a good approximation. Reference 
\cite{Guidal:2013rya} suggests that it is a 10$\%$ to 20$\%$ correction that 
can be applied to the data. 

\subsection{Experimental Status}
\label{sec:expover}
\begin{figure}[htb]
  \begin{center}
    \includegraphics[width=0.65\linewidth]{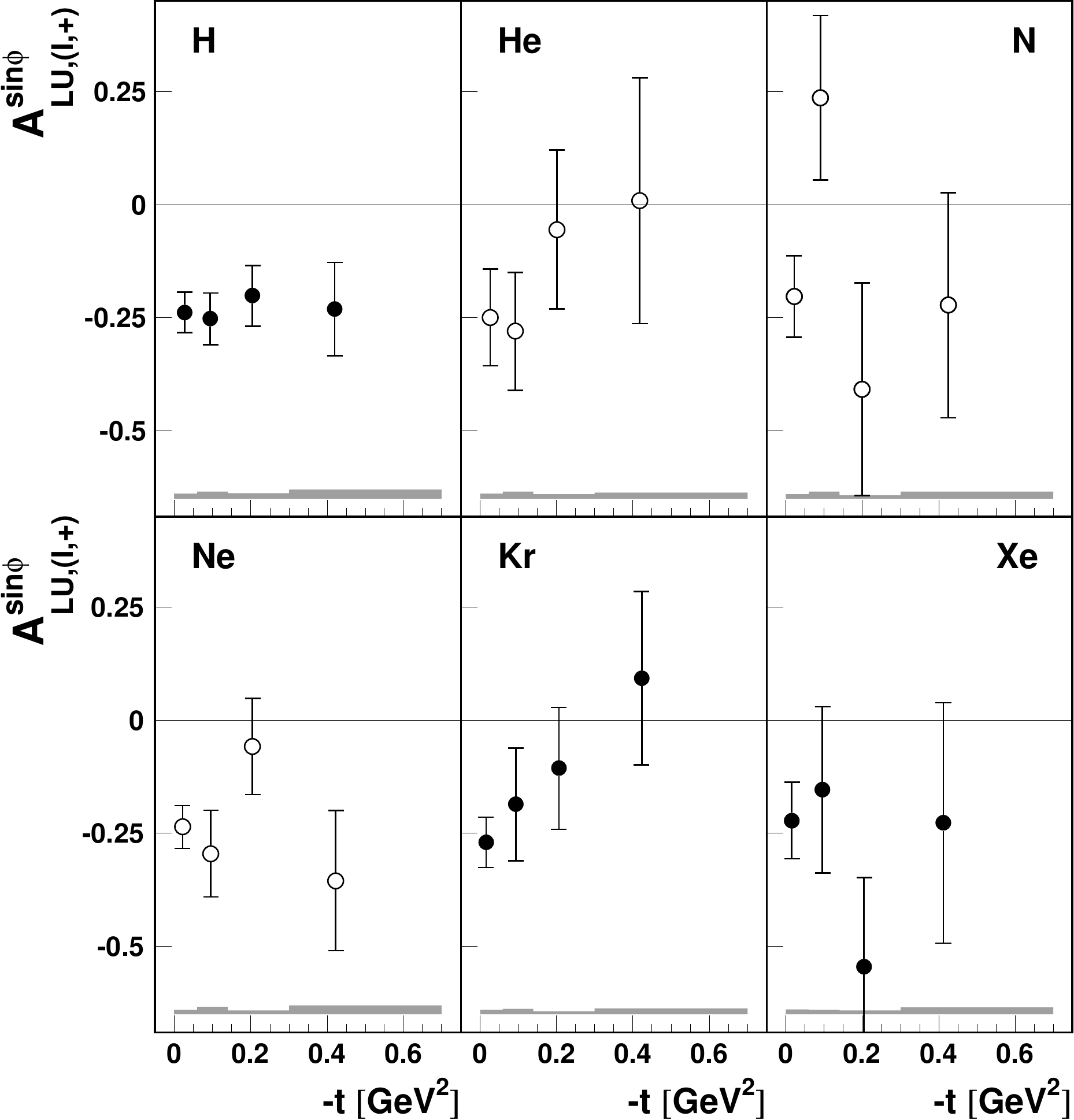}
    \caption{The $t$-dependence of the BSA on H, He, N, Ne, Kr, and Xe expressed in 
      terms of the coefficient $A_{LU}^{\sin(\phi)}$ of the $\sin(\phi)$ 
      contribution to $A_{LU}$~\cite{Airapetian:2009cga}; we note that in the 
      context of the HERMES fitting procedure $A_{LU}^{\sin(\phi)} \equiv A_{LU}$ 
    i.e. the denominator of Eq.~(\ref{eq:A_LU-coh}) was neglected.}
    \label{fig:her-alu-h-d-he-n}
  \end{center}
\end{figure}

The study of coherent nuclear DVCS is still in its infancy due to the 
challenging detection of the low energy recoil nucleus. The deuterium was 
investigated at HERMES~\cite{Airapetian:2009cga} and JLab Hall 
A~\cite{:2007vj}, and the HERMES experiment was the only one to study heavier 
nuclei ($^4$He, N, Ne, Kr, and Xe)~\cite{Airapetian:2009cga}. In the latter, 
the DVCS process was measured by identifying the scattered lepton and the real 
photon in the forward spectrometer. Sizable asymmetries 
(Figure~\ref{fig:her-alu-h-d-he-n}) have been reported in the missing mass region 
-1.5$< M_X <$ 1.7~GeV mass, while they generally vanish at higher 
masses~\cite{Airapetian:2009cga}.
%
%
These asymmetries are further separated into coherent and incoherent 
asymmetries taking advantage of the different $t$-dependence of the 
electromagnetic form factors: in the $^{4}$He case, for example, the coherent 
channel was assumed to dominate below -$t$~=~0.05~GeV$^2$. The selection of the 
different regions in $t$ (below and above) is then used to define coherent 
enriched and incoherent enriched data samples. The $A$-dependence of the ratio 
of the nuclear BSA to the proton BSA, over all the measured nuclei, is reported 
to be 0.91$\pm$0.19. Within the precision of the measurements, no obvious 
$A$-dependence of the BSA is observed: the coherent enriched ratio is 
compatible with unity, which is contradicting the predictions of 
different models~\cite{Liuti:2005gi,Guzey:2003jh,Belitsky:2000vk}. The 
incoherent enriched ratio, 0.93$\pm$0.23, is also compatible with unity as one 
would expect from an impulse approximation approach~\cite{Guzey:2003jh}.

\begin{figure}[H]
\centering
\includegraphics[scale=0.8]{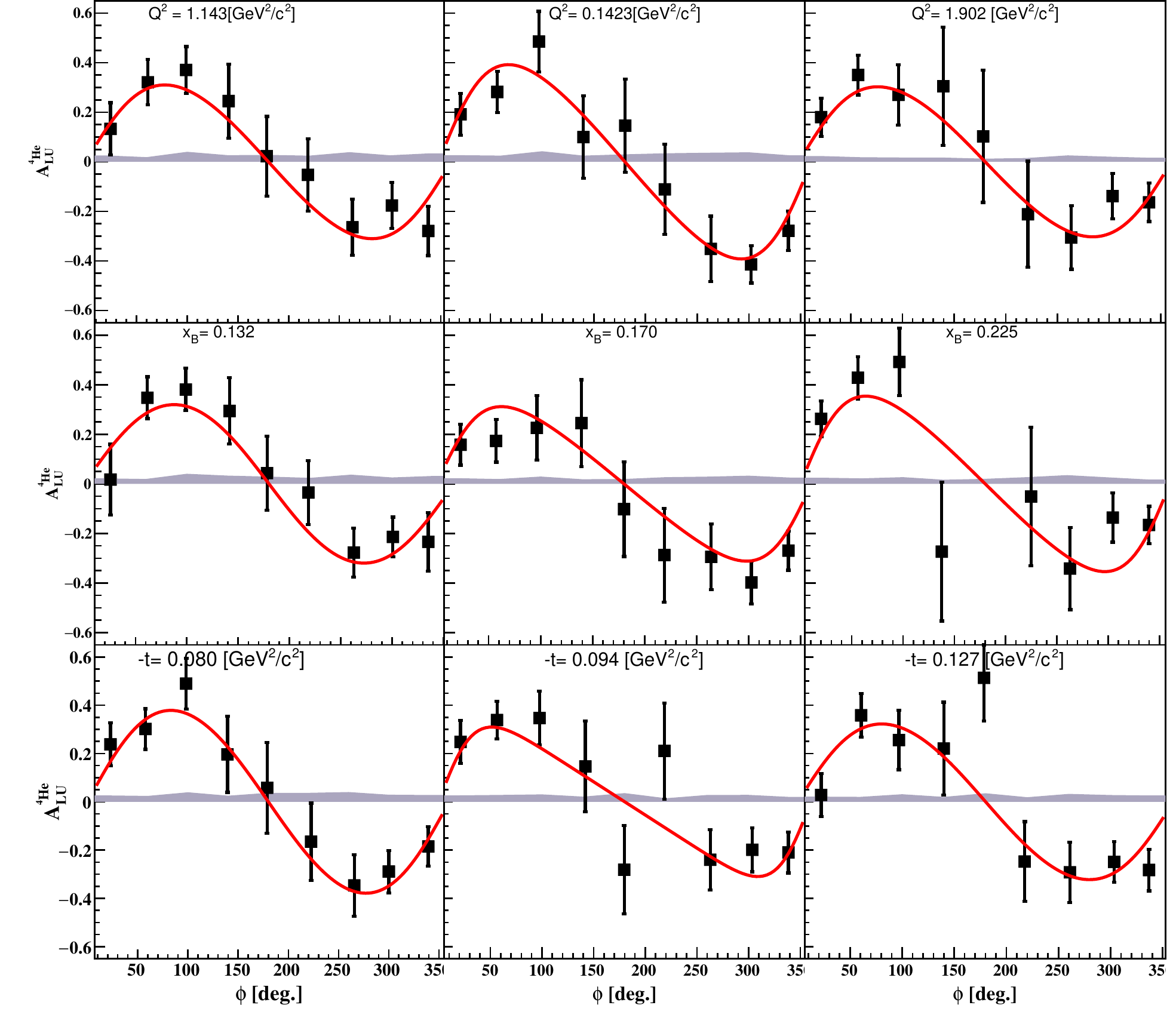}
\caption{[PRELIMINARY] The measured coherent $^4$He DVCS $A_{LU}$, from EG6 
   experiment, as a function of $\phi$ and Q$^2$ (top panel), $x_{B}$ (middle 
   panel), and $-t$~(bottom panel) bins \cite{eg6_note}. The error bars 
   represent the statistical uncertainties. The gray bands represent the 
   systematic uncertainties. The red curves are the results of the fits with 
   the form of equation \ref{eq:A_LU-coh}.}
\label{fig:bsa_coh_q2_bins}
\end{figure}

The CLAS collaboration has performed a new measurement (E08-024) of 
coherent exclusive DVCS on $^4$He, where all the products of the reaction have 
been detected including the low energy recoil $^4$He nucleus. This measurement was 
possible due to the high luminosity available at JLab, the large acceptance of 
CLAS spectrometer enhanced with the inner calorimeter (IC) and the addition of 
the newly built GEM based radial time projection chamber (RTPC). The IC was 
used to extend the photon detection to smaller angles and the RTPC was used to 
detect the recoil $^4$He nucleus. The data analysis and the corresponding 
internal review by the CLAS collaboration are completed \cite{eg6_note}, and a 
first publication draft is being finalized. The results indicate that the 
collaboration has been successful in measuring the exclusive DVCS both for the 
coherent and incoherent channels. Figure~\ref{fig:bsa_coh_q2_bins} shows the 
BSA $A_{LU}$ as a function of the azimuthal angle $\phi$ for different bins in 
$Q^2$ (top panel), $x_B$ (middle panel) and $-t$ (lower panel). These 
asymmetries are sizable indicating a strong nuclear DVCS signal.

\begin{figure}[H]
  \centering\includegraphics[scale=0.8]{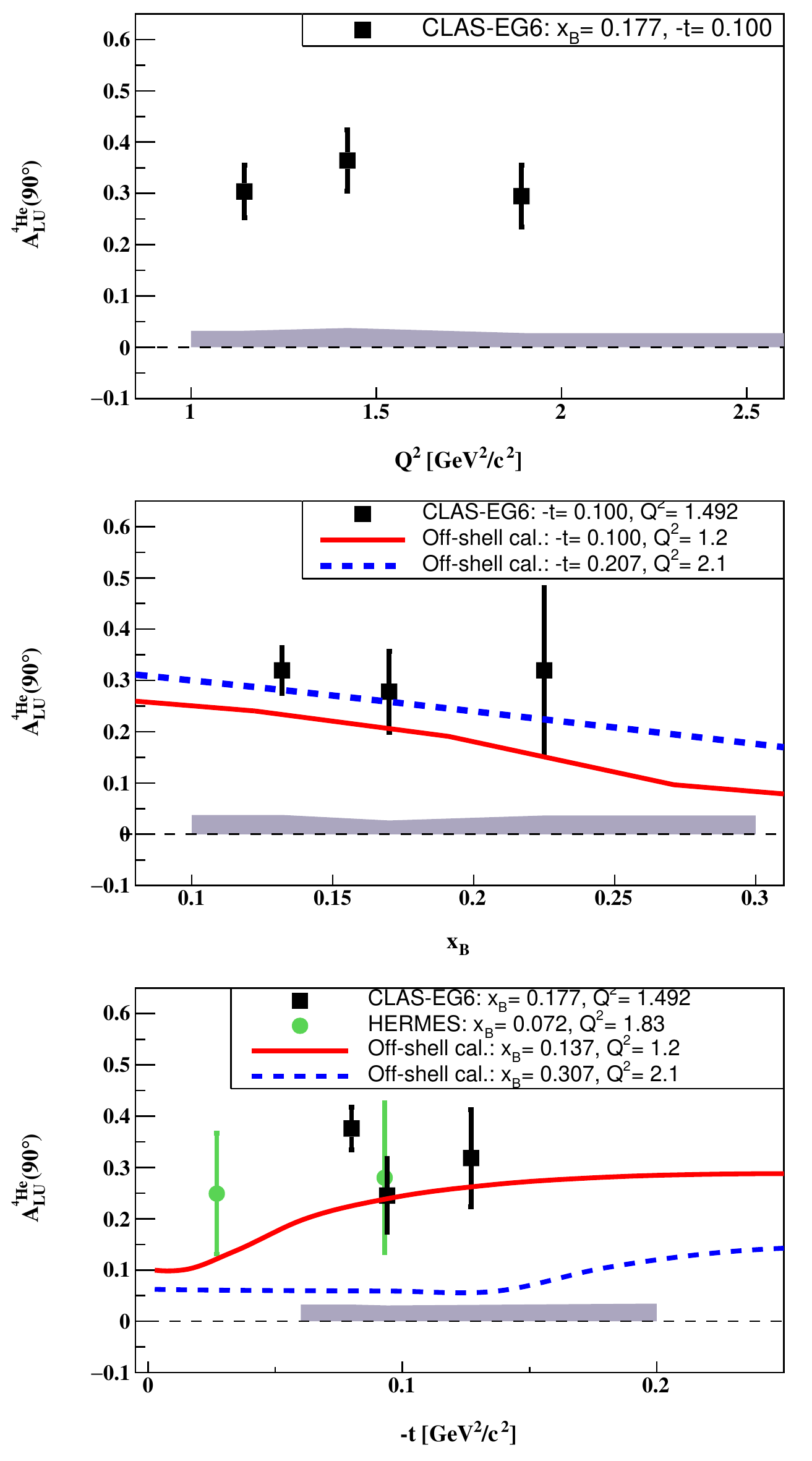} 
  \caption{[PRELIMINARY] From EG6 experiment, the $Q^{2}$-dependence (top panel), 
    the $x_{B}$ and the $t$-dependencies~(bottom panel) of the fitted coherent 
    $^4$He DVCS $A_{LU}$ asymmetry at $\phi$= 90$^{\circ}$ (black squares) 
    \cite{eg6_note}.  The curves are theoretical predictions from 
    \cite{Liuti:2005qj} for two values of $-t$. The green circles are the HERMES 
    $-A_{LU}$ (a positron beam was used) inclusive 
  measurements\cite{Airapetian:2009cga}.} \label{fig:coh_Q2_xB_t_ALU}
\end{figure}

Figure~\ref{fig:coh_Q2_xB_t_ALU} shows the $\sin\phi$ contribution to the 
coherent BSA $A_{LU}$, which also correspond to the coefficient 
$\frac{\alpha_0}{\alpha_1}$ in equation~\ref{eq:A_LU-coh} as a function of 
$Q^2$, $x_B$ and $-t$. It is clear the kinematic coverage and the statistics 
are limited, which made multidimensional binning impossible. Within the 
statistical uncertainties, CLAS data are in reasonable agreement with the 
model by Liuti et al.~\cite{Liuti:2005qj} for both the $x_B$ and $t$ 
dependencies, although a better comparison should be made with similar binning 
in $x_B$, $Q^2$ and $t$.  The Liuti at al. model includes dynamical 
off-shellness of the nucleons taking into account medium modifications beyond 
the conventional Fermi motion and binding effects, which are included in their 
spectral function. The model also appears to be consistently giving slightly 
smaller asymmetries than the data, which might indicate that some of the 
nuclear effects are still missing in this calculation. The CLAS measurements 
also agree with the HERMES data, considering HERMES large uncertainties.

\begin{figure}[H]
  \centering \includegraphics[scale=0.8]{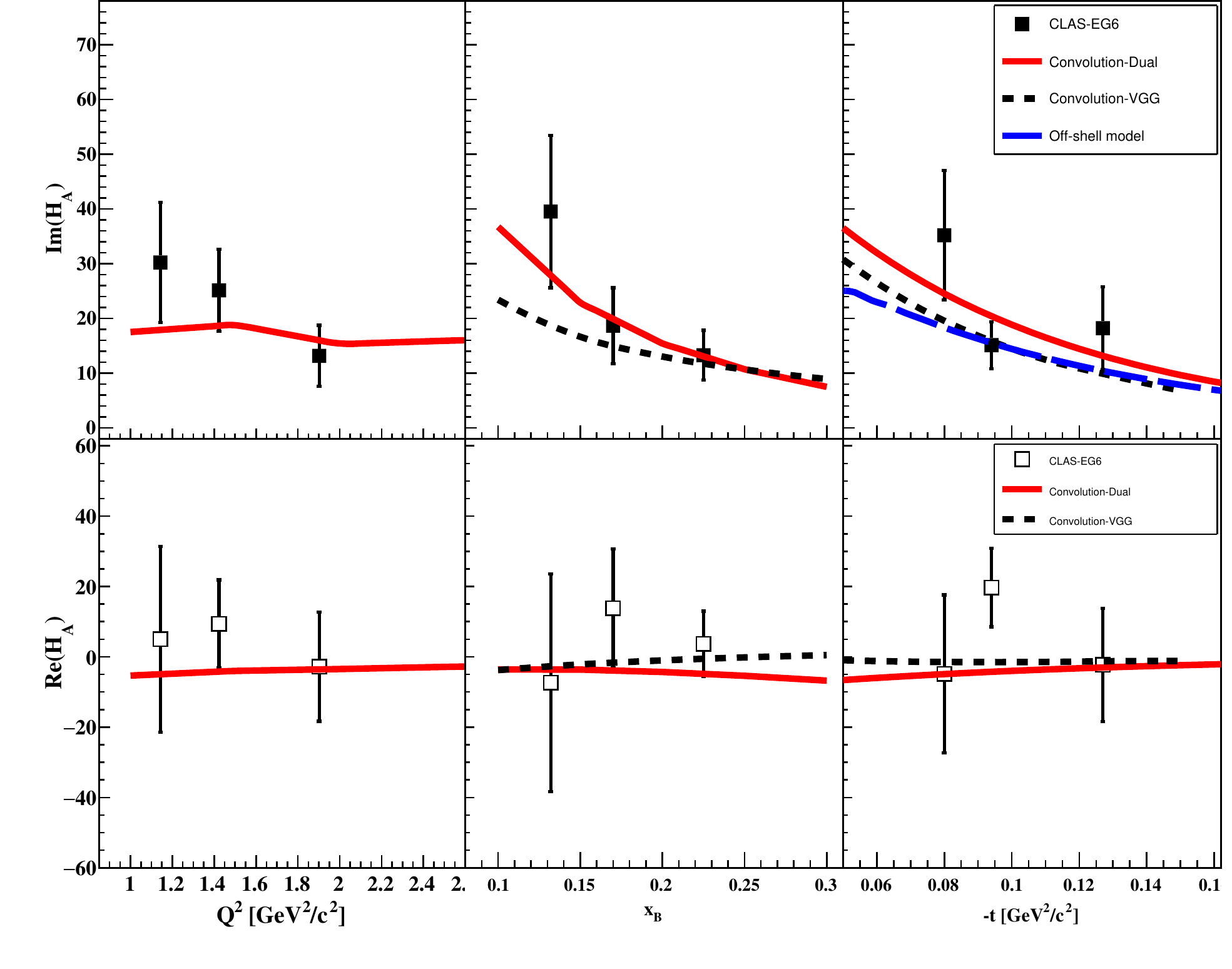}
  \caption{[PRELIMINARY] Model-independent extraction of the imaginary (top 
    panel) and real (bottom panel) parts of the $^4$He CFF $\mathcal{H}_A$, from 
    EG6 experiment \cite{eg6_note}, as functions of $Q^{2}$ (right panel), $x_B$ 
    (middle panel), and $t$ (left panel). The full red curves are calculations 
    based on an on-shell model from \cite{Guzey:2008th}.  The black-dashed 
    curves are calculations from a convolution model based on the VGG model for the 
    nucleons' GPDs \cite{Guidal:2004nd}. The blue long-dashed curve on the 
  top-right plot is from an off-shell model based on \cite{PhysRevC.88.065206}. }
  \label{fig:HA_CFF}
\end{figure}

As shown in equation \ref{eq:A_LU-coh}, one can extract both real and imaginary 
parts of the $^4$He CFF $\mathcal{H}_A$ from fitting the beam-spin asymmetry 
signals. This extraction is fully model-independent at leading twist and, in 
contrast with the proton's GPD extraction, does not necessitate any assumption on 
additional GPDs.  Figure \ref{fig:HA_CFF} presents the first ever experimental 
extraction of $\mathcal{H}_A$ from exclusive measurements as a function of 
$Q^{2}$, $x_B$, and $-t$. More theoretical effort is needed to develop 
predictions for $\mathcal{H}_A$. One can see a difference between the precision 
of the extracted real and imaginary parts, indicating the fact that the 
beam-spin asymmetry is mostly sensitive to the imaginary part of the  CFF 
$\mathcal{H}_A$.

These challenging CLAS measurements were a first step toward a promising 
program dedicated to nuclear QCD studies. With the 12 GeV upgrade and CLAS12 
augmented with the ALERT detector, exclusive nuclear DVCS and DVMP measurements 
in addition to tagged EMC and tagged DVCS experiments will allow our 
understanding of nuclear structure and nuclear effects to reach a new frontier.  

\section{Coherent \texorpdfstring{$\phi$}{phi} Production}\label{sec:DVMPFormalism}

\subsection{Accessing the Gluon GPD}\label{sec:sigLFormalism}

The gluon GPDs can be accessed in coherent $\phi$ production through a 
measurement of the longitudinal part of the differential cross section.
The gluon GPDs for the nucleon are related to the longitudinal differential 
cross-section for coherent vector meson production \cite{Girod:2012PR, 
Aktas:2005tz, Goloskokov:2007nt, Diehl:2005gn}:
\begin{equation}\label{eq:sigLproton}
  \frac{d\sigma_L}{dt} (\mathrm{proton}) = 
  \frac{\alpha_{em}}{Q^2}\frac{x_B^2}{1 - x_B}[(1 - \xi^2)|\langle H_g \rangle 
  |^2 + \mathrm{terms\,\,in} \langle E_g \rangle],
\end{equation}
where $\alpha_{em}$ is a QED coupling constant, $\xi$ is the skewness, and
the nucleon GPDs $H_g$ and $E_g$ are relatively unconstrained. Note that
the bracket notation $\langle H_g \rangle$ indicates an analog of the CFF
for the DVMP, see \cite{Favart:2015umi} for complete expressions. However for a 
spin-0 nucleus, such as $^4$He, with only one leading-twist gluon GPD, the 
extraction of the gluon GPD greatly simplifies
\begin{equation}\label{eq:sigLHe4}
  \frac{d\sigma_L}{dt} (^4\mathrm{He}) \propto |\langle H_g \rangle |^2.
\end{equation}
where $H_g$ is the only unknown on the right hand side.

The technique used to determine $\sigma_L$, which we quickly outline, is found 
in \cite{Schilling:1973ag,Schilling:1969um}. First, the angular distribution of 
the kaons decay is measured. This angular distribution is used to extract the 
spin-density matrix element. The angular distribution in the helicity frame of 
the vector meson is
\begin{eqnarray}\label{eqn:phi_r0400}
  W(\cos\theta_H) = \frac{3}{4}\left[(1 - r_{00}^{04}) + (3r_{00}^{04} -1) \cos^2 \theta_H\right]
\end{eqnarray}
where $r_{00}^{04}$ is a spin-density matrix element, and $\theta_H$ is the 
decay angle in the rest frame of the $\phi$ where the z-direction is aligned 
with the $\phi$ momentum in the center of momentum system.
Equation (\ref{eqn:phi_r0400}) is a result of s-channel helicity conservation 
and $r_{00}^{04}$ is extract by fitting its $\cos^2 \theta_H$ angular 
dependence. Next, the spin-density matrix element is used to determine the 
ratio $R = \sigma_L/\sigma_T$, which is the ratio of longitudinal to transverse 
cross-sections,
\begin{eqnarray}
  R  = \frac{r_{00}^{04}}{\epsilon(1-r_{00}^{04})},
\end{eqnarray}
where $\epsilon$ is the virtual photon polarization.

With $R$ determined from decay distribution of the vector meson, the measured 
differential cross-section is then used to extract the longitudinal part as
\begin{eqnarray}
  \frac{d\sigma_L}{dt} = \frac{1}{(\epsilon + 1/R)\Gamma(Q^2, x_B, 
  E)}\frac{d^3\sigma}{dQ^2 dx_B dt},
\end{eqnarray}
where $\Gamma$ is the virtual photon-flux.  Now with $d\sigma_L/dt$ extracted, 
we can use it in equation (\ref{eq:sigLHe4}) to study the gluon distribution in 
$^4He$.


\subsection{Experimental Status}

\begin{figure}[htb]
  \centering
  \includegraphics[width=0.5\textwidth,clip,trim=7mm 0mm 7mm 
  5mm]{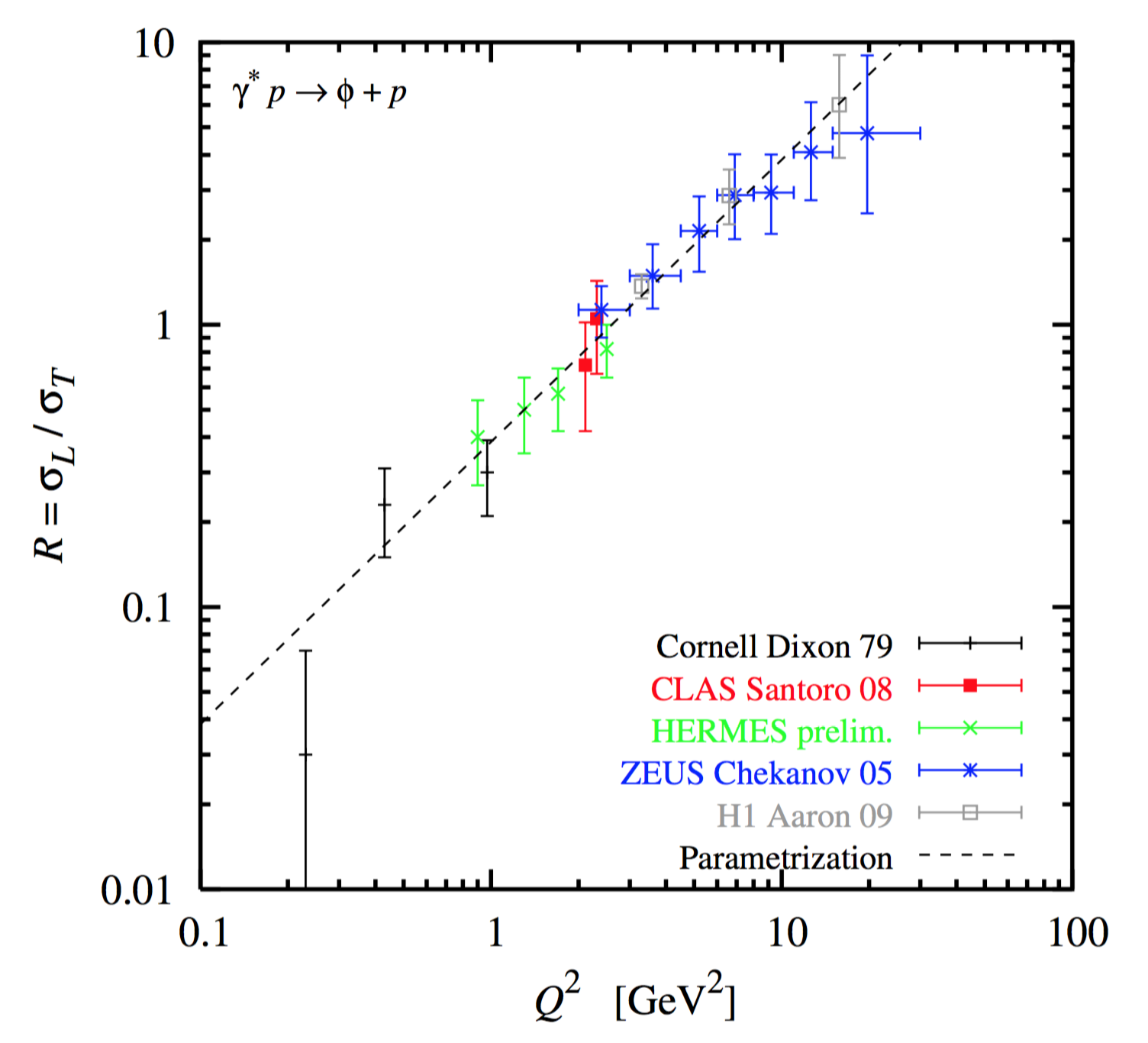}
  \caption{Figure come directly from PR12-12-007 \cite{Girod:2012PR}. The 
    parametrization of $R$ used to calculate $\phi$ production off a proton 
    target plotted vs Q$^2$ against world data. For more information on the 
    world data. See references: CLAS \cite{Lukashin:2001sh, Santoro:2008ai}, 
    Cornell \cite{Dixon:1978vy, Cassel:1981sx}, HERMES \cite{Borissov:2000zz}, 
    NMC \cite{Arneodo:1994id}, ZEUS \cite{Chekanov:2005cqa}, and H1 
  \cite{Aaron:2009xp}.} \label{fig:PR1207_verify}
\end{figure}

\begin{figure}[htb]
  \centering
  \includegraphics[width=0.48\textwidth,clip,trim=7mm 0mm 7mm 
  2mm]{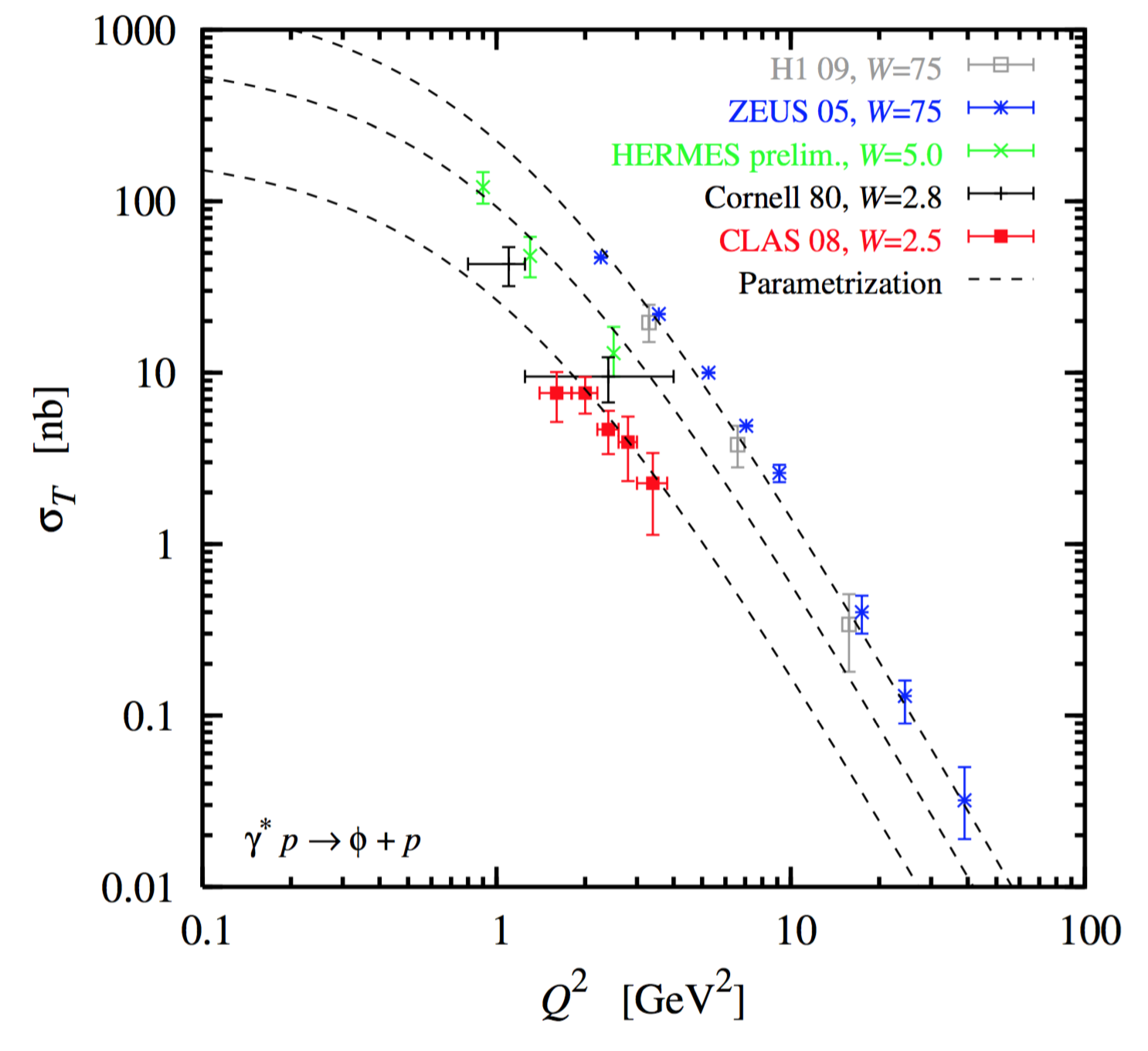}
  \includegraphics[width=0.48\textwidth,clip,trim=5mm 0mm 7mm 
  5mm]{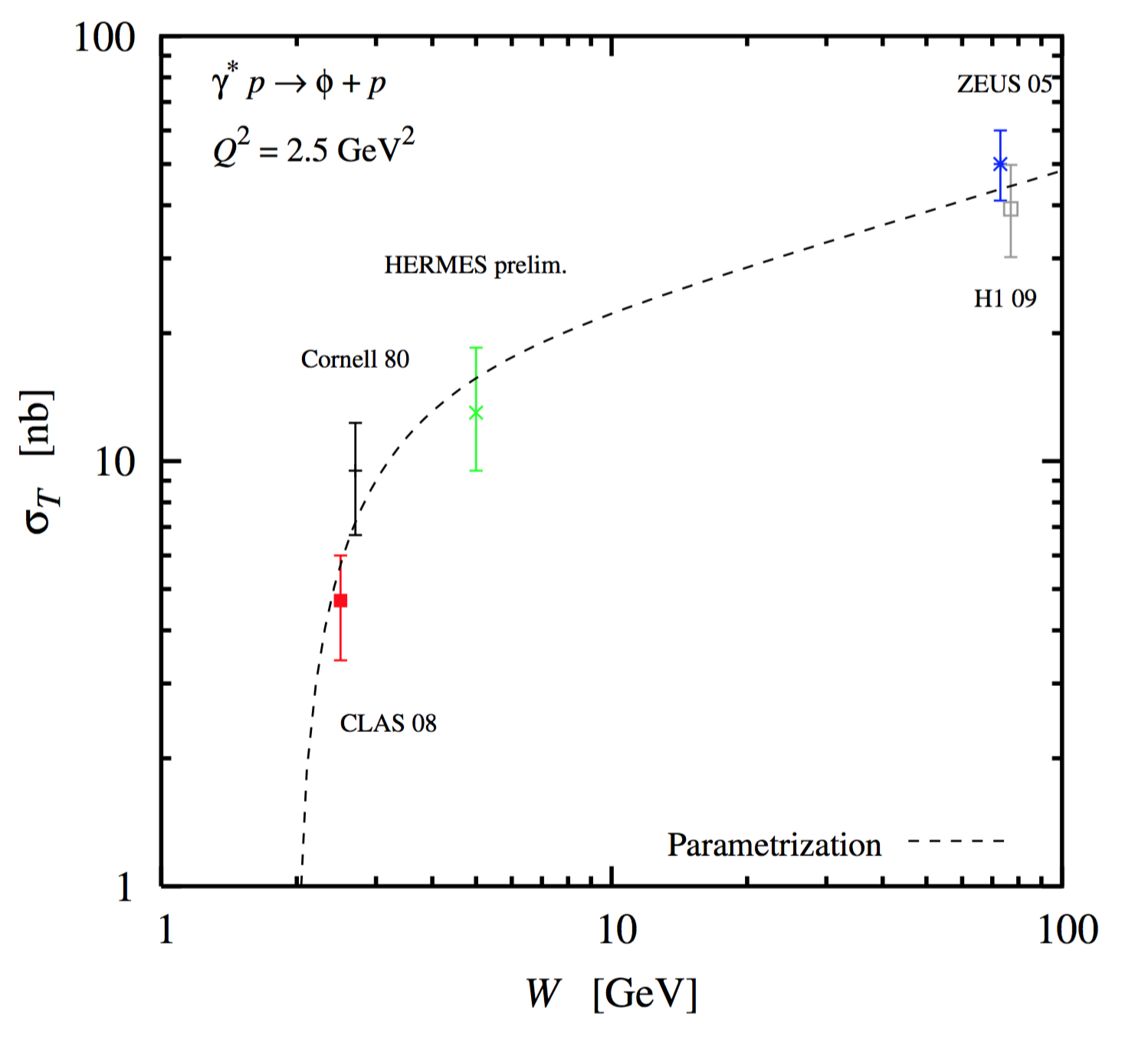}
  \caption{Figures come directly from PR12-12-007 \cite{Girod:2012PR}. The 
    parametrization in W and Q$^2$ used for cross-section calculation for $\phi$ 
    production off a proton target plotted against world data. For more 
    information on the world data, see references: CLAS \cite{Lukashin:2001sh, 
    Santoro:2008ai}, Cornell \cite{Dixon:1978vy, Cassel:1981sx}, HERMES 
    \cite{Borissov:2000zz}, NMC \cite{Arneodo:1994id}, ZEUS 
  \cite{Chekanov:2005cqa}, and H1 \cite{Aaron:2009xp}.  Reproduced from 
\cite{Girod:2012PR} \label{fig:PR1207_verify2}}
\end{figure}

Like the case with coherent DVCS, the experimental status of coherent $\phi$ 
production on nuclear targets is lacking. We are proposing the first 
measurement of exclusive electroproduction of the $\phi$ on $^4$He.  However, 
exclusive electroproduction on the nucleon does provide a very useful starting 
point. We will use the existing data on the proton, which is shown in 
Figures~\ref{fig:PR1207_verify} and~\ref{fig:PR1207_verify2}, to build up a 
reasonable model (see \ref{sec:phiEG}) which can be used to estimate production 
rates.

%

\setlength\parskip{\baselineskip}%
\chapter{Experimental Setup}
\label{chap:setting}
All the different measurements of the ALERT run group require, in addition to 
a good scattered electron measurement, the detection of low energy nuclear 
recoil fragments with a large kinematic coverage. Such measurements have been performed 
in CLAS (BONuS and eg6 runs), where the adequacy of a small additional detector
placed in the center of CLAS right around the target has shown to be the best 
solution. We propose here a similar setup using the CLAS12 spectrometer 
augmented by a low energy recoil detector. 

We summarize in Table~\ref{tab:req} the requirements for the different 
experiments proposed in the run group. By comparison with previous similar 
experiments, the proposed tagged measurements necessitate a 
good particle identification. Also, CLAS12 will be able to handle higher 
luminosity than CLAS so it will be key to exploit this feature in the future 
setting in order to keep our beam time request reasonable.

\begin{table}[ht!]
\centering
\footnotesize
\begin{tabu}{lccc}
\tabucline[2pt]{-}
Measurement  & Particles detected & $p$ range       & $\theta$ range                \rule[-7pt]{0pt}{20pt} \\
\tabucline[1pt]{-}                                                   
Nuclear GPDs & $^4$He             & $230 < p < 400 MeV/c$ & $\pi/4 < \theta < \pi/2$ rad  \rule[-7pt]{0pt}{20pt} \\
Tagged EMC   & p, $^3$H, $^3$He   & As low as possible    & As close to $\pi$ as possible \rule[-7pt]{0pt}{20pt} \\
Tagged DVCS  & p, $^3$H, $^3$He   & As low as possible    & As close to $\pi$ as possible \rule[-7pt]{0pt}{20pt} \\
\tabucline[2pt]{-}
\end{tabu}
\caption{Requirements for the detection of low momentum spectator fragments of the proposed measurements.}
\label{tab:req}
\end{table}

This chapter will begin with a brief description of CLAS12.  
After presenting the existing options for recoil detection and recognize that 
they will not fulfill the needs laid out above, we will describe the design of
the proposed new recoil detector ALERT. We will then present the reconstruction 
scheme of ALERT and show the first prototypes built by our technical 
teams. Finally, we specify the technical contributions of the different 
partners.

%
\section{The CLAS12 Spectrometer}
\begin{figure}
  \begin{center}
    \includegraphics[angle=0, width=0.75\textwidth]{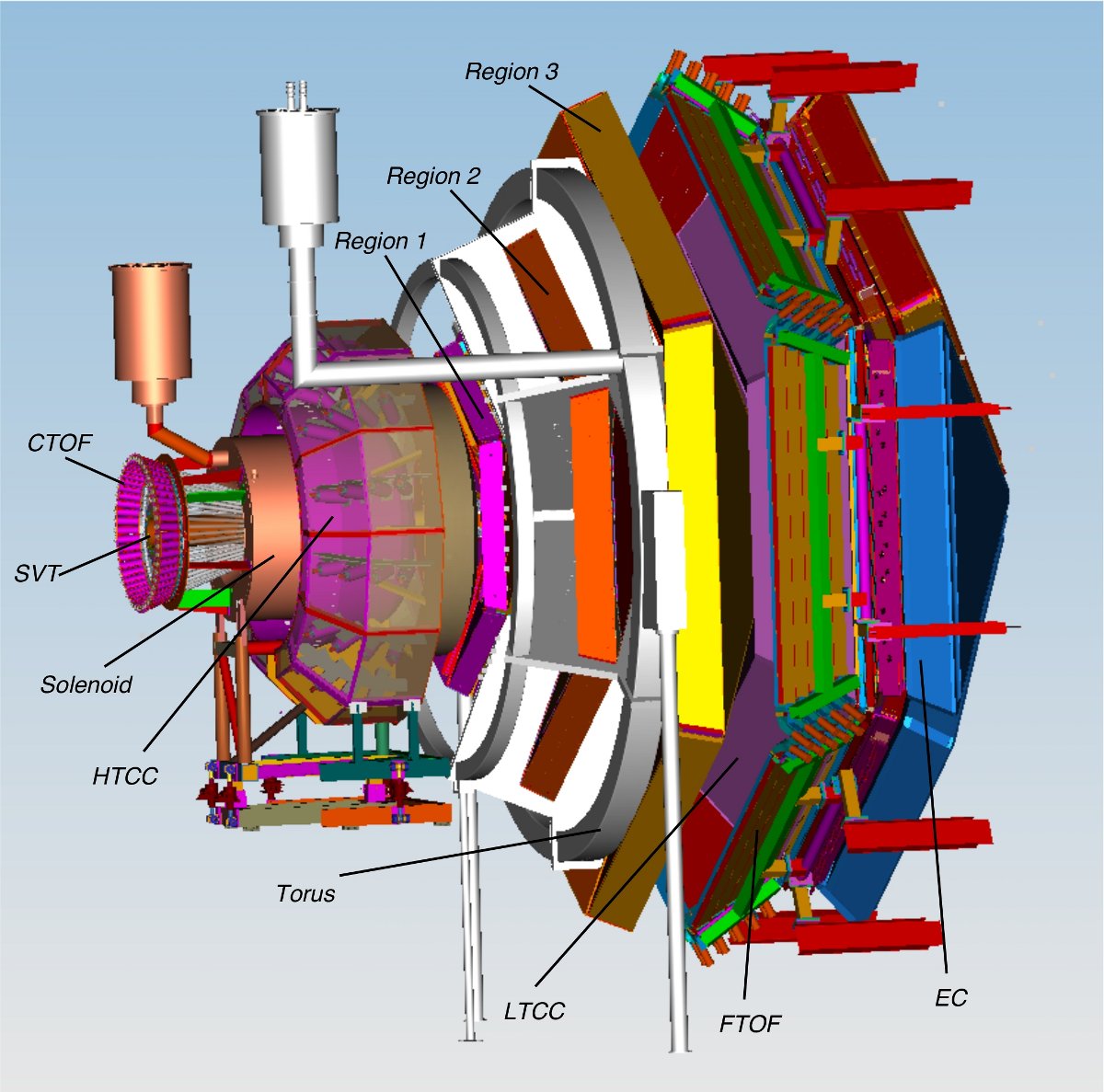}
    \caption{The schematic layout of the CLAS12 baseline design.}
    \label{fig:fd}
  \end{center}
\end{figure}

The CLAS12 detector is designed to operate with 11~GeV beam at an 
electron-nucleon luminosity of $\mathcal{L} = 
1\times10^{35}~$cm$^{-2}$s$^{-1}$. The baseline configuration of the CLAS12 
detector consists of the forward detector and the central detector 
packages~\cite{CD} (see Figure~\ref{fig:fd}). We use the forward detector
for electron detection in all ALERT run group proposals, while DVCS centered
proposals also use it for photon detection. The central
detector's silicon tracker and micromegas will be removed to leave room for
the recoil detector. 

The scattered electrons and photons will be detected in the forward detector which consists 
of the High Threshold Cherenkov Counters (HTCC), Drift Chambers (DC), the Low 
Threshold Cherenkov Counters (LTCC), the Time-of-Flight scintillators (TOF), 
the Forward Calorimeter and the Preshower Calorimeter. The charged particle 
identification in the forward detector is achieved by utilizing the combination 
of the HTCC, LTCC and TOF arrays with the tracking information from the Drift 
Chambers. The HTCC together with the Forward Calorimeter and the Preshower 
Calorimeter will provide a pion rejection factor of more than 2000 up to a 
momentum of 4.9~GeV/c, and a rejection factor of 100 above 4.9 GeV/c. The photons
are detected using the calorimeters.

\section{Available options for a Low Energy Recoil Detector}
We explored available solutions for the low-energy recoil tracker with 
adequate momentum and spatial resolution, and good particle identification for 
recoiling light nuclei (p, $^3$H and $^3$He). After investigating the 
feasibility of the proposed measurements using the CLAS12 Central Detector and 
the BONuS Detector~\cite{bonus6,bonus12}, we concluded that we needed to build 
a dedicated detector. We summarize in the
following the facts that led us to this conclusion.
\subsection{CLAS12 Central Detector}
The CLAS12 Central Detector~\cite{CD} is designed to detect various charged 
particles over a wide momentum and angular range. The main detector package 
includes:
\begin{itemize}
 \item Solenoid Magnet: provides a central longitudinal magnetic field up to 
5~Tesla, which serves to curl emitted low energy M{\o}ller electrons and determine 
particle momenta through tracking in the central detector.
 \item Central Tracker: consists of 3 double layers of silicon strips and 6 
    layers of Micromegas. The thickness of a single silicon layer is  
    \SI{320}{\um}.
 \item Central Time-of-Flight: an array of scintillator paddles with a 
cylindrical geometry of radius 26 cm and length 50 cm; the thickness of the 
detector is 2 cm with designed timing resolution of $\sigma_t = 50$ ps, used 
to separate pions and protons up to 1.2 GeV/$c$.
\end{itemize}

The current design, however, is not optimal for low energy particles 
($p<300$~MeV/$c$) due to the energy loss in the first 2 silicon strip layers. 
The momentum detection threshold is $\sim 200$ MeV/$c$ for protons, $\sim 
350$~MeV/$c$ for deuterons and even higher for $^3$H and $^3$He. These values 
are significantly too large for any of the ALERT run group proposals.

\subsection{BONuS12 Radial Time Projection Chamber}
The original BONuS detector was built for Hall B experiment E03-012 to study 
neutron structure at high $x_B$ by scattering electrons off an almost on-shell 
neutron inside deuteron. The purpose of the detector was to tag the low energy 
recoil protons ($p>60$ MeV/$c$). The key component for detecting the slow 
protons was the Radial Time Projection Chamber (RTPC) based on Gas Electron 
Multipliers (GEM). A later run period (eg6) used a 
newly built RTPC with a new design to detect recoiling $\alpha$ particles in 
coherent DVCS scattering. The major improvements of the eg6 RTPC were full 
cylindrical coverage and a higher data taking rate.

The approved 12~GeV BONuS (BONuS12) experiment is planning to use a similar 
device with some upgrades. The target gas cell length will be doubled, and the 
new RTPC will be longer as well, therefore doubling the luminosity and 
increasing the acceptance. Taking advantage of the larger bore ($\sim 700$ mm) of 
the 5~Tesla solenoid magnet, the maximum radial drift length will be increased 
from the present 3 cm to 4 cm, improving the momentum resolution by 
50\%~\cite{bonus12} and extending the momentum coverage. The main features of 
the proposed BONuS12 detector are summarized in Table~\ref{tab:comp}.

\begin{table}[tbp]
\bgroup
\def\arraystretch{1.1}%
\tabulinesep=1.2mm
\begin{tabu}{lcc}
\tabucline[2pt]{-}                                                   
\textbf{Detector Property}  & \textbf{RTPC}        & \textbf{ALERT}\\
\tabucline[1pt]{-}                                                   
Detection region radius & 4 cm                & 5 cm\\
Longitudinal length & $\sim$ 40 cm         & $\sim$ 30 cm \\
Gas mixture         & 80\% helium/20\% DME & 90\% helium/10\% isobutane \\
Azimuthal coverage  & 360$^{\circ}$               & 340$^{\circ}$\\
Momentum range      & 70-250 MeV/$c$ protons & 70-250 MeV/$c$ protons\\
Transverse mom. resolution & 10\% for 100~MeV/c protons & 10\% for 100~MeV/c protons\\
$z$ resolution & 3~mm & 3~mm \\
Solenoidal field    & $\sim 5$ T           & $\sim 5$ T \\
ID of all light nuclei & No                    & Yes \\
Luminosity      &$3\times10^{33}$ nucleon/cm$^{2}$\!/s & $6\times10^{34}$ nucleon/cm$^{2}$\!/s\\
Trigger             & can not be included  & can be included \\
\tabucline[1pt]{-}                                                   
\end{tabu}
\egroup
\caption{\label{tab:comp}Comparison between the RTPC (left column) and the new tracker (right column).}
\end{table}

\begin{figure}
  \begin{center}
    \includegraphics[angle=0, width=0.5\textwidth]{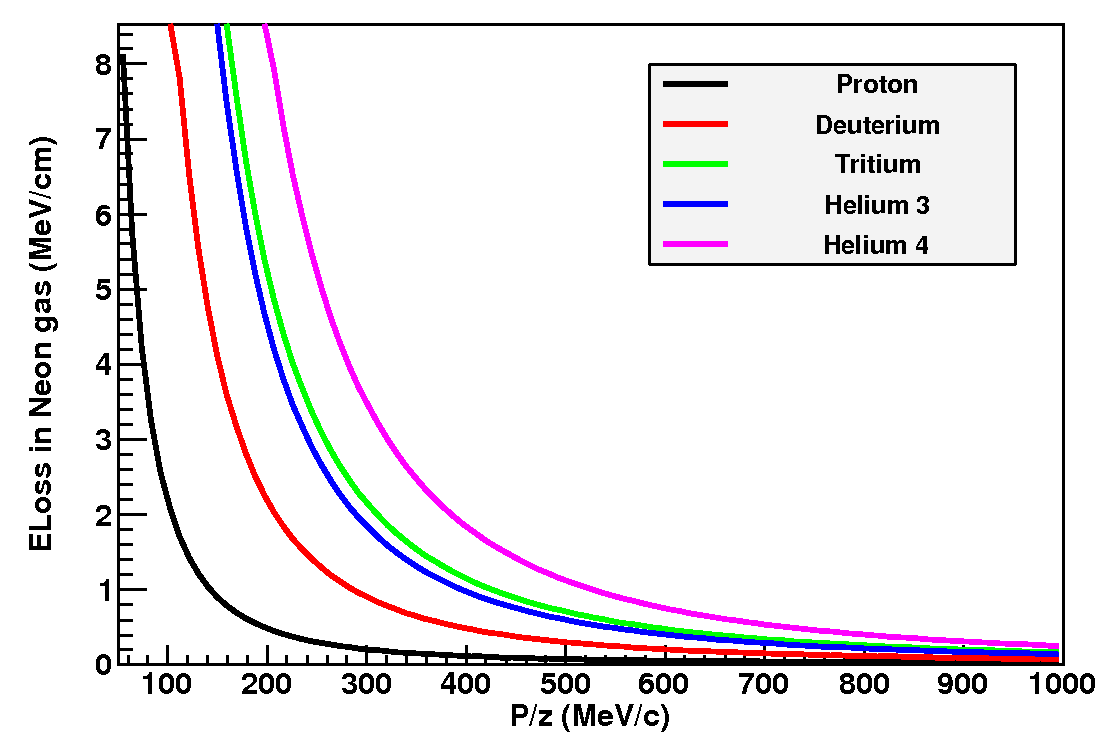}
    \caption{Calculation of energy loss in Neon gas as a function of the particle 
momentum divided by its charge for different nuclei. }
    \label{fig:eloss}
  \end{center}
\end{figure}

In principle, particle identification can be obtained from the RTPC through the 
energy loss $dE/dx$ in the detector as a function of the particle momentum (see 
Figure~\ref{fig:eloss}). However, with such a small difference between $^3$H and 
$^3$He, it is nearly impossible to discriminate between them
on an event by event basis because of the intrinsic width of the $dE/dx$ 
distributions. This feature is not problematic when using deuterium target, 
but makes the RTPC no longer a viable option for our tagged EMC and tagged DVCS 
measurements which require a $^4$He target and the differentiation of $^4$He, 
$^3$He, $^3$H, deuterons and protons.

Another issue with the RTPC is its slow response time due to a long drift 
time ($\sim5~\mu$s). If a fast recoil detector could be included in the trigger 
it would have a significant impact on the background rejection. Indeed, in
about 90\% of DIS events on deuteron or helium, the spectator fragments have too low energy 
or too small angle to get out of the target and be detected. By including
the recoil detector in the trigger, we would not be recording these events anymore.
Since the data acquisition speed was the main limiting factor for 
both BONuS and eg6 runs in CLAS, this 
would be a much needed reduction of the pressure on the DAQ.

\subsection{Summary}
In summary, we found that the threshold of the CLAS12 inner 
tracker is significantly too high to be used for our measurements. On the other hand, the 
recoil detector planned for BONuS12, a RTPC, is not suitable due to its 
inability to distinguish all kind of particles we need to measure.  
Moreover, as the RTPC cannot be efficiently included in the trigger, a lot of  
background events are sent to the readout electronics, which will cause its saturation and 
limit the maximum luminosity the detector can handle. Therefore, we propose 
a new detector design.

\section{Design of the ALERT Detector}
We propose to build a low energy recoil detector consisting of two sub-systems: 
a drift chamber and a scintillator hodoscope.
The drift chamber will be composed of 8 layers of sense wires to provide tracking 
information while the scintillators will provide particle 
identification through time-of-flight and energy measurements. To reduce the 
material budget, thus reducing the threshold to detect recoil particles 
at as low energy as possible, the scintillator 
hodoscope will be placed inside the gas chamber, just outside of the last 
layer of drift wires.

The drift chamber volume will be filled with a light gas mixture (90\% He and 
10\% C$_4$H$_{10}$) at atmospheric pressure. The amplification potential will
be kept low enough in order to not be sensitive to relativistic particles 
such as electrons and pions. Furthermore, a light 
gas mixture will increase the drift speed of the electrons from 
ionization. This will allow the chamber to withstand higher rates and 
experience lower hit occupancy. The fast signals from the chamber and 
the scintillators will be used in coincidence with electron trigger 
from CLAS12 to reduce the overall DAQ trigger rate and 
allow for operation at high luminosity.

\begin{figure}[tbp]
  \begin{center}
    \includegraphics[angle=0, width=0.75\textwidth]%
                    {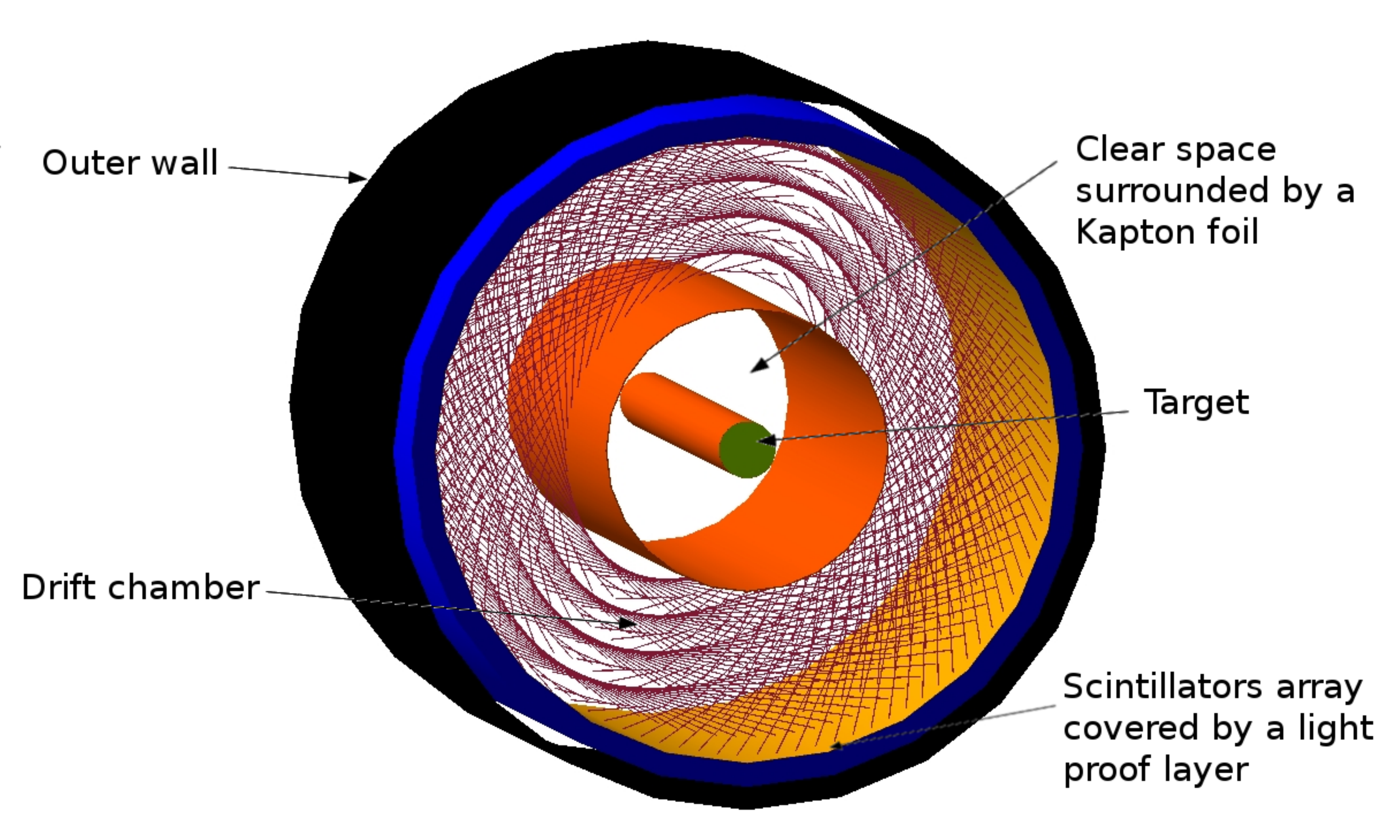}
    \caption{The schematic layout of the ALERT detector design, viewed 
from the beam direction.}
    \label{fig:new_lay}
  \end{center}
\end{figure}

The detector is designed to fit inside the central TOF of CLAS12; the 
silicon vertex tracker and the micromegas vertex tracker (MVT) will be 
removed. The available space has thus an outer radius of slightly more 
than 20~cm. A schematic 
layout of the preliminary design is shown in Figure~\ref{fig:new_lay} and its
characteristics compared to the RTPC design in Table~\ref{tab:comp}. The 
different detection elements are covering about $340^{\circ}$ of the polar 
angle to leave room for mechanics, and are 30~cm long with an effort made to 
reduce the particle energy loss through the materials. From the inside out,
it is composed of:
\begin{itemize}
\item a 30~cm long cylindrical target with an outer radius of 6~mm and target 
   walls \SI{25}{\um} Kapton filled with 3~atm of helium;
\item a clear space filled with helium to reduce secondary scattering from
   the high rate M\o{}ller electrons with an outer radius of 30~mm;
\item the drift chamber, its inner radius is 32~mm and its outer radius is 
85~mm;
\item two rings of plastic scintillators placed inside the gaseous chamber, 
   with total thickness of roughly 20~mm.
\end{itemize}

\subsection{The Drift Chamber}
While drift chambers are very useful to cover large areas at a moderate price, 
huge progress has been made in terms of their ability to withstand higher rates 
using better electronics, shorter distance between wires and optimization of 
the electric field over pressure ratio. Our design is based on other chambers 
developed recently. For example for the dimuon arm of ALICE at CERN, drift 
chambers with cathode planes were built in Orsay~\cite{AliceMuonArmChamber}. 
The gap between sense wires is 2.1~mm and the distance between two cathode 
planes is also 2.1~mm, the wires are stretched over about 1~m. Belle II is 
building a cylindrical drift chamber very similar to what is needed for this 
experiment and for which the space between wires is around 
2.5~mm~\cite{BelleIItdr}. Finally, a drift chamber with wire gaps of 1~mm is 
being built for the small wheel of ATLAS at CERN~\cite{ATLASChamber}. The 
cylindrical drift chamber proposed for our experiment is 300~mm long, and we 
therefore considered that a 2~mm gap between wires is technically a rather 
conservative goal. Optimization is envisioned based on experience with 
prototypes. 

The radial form of the detector does not allow for 90 degrees x-y wires in the 
chamber. Thus, the wires of each layer are at alternating angle of $\pm$ 
10$^{\circ}$, called the stereo-angle, from the axis of the drift chamber.  We 
use stereo-angles between wires to determine the coordinate along the beam 
axis ($z$). This setting makes it possible to use a thin forward end-plate to 
reduce multiple scattering of the outgoing high-energy electrons. A rough 
estimate of the tension due to the $\sim$2600 wires is under 600~kg, 
which appears to be reasonable for a composite end-plate. 

\begin{figure}
  \begin{center}
    \includegraphics[angle=0, width=0.7\textwidth,clip, trim=0mm 0mm 4cm 10cm]{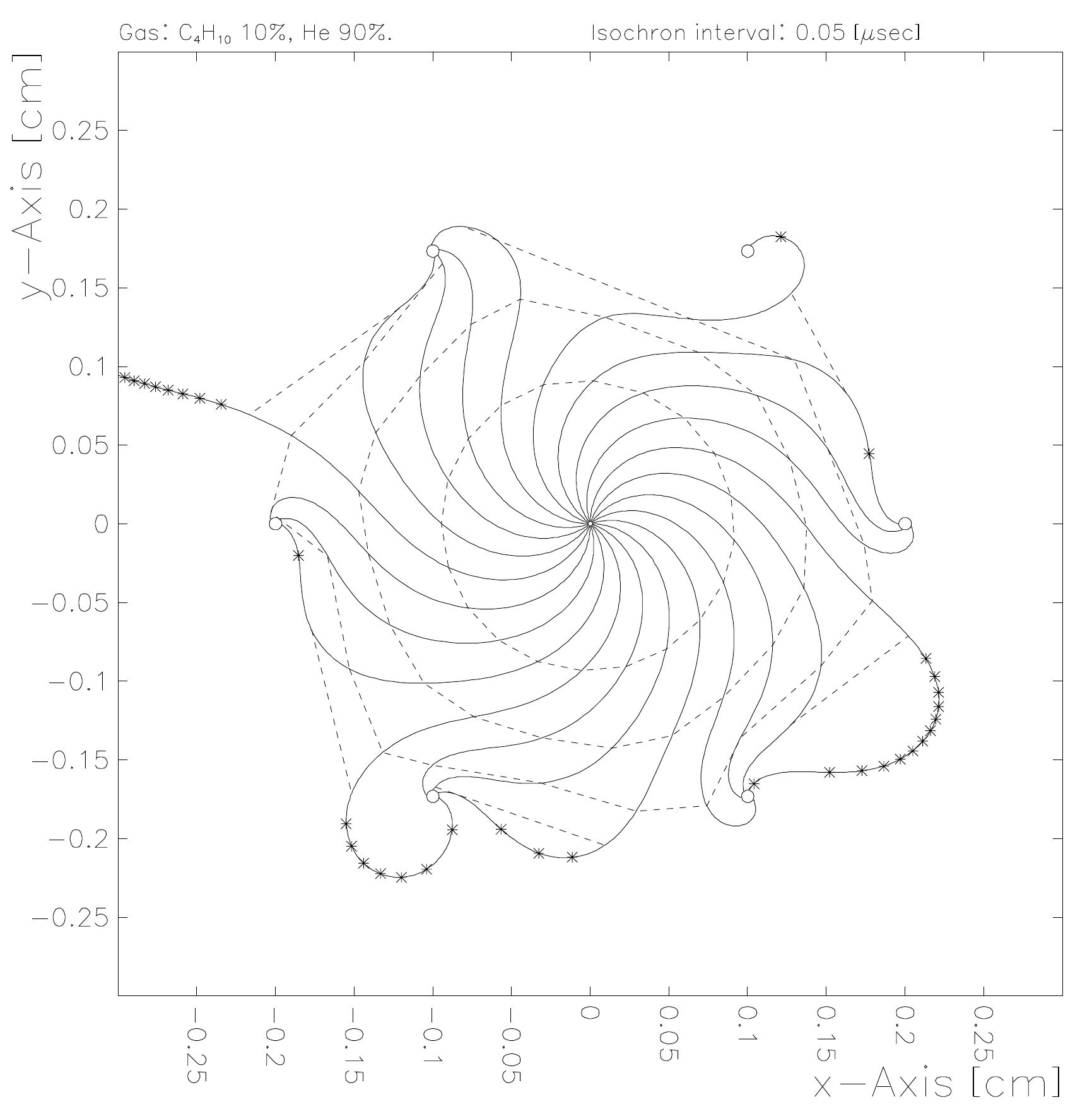}
    \caption{Drift lines simulated using MAGBOLTZ \cite{Magboltz} for one 
sense wire (at the center) surrounded by 6 field wires. The two electric field 
lines leaving the cell disappear when adjusting the voltages on the wires. 
Dashed lines are isochrones spaced by 50~ns. This shows that the maximum drift 
time is about 250~ns.}
    \label{fig:drift_cell}
  \end{center}
\end{figure}

The drift chamber cells are composed of one sense wire made of gold plated 
tungsten surrounded by field wires, however the presence of the 5~T magnetic 
field complicates the field lines. Several cell configurations have been studied with 
MAGBOLTZ~\cite{Magboltz}, we decided to choose a conservative 
configuration as shown in Figure~\ref{fig:drift_cell}. The sense wire is 
surrounded by 6 field wires placed equidistantly from it in a hexagonal 
pattern. The distance between the sense and field wires is constant and equal 
to 2~mm. Two adjacent cells share the field wires placed between them. The 
current design will have 8 layers of cells of similar radius. 
The simulation code MAGBOLTZ is calculating the drift speed and drift paths of 
the electrons (Figure~\ref{fig:drift_cell}). With a moderate electric field, the 
drift speed is around 10~microns/ns, the average drift time expected is thus 
250~ns (over 2~mm). Assuming a conservative 10~ns time resolution, the spatial 
resolution is expected to be around 200~microns due to field distortions and 
spread of the signal.

\begin{figure}
  \begin{center}
    \includegraphics[angle=0, width=0.5\textwidth, trim=5mm 5mm 5mm 15mm, 
    clip]{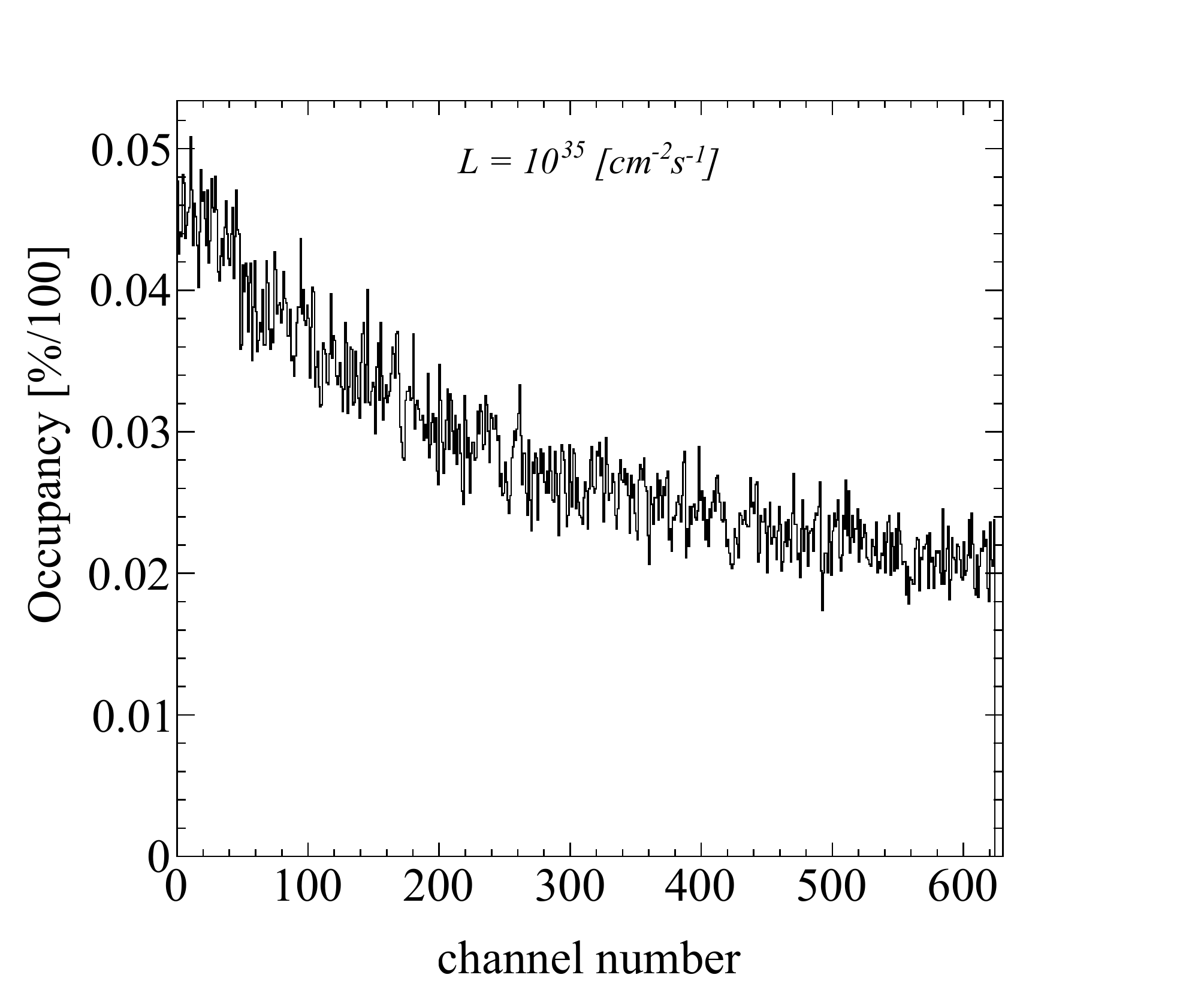}
    \caption{\label{fig:RCoccupancy}A full Geant4 simulation of the ALERT drift 
       chamber hit occupancy
       at a luminosity of $10^{35}$ cm$^{-2}$s$^{-1}$. The channel numbering 
    starts with the inner most wires and works outwards.}
  \end{center}
\end{figure}

\begin{figure}
  \begin{center}
    \includegraphics[angle=0, width=0.7\textwidth, trim=5mm 5mm 5mm 10mm, 
    clip]{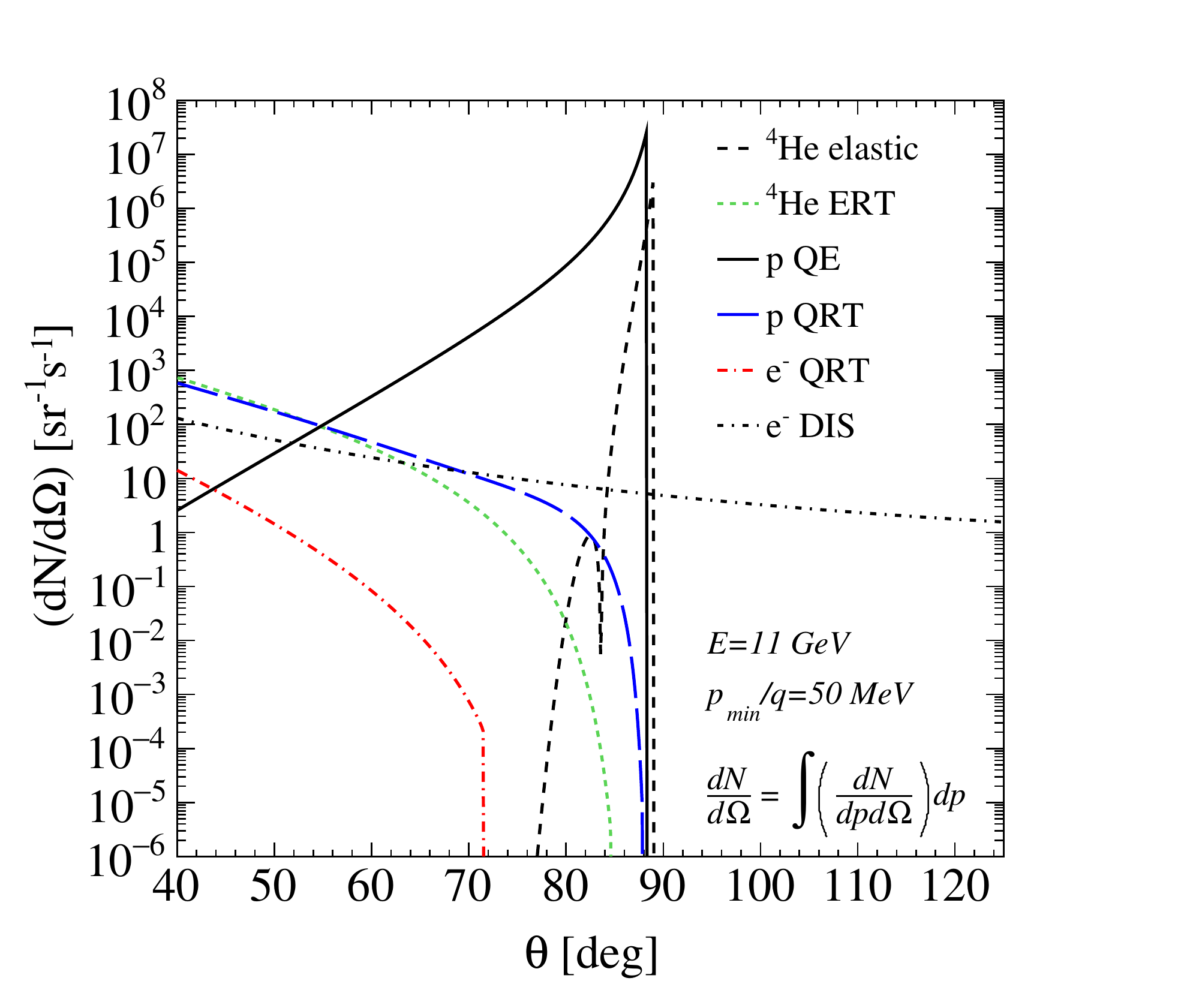}
    \caption{\label{fig:ALERTrates}The rates for different processes as 
    function of angle. The quasi-elastic radiative tails (QRT), $^4$He elastic 
 radiative tail (ERT), and DIS contributions have been integrated over momenta 
 starting at $p/q$ = 50~MeV/c, where $q$ is the electric charge of the particle 
 detected.}
  \end{center}
\end{figure}

The maximum occupancy, shown in Figure~\ref{fig:RCoccupancy},
is expected to be around 5\% for the inner most wires at $10^{35}$~cm$^{-2}$s$^{-1}$
(including the target windows). This is the maximum available luminosity for the 
baseline CLAS12 and is obtained based on the physics channels depicted 
in Figure~\ref{fig:ALERTrates}, assuming an integration time of 200~ns and 
considering a readout wire separation of 4~mm. This amount of accidental hits 
does not appear to be reasonable for a good tracking quality, we therefore 
decided to run only at half this luminosity for our main production runs. This 
will keep occupancy below 3\%, which is a reasonable amount for a drift chamber 
to maintain high tracking efficiency. When running the coherent processes with 
the $^4$He target, it is not necessary to detect the protons\footnote{This 
   running condition is specific to the proposal ``Partonic Structure of Light 
Nuclei'' in the ALERT run group.}, so the rate of accidental hits can then be 
highly reduced by increasing the detection threshold, thus making the chamber 
blind to the protons\footnote{The CLAS {\it eg6} run period was using the RTPC in 
the same fashion.}. In this configuration, considering that our main 
contribution to occupancy are quasi-elastic protons, we are confident that the 
ALERT can work properly at $10^{35}$~cm$^{-2}$s$^{-1}$.

We are currently planning to use the electronics used by the MVT of CLAS12, 
known as the DREAM chip \cite{7097517}. Its dynamic range and time resolution 
correspond to the needs of our drift chamber. To ensure that it is the 
case, tests with a prototype will be performed at the IPN Orsay (see 
section~\ref{sec:proto}).

\subsection{The Scintillator Array} \label{sec:scint}
The scintillator array will serve two main purposes. First, it will provide a 
useful complementary trigger signal because of its very fast response time, 
which will reduce the random background triggers. Second, it will provide 
particle identification, primarily through a time-of-flight measurement, but 
also by a measurement of the particle total energy deposited and path length in 
the scintillator which is important for doubly charged ions.

The length of the scintillators cannot exceed roughly 40~cm to keep the time 
resolution below 150~ps. It must also be segmented to match with tracks 
reconstructed in the drift chamber. Since $^3$He and $^4$He will travel at 
most a few mm in the scintillator for the highest anticipated momenta 
($\sim$~400~MeV/c), a multi-layer scintillator design provides an extra handle on 
particle identification by checking if the range exceeded the thickness of 
the first scintillator layer.

The initial scintillator design consists of a thin (2~mm) inner layer of 60 
bars, 30~cm in length, and 600 segmented outer scintillators (10 segments 
3~cm long for each inner bar) wrapped around the drift chamber. Each of these 
thin inner bars has SiPM\footnote{SiPM: silicon photomultiplier.} detectors 
attached to both ends. A thicker outer layer (18~mm) will be further segmented 
along the beam axis to provide position information and maintain good time 
resolution.

\begin{figure}
  \begin{center}
    \includegraphics[width=0.48\textwidth]{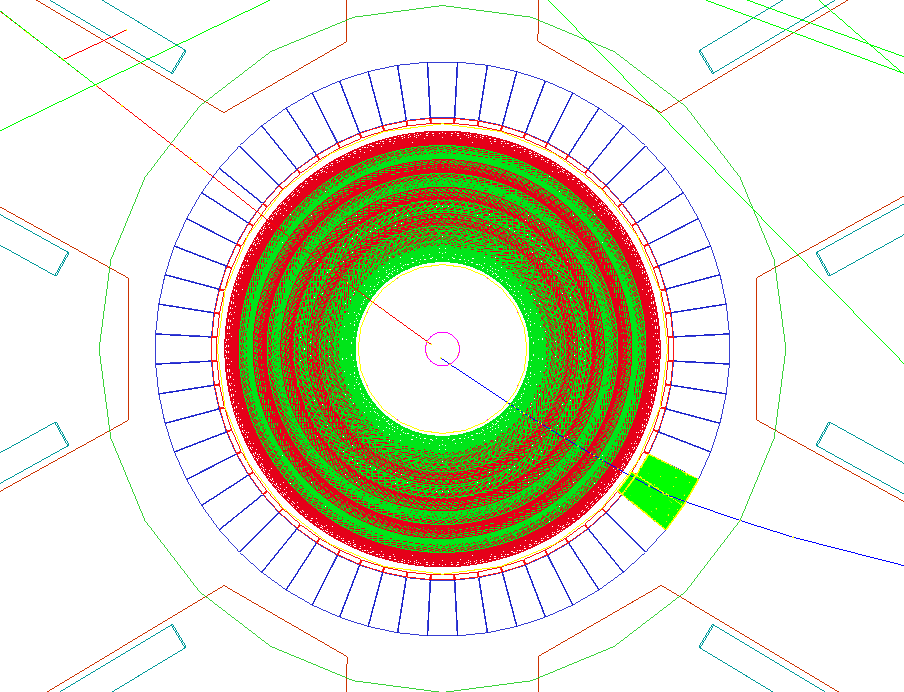}
    \includegraphics[width=0.48\textwidth]{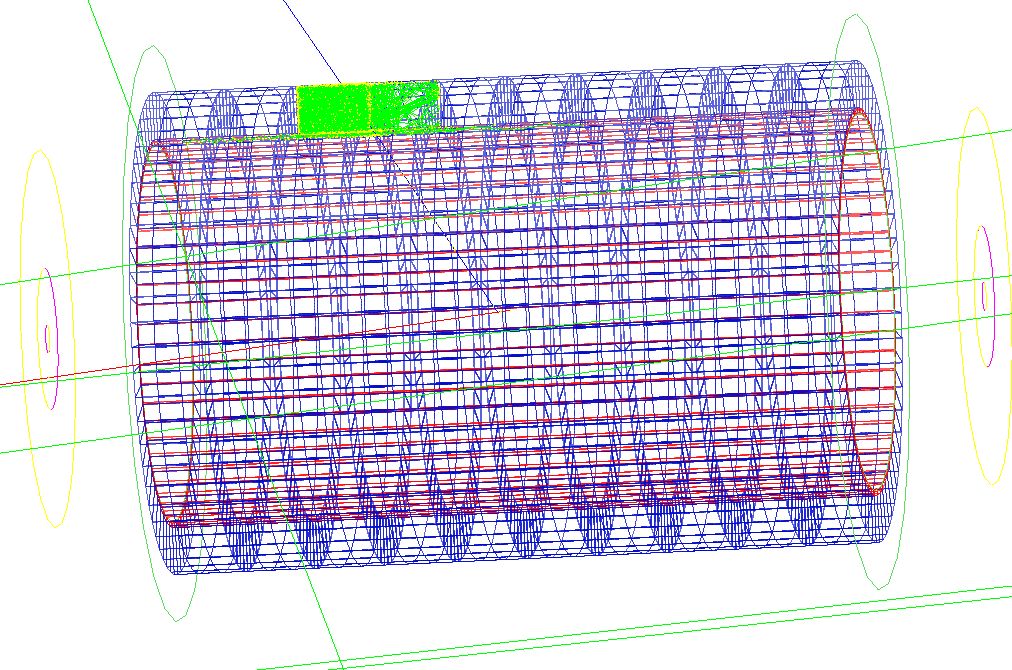}
    \caption{\label{fig:scintHodoscopeDesign}Geant4 simulation of a proton 
    passing through the recoil drift chamber and scintillator hodoscope. The 
 view looking downstream (left) shows the drift chamber's eight alternating 
 layers  of wires (green and red) surrounded by the two layers of scintillator 
 (red and blue). Simulating a proton through the detector, photons (green) are 
 produced in a few scintillators. On the right figure, the dark blue rings are graphical feature showing the contact between the adjacent outer scintillators.}
  \end{center}
\end{figure}

For the outer layer, a dual ended bar design and a tile design with embedded 
wavelength shifting fiber readouts similar to the forward tagger's hodoscope for 
CLAS12~\cite{FThodo} were considered. After simulating these designs, it was 
found that the time resolution was insufficient except only for the smallest 
of tile designs (15$\times$15$\times$7~mm$^3$). Instead of using fibers, a 
SiPM will be mounted directly on the outer layer of a keystone shaped 
scintillator that is 30~mm in length and 18~mm thick. This design can be seen 
in Figure~\ref{fig:scintHodoscopeDesign} which shows a full Geant4 simulation of 
the drift chamber and scintillators. By directly mounting the SiPMs to the 
scintillator we collect the maximum signal in the shortest amount of time.  
With the large number of photons we expect, the time resolution of SiPMs will 
be a few tens of ps, which is well within our target.

\begin{figure}
  \begin{center}
    \includegraphics[angle=0, 
    width=0.48\textwidth]{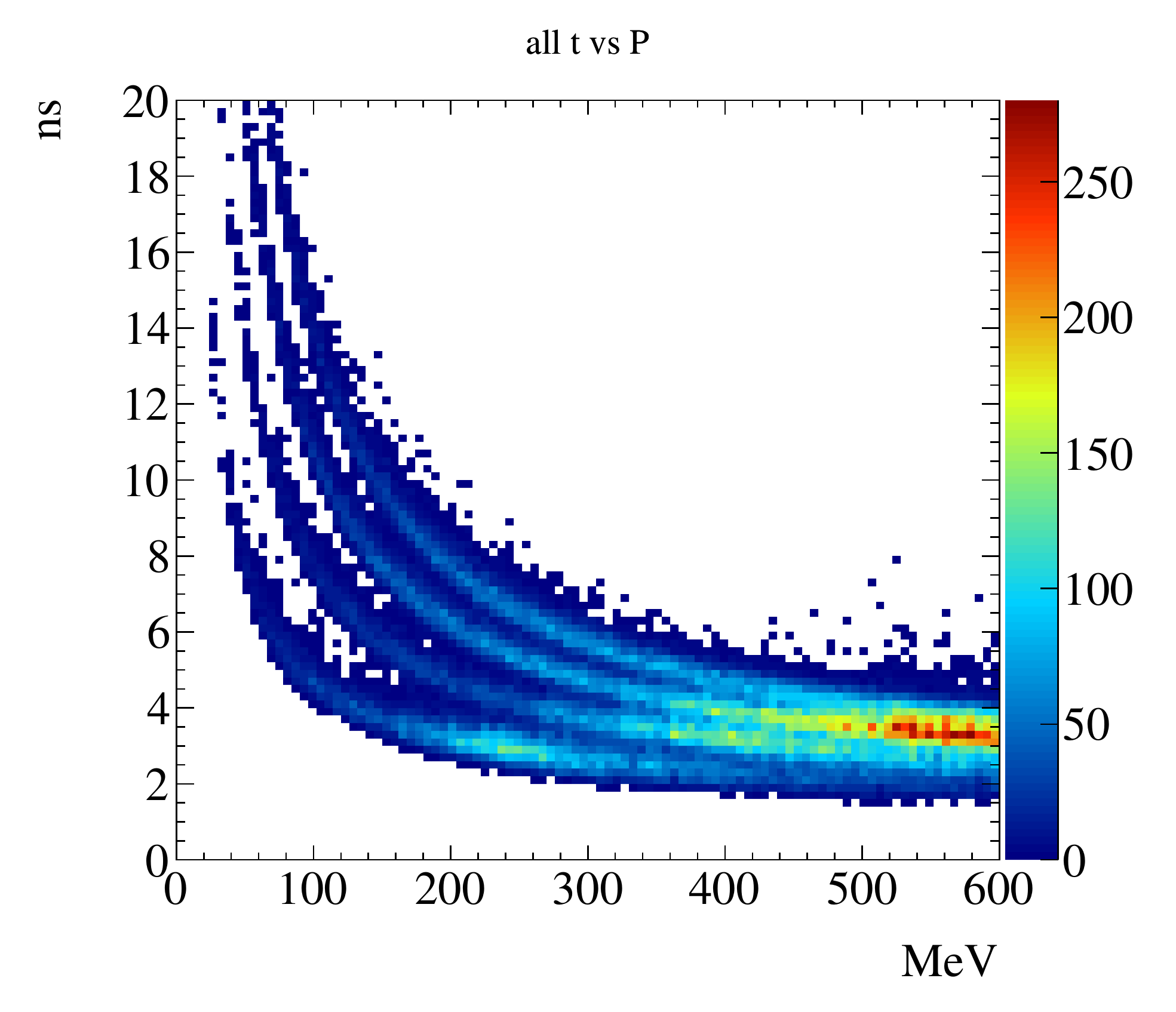}
    \includegraphics[angle=0, 
    width=0.48\textwidth]{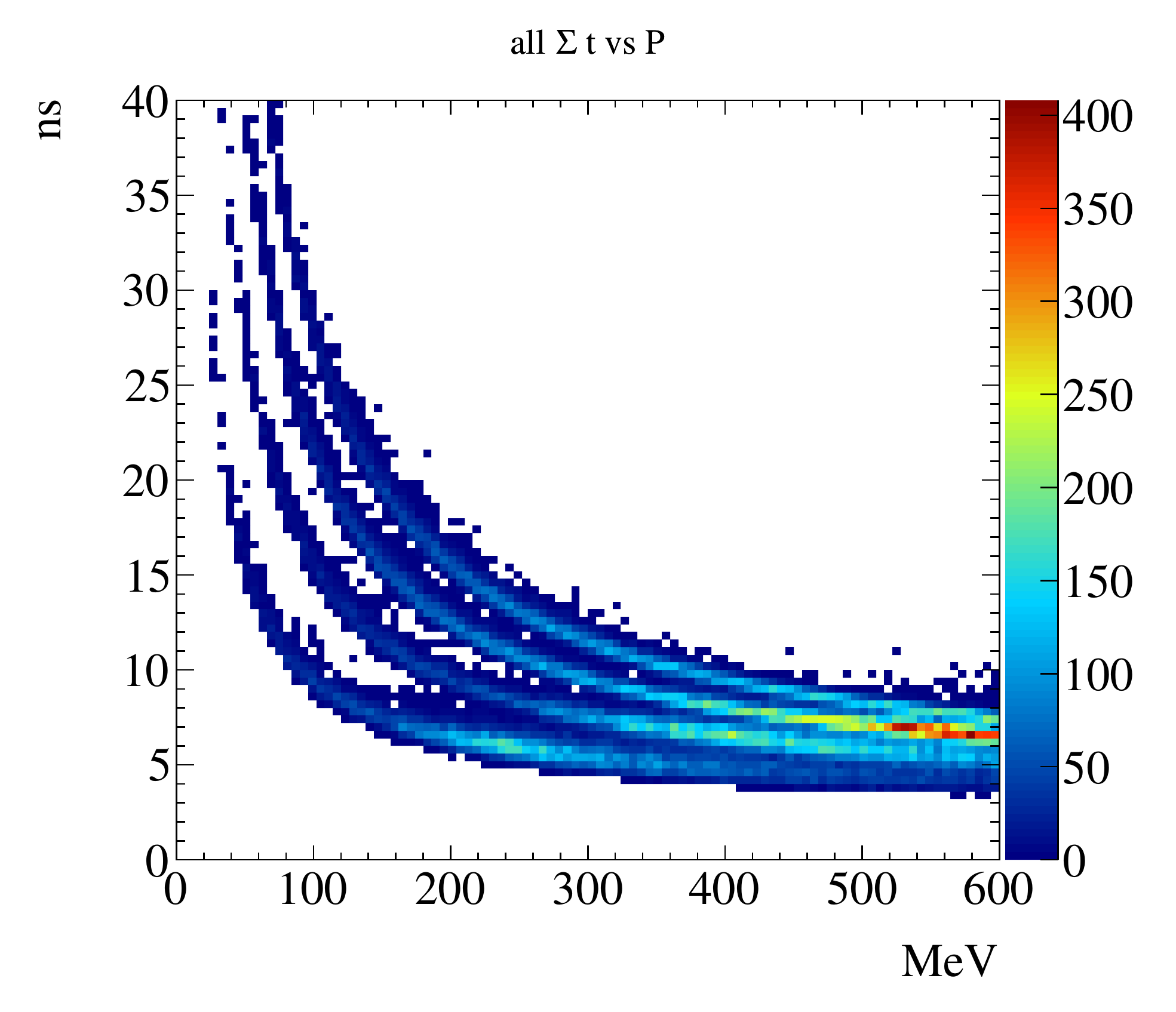}
    \caption{
       \label{fig:scintTimeVsP}Simulated TOF for the various recoil particles 
    vs Momentum. The TOF from just a single readout is shown on the left and 
 the sum of the dual ended readout is shown on the right.   }
  \end{center}
\end{figure}

The advantage of a dual ended readout is that the time sum is proportional to 
the TOF plus a constant. The improved separation of different particles can 
be seen in Figure~\ref{fig:scintTimeVsP}. Reconstructing the position of a hit 
along the length of a bar in the first layer is important for the doubly 
charged ions because they will not penetrate deep enough to reach the second 
layer of segmented scintillator.

The front-end electrons for the SiPMs will include preamplifiers and ASICs\footnote{ASIC: application-specific integrated circuit.}
which provide both TDC and ADC readouts. The PETIROC-2A\cite{PETIROC} ASIC 
provides excellent time resolution ($18$~ps on trigger output with 4 
photoelectrons detected) and a maximum readout rate at about 40k events/s.
Higher readout rates can be handled by using external digitizers by using the 
analog mode of operation and increase this rate by an order of magnitude. The 
ASIC also has the advantage of being able to tune the individual over-bias 
voltages with an 8-bit DAC.

The expected radiation damage to the SiPMs and scintillator material is found 
to be minimal over the length of the proposed experiment. We used the CLAS12 
forward tagger hodoscope technical design report~\cite{FThodo} as a very 
conservative baseline for this 
comparison. We arrived at an estimated dose of 1 krad after about 4.5 months of 
running. The damage to the scintillator at 100 times these radiation levels  
would not be problematic, even for the longest lengths of scintillator 
used~\cite{Zorn:1992ew}.
Accumulated dose on the SiPMs leads to an increased dark current. Similarly
than for scintillators, we do not expect it to be significant over the length of the 
experiment. The interested reader is referred to the work on
SiPMs for the Hall-D detectors~\cite{Qiang:2012zh,Qiang:2013uwa}. A front-end 
electronics prototype will be tested for radiation hardness but we expect  any 
damage to negligible~\cite{commPETIROC}.


\subsection{Target Cell}\label{sec:targetCell}

\begin{figure}
  \begin{center}
    \includegraphics[angle=0, trim={0 0 15cm 0}, clip,
    width=0.99\textwidth]{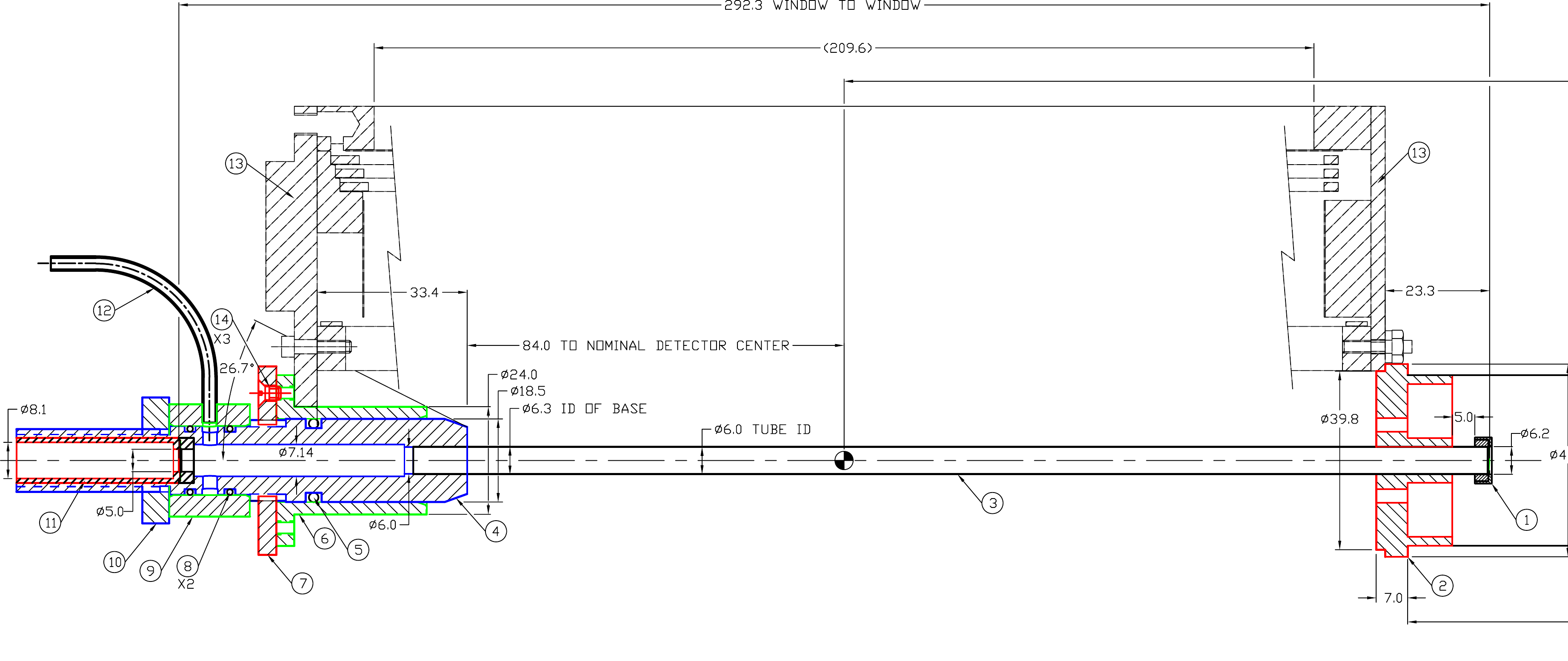}
    \caption{ \label{fig:eg6TargetDrawing}The eg6 target design drawing.}
  \end{center}
\end{figure}

The design of the proposed ALERT target will be very similar to the eg6 target shown 
in Figure~\ref{fig:eg6TargetDrawing}.
The target parameters are shown in Table~\ref{tab:target} 
with the parameters of other existing and PAC approved targets.
Note that, the proposed target has an increased radius of 6~mm compared to all the 
others which have 3~mm radius. This increase compared to the previous CLAS targets has been made
in order to compensate for the expected increase of beam size at 11~GeV. The BONuS12
target is still presently proposed to be 3~mm in radius, if such a target
is operated successfully in JLab, we will definitely consider using a
smaller radius as well, but we prefer to propose here a safer option that we know
will work fine.

\begin{table}
\centering
\caption{Comparison of various straw targets used at JLab.The 
"JLab test targets" correspond
to recent tests performed in JLab for the BONuS12 target, they have
been tested for pressure but have never been tested with beam.
}
\newcolumntype Y{S [ group-four-digits=true,
round-mode=places,
round-precision=1,
round-integer-to-decimal=true,
per-mode=symbol ,
detect-all]}
\tabucolumn Y
\label{tab:target}
\bgroup
\def\arraystretch{1.2}%
\tabulinesep=1mm
\begin{tabu}{l C{1.5cm}C{3.0cm}C{2cm}}
\tabucline[2pt]{-}
\textbf{Experiment} & \textbf{Length} & \centering \textbf{Kapton wall thickness} &
\textbf{Pressure} \\ \tabucline[1pt]{-}
CLAS target (eg6)           & 30~cm & \SI{27}{\um} & 6.0~atm   \\
BONuS12 (E12-06-113) target & 42~cm & \SI{30}{\um} & 7.5~atm \\
JLab test target 1          & 42~cm & \SI{30}{\um} & 3.0~atm   \\
JLab test target 2          & 42~cm & \SI{50}{\um} & 4.5~atm \\
JLab test target 3          & 42~cm & \SI{60}{\um} & 6.0~atm   \\
ALERT proposed target       & 35~cm & \SI{25}{\um} & 3.0~atm   \\
\tabucline[2pt]{-}
\end{tabu}
\egroup
\end{table}

%
\section{Simulation of ALERT and reconstruction} \label{sec:sim}
The general detection and reconstruction scheme for ALERT is as follows. We fit 
the track with the drift chamber and scintillator position information to
obtain the momentum over the charge. Next, using the 
scintillator time-of-flight, the particles are separated and identified by 
their mass-to-charge ratio, therefore leaving a degeneracy for the deuteron and 
$\alpha$ particles.
The degeneracy between deuteron and $\alpha$ particles can be resolved in a few 
ways.  The first and most simple way is to observe that an $\alpha$ will almost 
never make it to the second layer of scintillators and therefore the absence (presence) of a 
signal would indicate the particle is an $\alpha$~(deuteron). Furthermore, as 
will be discussed below, the measured dE/dx will differ for $^4$He and $^2$H, 
therefore, taking into account energy loss in track fitting alone can provide 
separation. Additionally taking further advantage of the measured total energy 
deposited in the scintillators can help separate the $\alpha$s and deuterons.

\subsection{Simulation of ALERT}
\begin{figure}[tbp]
    \begin{center}
        \includegraphics[width=0.45\textwidth]{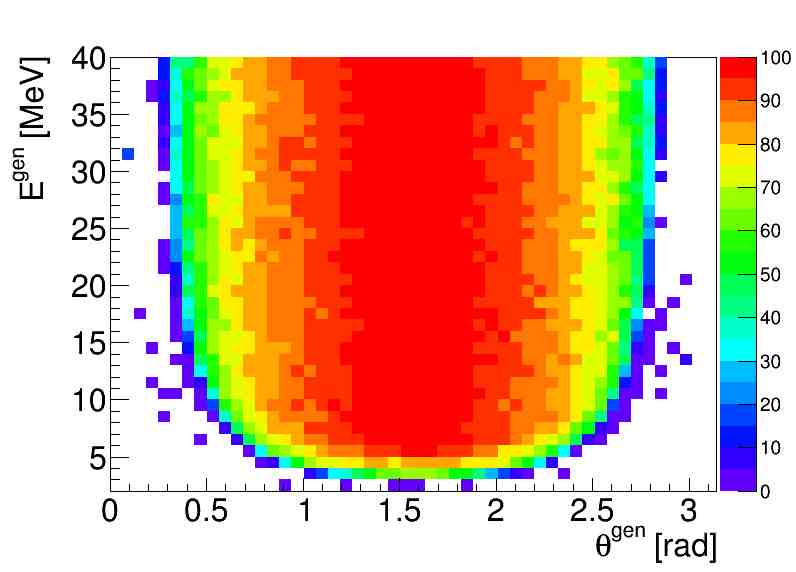}
        \includegraphics[width=0.45\textwidth]{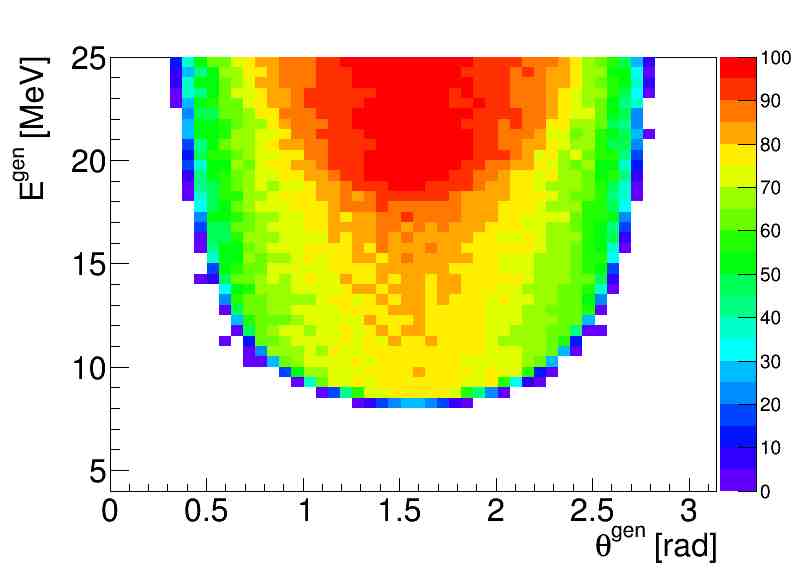}
        \caption{Simulated recoil detector acceptance percentage, for protons (left) and 
$^4$He (right), when requiring energy deposition in the scintillators arrays. 
\label{fig:acceptance}}
    \end{center}
\end{figure}

The simulation of the recoil detector has been implemented with the full 
geometry and material specifications in GEANT4. It includes a 5~Tesla homogeneous 
solenoid field and the entire detector filled with materials as described in the 
previous section. In this study all recoil species are generated with the same 
distributions: flat in momentum from threshold up to 40~MeV 
($\sim$~250~MeV/c) for protons and about 25~MeV for other particles; isotropic 
angular coverage; flat distribution in $z$-vertex; and a radial vertex 
coordinate smeared around the beam line center by a Gaussian distribution of 
sigma equal to the expected beam radius (0.2 mm).
For reconstruction, we require that the particle reaches the scintillator
and obtain the acceptance averaged over the $z$-vertex position shown in 
Figure~\ref{fig:acceptance}.

\subsection{Track Fitting}
\begin{figure}[tbp]
    \begin{center}
        \includegraphics[height=4.5cm, width=0.32\textwidth]{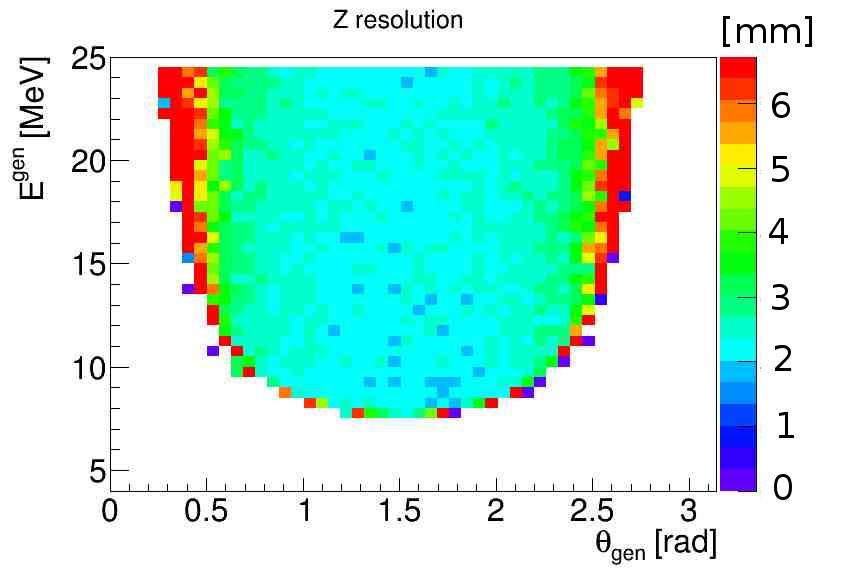}
        \includegraphics[height=4.5cm, width=0.32\textwidth]{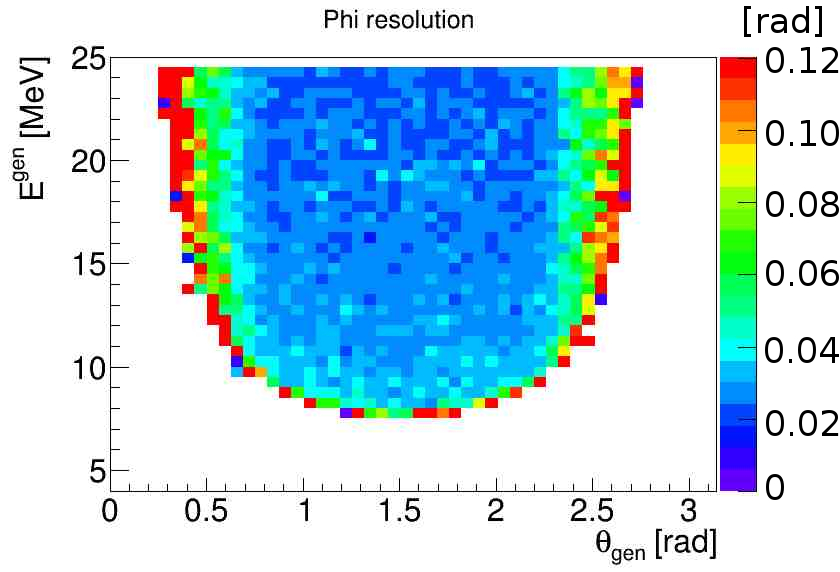}
        \includegraphics[height=4.5cm, width=0.32\textwidth]{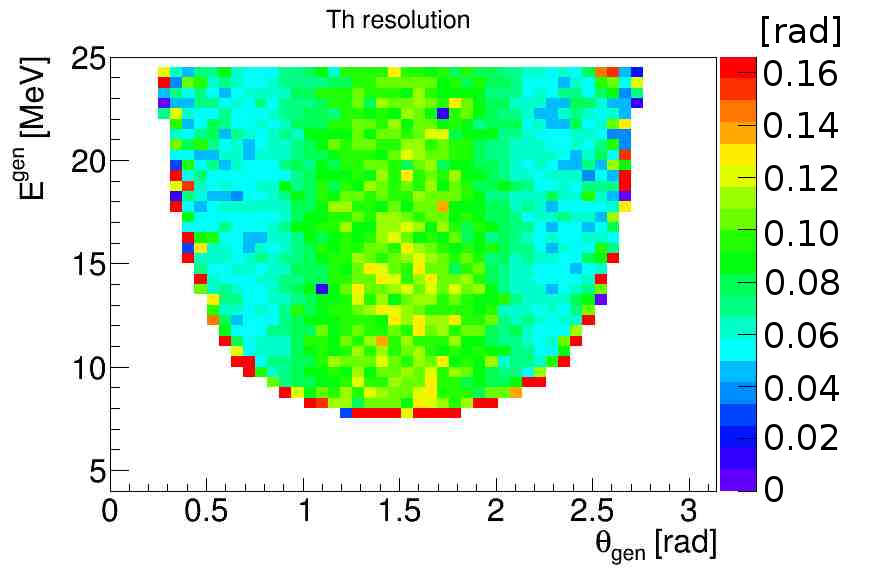}
        \caption{Resolutions for simulated $^4$He:  $z$-vertex resolution in mm (left), azimuthal (center) 
          and polar (right) angle resolutions in radians for the lowest energy
          regime when the recoil track reaches the scintillator.\label{fig:tracking}}
    \end{center}
\end{figure}

\begin{figure}[tbp]
    \begin{center}
        \includegraphics[width=0.45\textwidth]{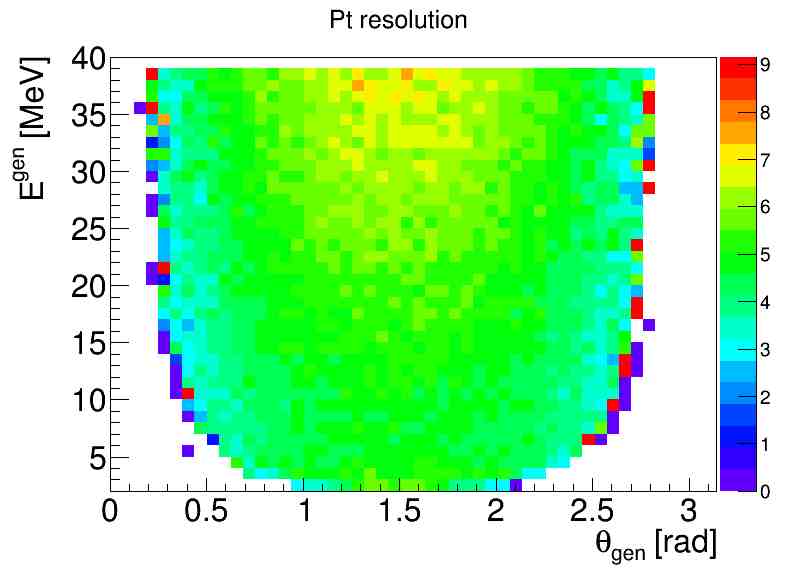}
        \includegraphics[width=0.45\textwidth]{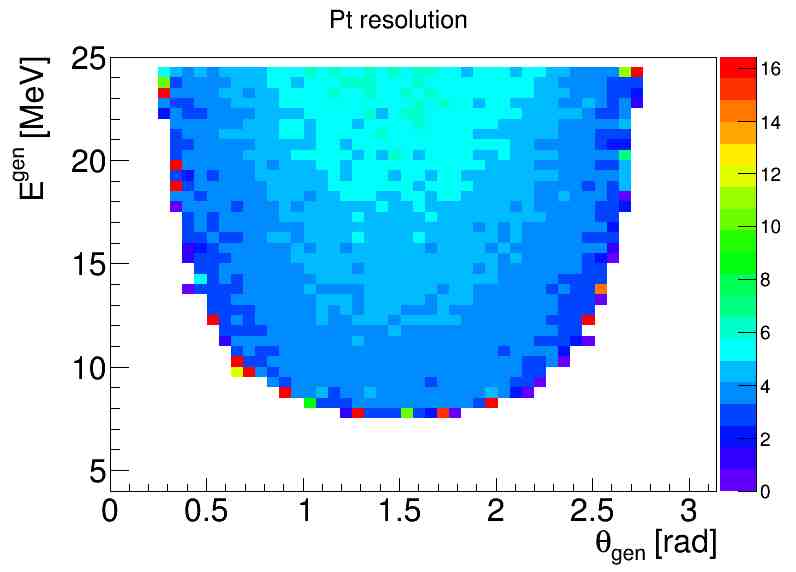}
        \caption{Simulated momentum resolutions (in \%) as a function of energy and 
                 polar angle for protons (left) and $^4$He 
                 (right) integrated over all $z$, when the recoil track reaches the scintillators 
                 array. \label{fig:presolution}}
    \end{center}
\end{figure}

The tracks are obtained using a helix fitter giving the coordinates of 
the vertex and the momentum of the particle. The energy deposited in 
the scintillators could also be used to help determine the kinetic energy of the 
nucleus, but is not implemented in the studies we performed here. 
The tracking capabilities of the recoil detector are investigated 
assuming a spatial resolutions of \SI{200}{\um} for the drift chamber. The wires 
are strung in the $z$-direction with a stereo angle of \ang{10}. The resulting difference between 
generated and reconstructed variables from simulation is shown in 
Figure~\ref{fig:tracking} for $^4$He particles. The momentum resolution for both protons and 
$^4$He is presented in Figure~\ref{fig:presolution}.


\subsection{Particle identification in ALERT}

\begin{figure}[tbp]
    \begin{center}
        \includegraphics[width=0.7\textwidth]{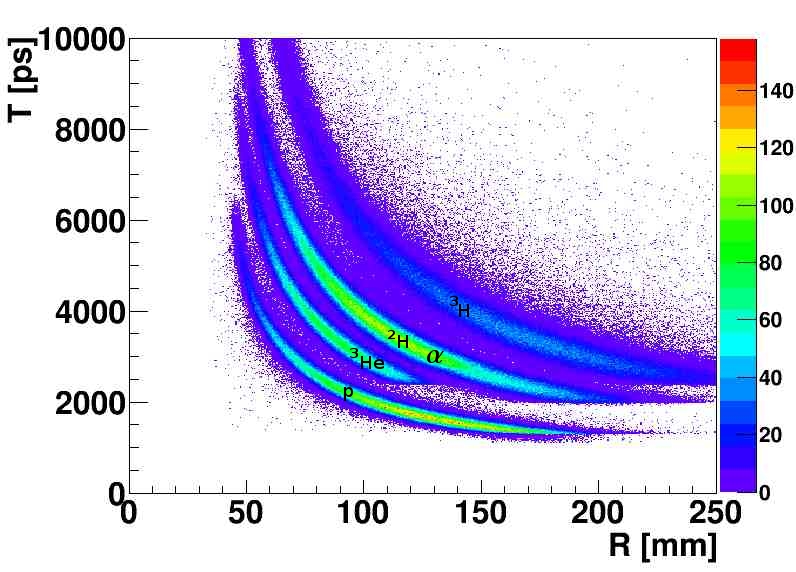}
        \caption{Simulated time of flight at the scintillator versus the 
reconstructed radius in the drift chamber. The bottom band corresponds to 
the proton, next band is the $^3$He nuclei, $^2$H and $\alpha$ are overlapping in 
the third band, the uppermost band is $^3$H\label{fig:SIMtof}. $^2$H and 
$\alpha$ are separated using dE/dx.}
    \end{center}
\end{figure}

The particle identification scheme is investigated using the GEANT4
simulation as well. The scintillators 
have been designed to ensure a 150~ps time resolution. To determine the dE/dx 
resolution, measurements will be necessary for the scintillators and for the 
drift chamber as this depends on the detector layout, gas mixture, 
electronics, voltages... Nevertheless, from \cite{Emi}, one can assume that 
with 8 hits in the drift chamber and the measurements in the 
scintillators, the energy resolution should be at least 10\%.
Under these conditions, a clean separation of three of the five nuclei is shown 
in Figure~\ref{fig:SIMtof} solely based on the time of flight measured by the 
scintillator compared to the reconstructed momentum from the drift chamber. 
We then separate $^2$H and $\alpha$ using dE/dx in the drift chamber and in the 
scintillators.

To quantify the separation power of our device, we simulated an equal quantity 
of each species. We obtained a particle identification efficiency of 99\% for 
protons, 95\% for $^3$He and 98\% for $^3$H and around 90\% for $^2$H and 
$\alpha$ with equally excellent rejection factors. It is important to note that 
for this analysis, only the energy deposited in the scintillators was used, not 
the energy deposited in the
drift chamber nor the path length in the scintillators, thus these numbers
are very likely to be improved when using the full information\footnote{The 
uncertainty remains important about the resolutions that will be achieved 
for these extra information. So we deemed more reasonable to ignore them
for now.}. This analysis indicates that the proposed reconstruction 
and particle identification schemes for this design are quite promising.  
Studies, using both simulation software and prototyping, are ongoing to 
determine the optimal detector parameters to minimize the detection threshold 
while maximizing particle identification efficiency. The resolutions presented 
above have been implemented in a fast Monte-Carlo used to evaluate their impact 
on our measurements.

\section{Drift chamber prototype}
\label{sec:proto}
Since the design of the drift chamber presents several challenges in term of
mechanical assembly, we decided to start prototyping early. The goal is to find a 
design that will be easy to install and to maintain if need be, while keeping the 
amount of material at a minimum. This section presents the work done in Orsay 
to address the main questions concerning the mechanics that needed to be answered:
\begin{itemize}
\item How to build a stereo drift chamber with a 2~mm gap between wires?
\item Can we have frames that can be quickly changed in case of a broken wire?
\item How to minimize the forward structure to reduce the multiple scattering,
while keeping it rigid enough to support the tension due to the wires?
\end{itemize}

\begin{figure}[tbp]
    \begin{center}
        \includegraphics[width=0.4\textwidth]{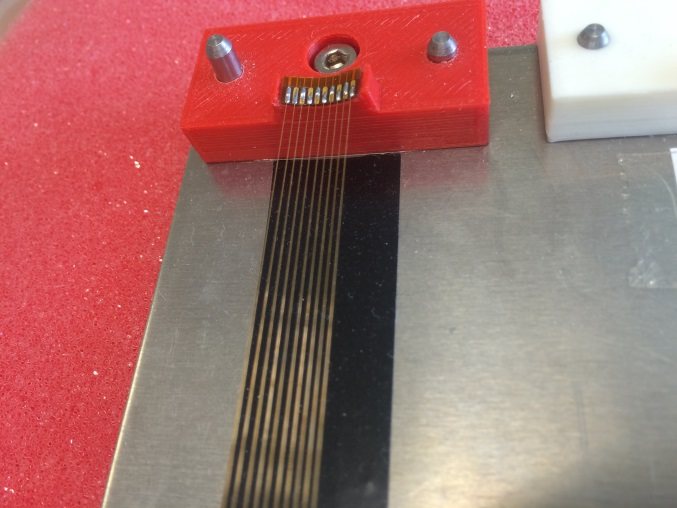}
        \caption{Welded wires on a curved structure with a 2~mm gap between each wire.}
        \label{soldOK}
    \end{center}
\end{figure}

For the first question, small plastic structures realized with a 3D printer 
were tested and wires welded on it, as shown in Figure \ref{soldOK}. This 
demonstrated our ability to weld wires with a 2~mm gap on a curved structure.
 
\begin{figure}[tbp]
    \begin{center}
        \includegraphics[width=0.40\textwidth]{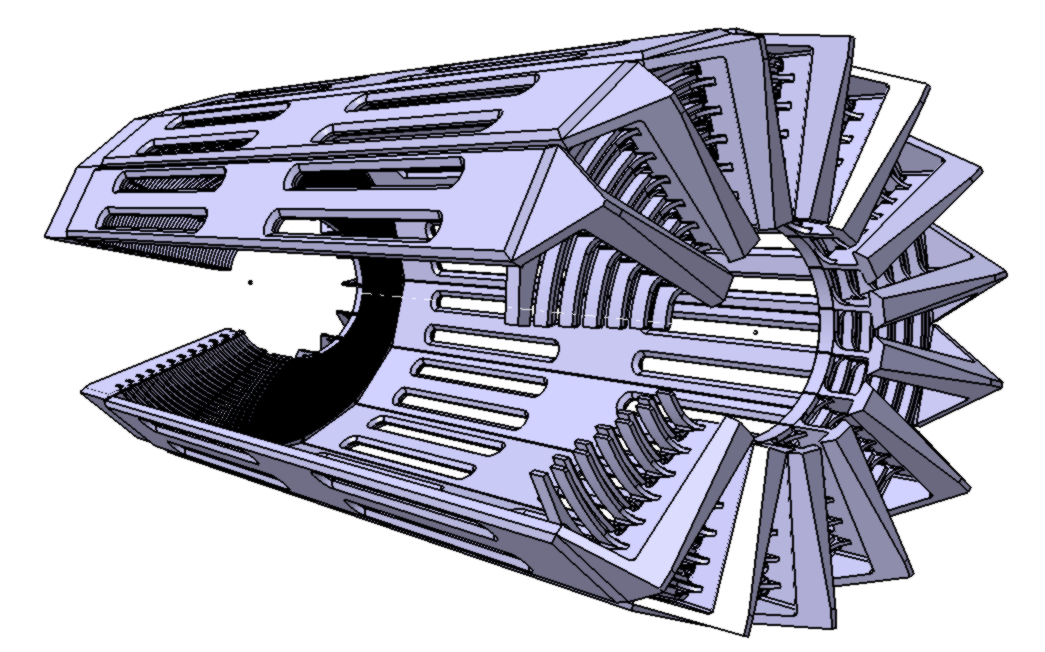}
        \includegraphics[width=0.40\textwidth]{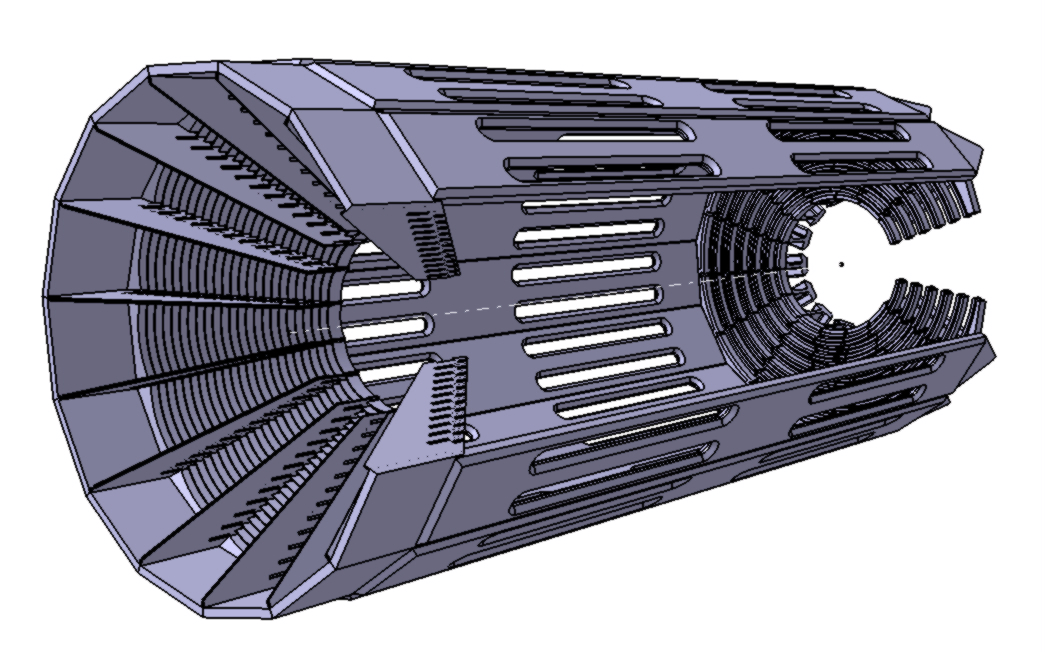}
        \caption{Upstream (left) and downstream (right) ends of the prototype 
        detector in computer assisted design (CAD) with all the sectors included.  \label{wholeView}}
    \end{center}
\end{figure}

\begin{figure}[tbp]
    \begin{center}
        \includegraphics[width=0.4\textwidth]{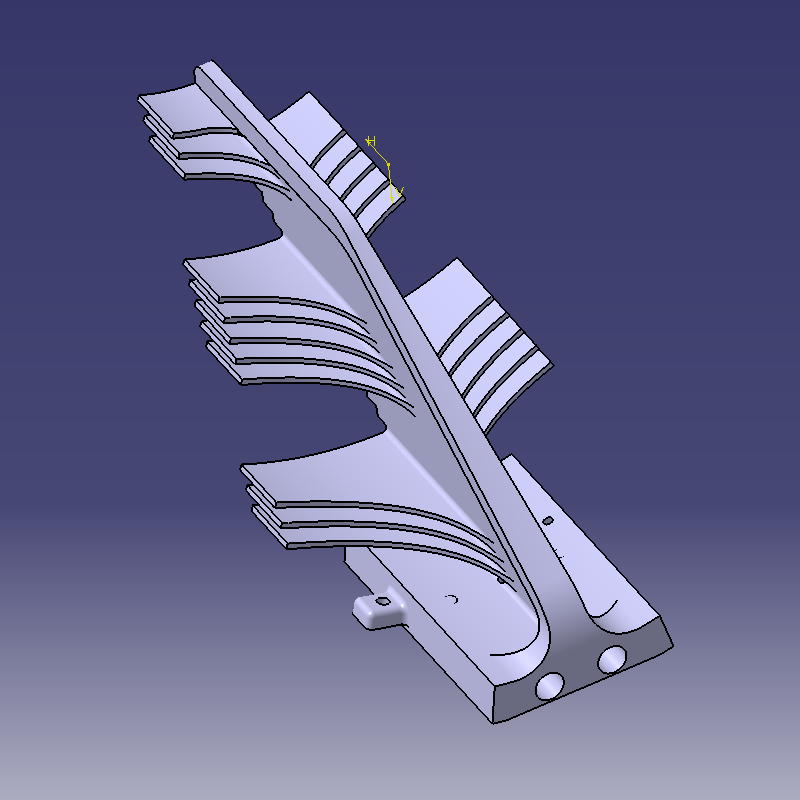}
        \includegraphics[width=0.4\textwidth]{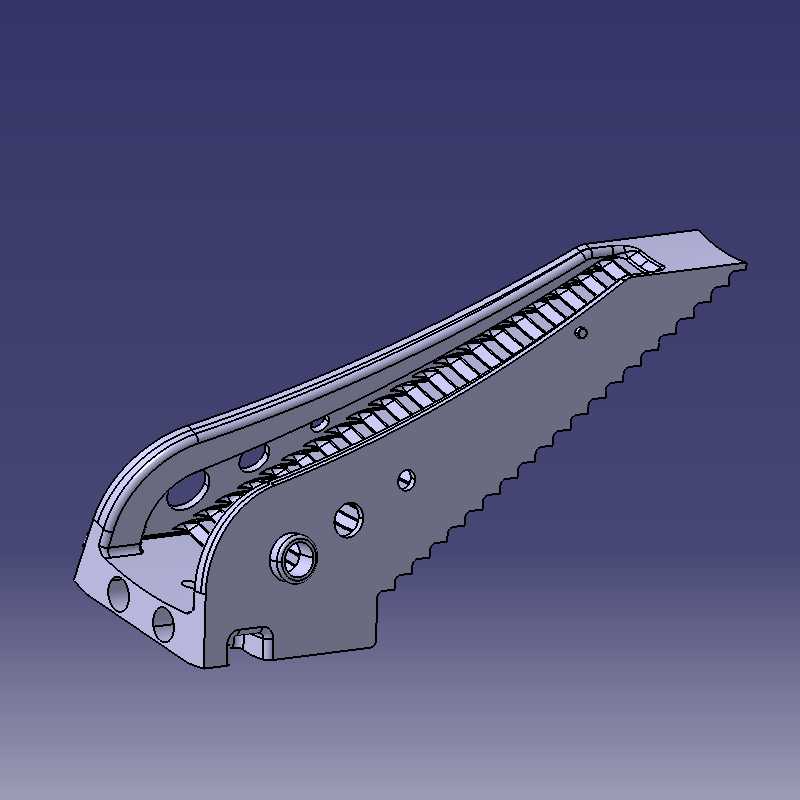}
        \caption{Close up on the CAD of the upstream piece (left) and downstream 
        piece(right) of the drift chamber. Note that the design of the pieces has been
        optimized in comparison of what is shown in Figure~\ref{wholeView}.}
        \label{fig:CAD}
    \end{center}
\end{figure}

To limit issues related to broken wires, we opted for a modular detector made of 
identical sectors. Each sector covers 20$^{\circ}$ of the azimuthal angle 
(Figure~\ref{wholeView}) and can be rotated around the beam axis to be separated 
from the other sectors. This rotation is possible due to the absence of one 
sector, leaving a 20$^{\circ}$ dead angle. Then, if a wire breaks, its sector 
can be removed independently and replaced by a spare. Plastic and metallic 
prototype sectors were made with 3D printers to test the assembling procedure and 
we have started the construction of a full size prototype of one sector.
The shape of each sector is constrained by the position of the wires. It has 
a triangular shape on one side and due to the stereo angle, the other side 
looks like a pine tree with branches alternatively going left and right from 
a central trunk (Figure~\ref{fig:CAD}).

\begin{figure}[tbp]
    \begin{center}
        \includegraphics[width=0.7\textwidth]{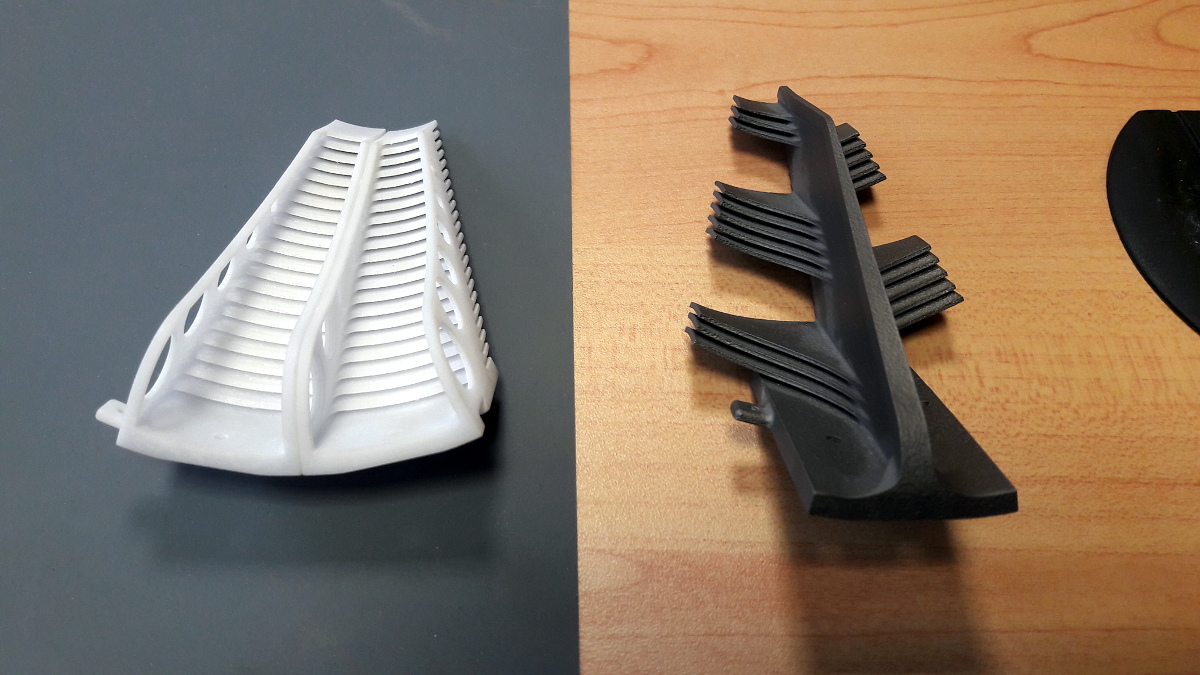}
        \caption{Prototypes for the mechanical parts of the drift chamber made out of plastic
        for the forward part and titanium for the backward.}
        \label{OneSector}
    \end{center}
\end{figure}

Finally, the material used to build the structure will be studied in details with 
future prototypes. Nevertheless, most recent plans are to use high rigidity plastic
in the forward region and metal for the backward structure (as in Figure~\ref{OneSector}). The 
prototypes are not only designed to check the mechanical requirements summarized above 
but also to verify the different cell configurations, and to test the DREAM 
electronics (time resolution, active range, noise). 

\section{Technical contributions from the research groups}
The effort to design, build and integrate the ALERT detector is led by four 
research groups, Argonne National Lab (ANL), 
Institut de Physique Nucl\'eaire d'Orsay (IPNO), Jefferson Lab and Temple 
University (TU). 

Jefferson Lab is the host institution. ANL, IPNO and TU have 
all contributed technically to CLAS12. ANL was involved in the construction of 
the high-threshold Cherenkov counters (HTCC) for CLAS12. ANL has a memorandum 
of understanding (MOU) with JLab on taking responsibility for the HTCC light 
collection system including testing the photomultipliers and the magnetic 
shielding. For the RICH detector for CLAS12, ANL developed full GEANT-4 
simulations in addition to the tracking software. ANL also developed the 
mechanical design of the detector support elements and entrance and exit 
windows in addition to the front-end electronics cooling system. IPNO took 
full responsibility for the design and construction of CLAS12 neutron detector 
(CND). The CND was successfully delivered to Jefferson Lab. TU played an 
important role in the refurbishment of the low threshold Cherenkov counters 
(LTCC), which was completed recently. All 216 photomultipliers have been coated 
with wavelength shifting material (p-Terphenyl) at Temple University, which 
resulted in a significant increase in the number of photoelectrons response.

The three institutions have already shown strong technical commitment to JLab 
12~GeV upgrade, with a focus on CLAS12 and this proposal is a continuation of 
this commitment.

\subsection{Argonne National Laboratory and Temple University}
The ANL medium energy group is responsible for the ALERT scintillator system, 
including scintillation material, light collection device and electronics. 
First results of simulations have led to the design proposed here. This work 
will continue to integrate the scintillator system with the wire chamber. ANL 
will collaborate closely with Temple University to test the light detection 
system. Both institutions will be responsible to assemble and test the detector.

Argonne will provide the electronics and technical support required to
integrate the scintillator detector system into the CLAS12 DAQ. The effort
will minimize the effort required on the part of the Hall B staff.

\subsection{Institut de Physique Nucl\'eaire d'Orsay}
The Institut de Physique Nucl\'eaire d'Orsay is responsible for the wire
chamber and the mechanical structure of the detector design and construction. 
As shown in the proposal, this work
has already started, a first prototype is being built 
to test different cell forms, wire material, wire thickness, 
pressure, etc. This experience will lead to a complete design of the ALERT detector 
integrating the scintillator built at ANL, the gas distribution system and the
electronic connections.

In partnership with {\it CEA Saclay}, IPN Orsay will also test the
use of the DREAM front-end chip for the wire chamber. Preliminary tests were
successful and will continue. The integration of the chip with CLAS12 is
expected to be done by the {\it CEA Saclay}, since they use the same chip to 
readout the CLAS12 MVT. Adaptations to the DAQ necessary when the MVT will be 
replaced by ALERT will be performed by the staff of IPN Orsay.

\subsection{Jefferson Laboratory}\label{sec:jlabContributions}
We expect Jefferson Lab to help with the configuration of the beam line.  
This will include the following items.

\paragraph{Beam Dump Upgrade}
The maximum beam current will be around 1000~nA for the production runs at 
$10^{35}$~cm$^{-2}$s$^{-1}$, which is not common for Hall-B.
To run above 500 nA the ``beam blocker'' will need to be upgraded to handle 
higher power. The beam blocker attenuates the beam seen by the Faraday cup.   
This blocker is constructed of copper and is water cooled. Hall B staff have 
indicated that this is a rather straightforward engineering task and has no 
significant associated costs~\cite{beamBlocker}. 

\paragraph{Straw Target}
We also expect JLab to design and build the target for the experiment as it 
will be a very similar target as the ones build for CLAS BONuS and eg6 runs.
See section \ref{sec:targetCell} for more details. 

\paragraph{Mechanical Integration}
We also expect Jefferson Laboratory to provide assistance in the detector 
installation in the Hall. This will include providing designers at ANL and IPNO 
with the technical drawings required to integrate ALERT with CLAS12. We will 
also need some coordination between designers to validate the mechanical 
integration. 

\paragraph{CLAS12 DAQ Integration}
We also will need assistance in connecting the electronics of ALERT to the 
CLAS12 data acquisition and trigger systems. This will also include help 
integrating the slow controls into the EPICs system.

\setlength\parskip{\baselineskip}%
\chapter{Proposed Measurements}
\label{chap:Measure}
The proposed measurements of coherent DVCS and $\phi$ electroproduction off 
$^4$He and their analyses are discussed in this chapter. For DVCS, the 
scattered electron, the real photon and the recoiling $^4$He nucleus will all 
be detected. For coherent $\phi$ production, we require the detection of the 
scattered electron, the recoiling $^4$He nucleus, and either a kaon pair for 
the identification of the $\phi$ meson through its invariant mass or a single 
kaon, and in that case, the missing kaon will be reconstructed through missing 
momentum and energy. For each process we first discuss an empirical model of 
the existing data and use this information in a Monte Carlo event generator.
The kinematic coverage and projected statistical and systematic uncertainties 
are presented. In section \ref{sec:impactParDists} the impact parameter 
distributions from these two models is presented.

\section{Exclusive Coherent DVCS}
DCVS is the hard exclusive production of a real photon in lepton scattering.  
For coherent production, we have
\begin{equation}
  e(\mathbf{P}_{\textnormal{e}}) + \textnormal{$^{4}$He}(\mathbf{P}_{\textnormal{$^{4}$He}}) \rightarrow 
  e(\mathbf{P'_{\textnormal{e}}}) +  \textnormal{$^{4}$He}(\mathbf{P'}_{\textnormal{$^{4}$He}}) + \gamma
  (\mathbf{P}_{\gamma})
  \label{coh_dvcs_equ}
\end{equation}
where $\mathbf{P}_{\textnormal{e}}(\mathbf{P'}_{\textnormal{e}})$ is the 
four-momentum of the incoming (outgoing) electron. $\mathbf{P}_{\gamma*} = 
\mathbf{P}_{\textnormal{e}} - \mathbf{P'}_{\textnormal{e}}$ is the 
four-momentum of the virtual photon and 
$\mathbf{P}_{\textnormal{$^{4}$He}}(\mathbf{P'}_{\textnormal{$^{4}$He}})$ is 
the four-momentum $^4$He nucleus in the initial (final) state. The photon 
virtuality is $Q^2 = 4EE^\prime \sin^2 (\theta/2)$, where $E$ and $E^\prime$ 
are the energy of the incoming and outgoing electron respectively. The 
four-momentum transfer to the nucleus is:
\begin{equation} 
  t=(\mathbf{P}_{\textnormal{$^{4}$He}} - \mathbf{P'}_{\textnormal{$^{4}$He}})^2 = ( \mathbf{P}_{\gamma*} - \mathbf{P}_{\gamma})^2.
  \label{eq:t-eq}
\end{equation}
Other variables of interest are $\phi$ the angle between the lepton scattering 
angle and photon production plane, $\nu = E -E'$ is the energy of the virtual 
photon. The kinematical cuts on the detected electron are:
\begin{itemize}
  \item $Q^{2} >$ 1 GeV$^{2}$: to ensure that the interaction occurs at the 
    partonic level and the applicability of factorization in the DVCS  
    handbag diagram.
  \item $ -t > -t_{min}$: the transferred momentum squared to the recoil 
    $^{4}He$ has to be greater than a minimum value defined by the kinematics 
    of the beam and the scattered electron as:
    \begin{equation}
      t_{min} = - Q^{2} \frac{2(1-x_{A})(1 -\sqrt{1+\epsilon ^{2}}) + 
      \epsilon^{2}}{4 x_{A}(1-x_{A})+ \epsilon ^{2}},
    \end{equation}
\end{itemize}
For all events, the scattered electron and the real photon will be detected in 
the CLAS12 spectrometer while the recoiling $^4$He nucleus will be detected in 
the ALERT detector. For both CLAS12 and ALERT, we use a FastMC package based on 
GEANT4 to simulate detector's acceptance. The different CLAS12 detector's 
resolutions were taken from CLAS12 fastMC. While for ALERT, we used 
parametrization of the resolutions obtained from the GEANT4 simulations 
described in section~\ref{sec:sim}. 

\begin{figure}[htb]
  \includegraphics[scale=0.4]{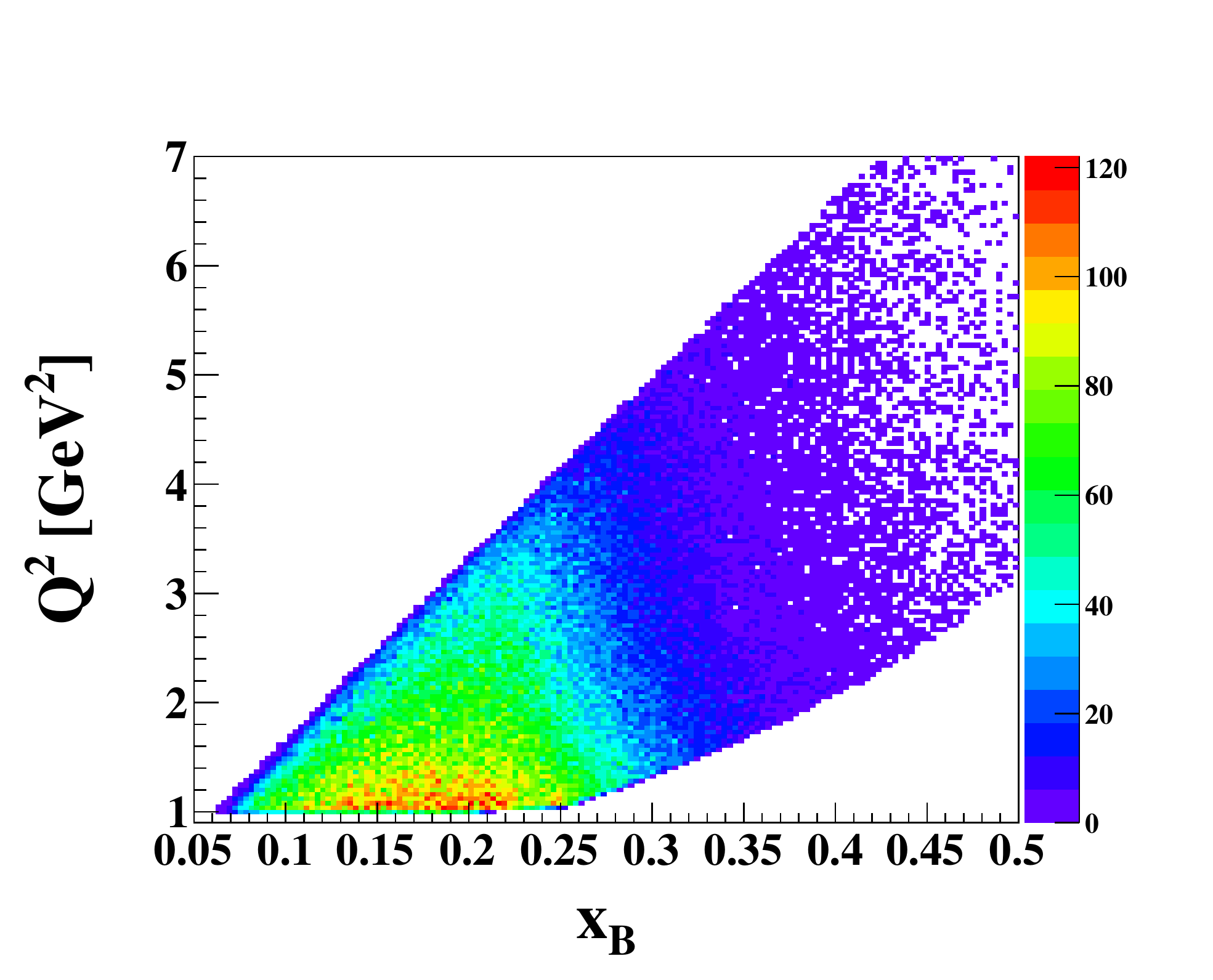}
  \includegraphics[scale=0.4]{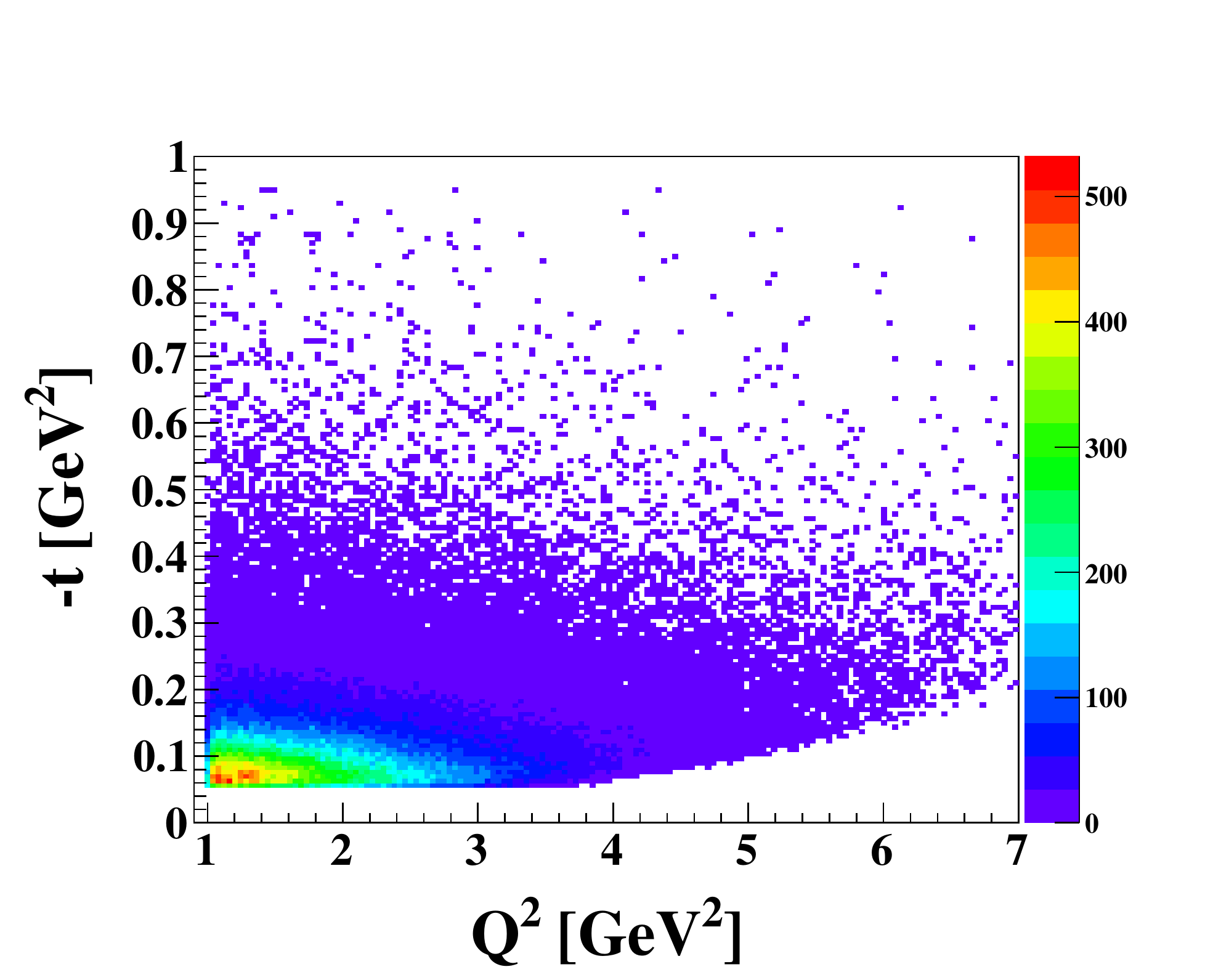}
  \caption{Left: correlation between $Q^{2}$ and $x_{B}$. Right: correlation 
  between $-t$ and $Q^{2}$ for coherent DVCS off $^{4}$He.\label{fig:Q2-xB}}
\end{figure}

\begin{figure}[htb]
  \centering
  \includegraphics[width=0.49\textwidth,clip,trim=10mm 5mm 10mm 20mm]{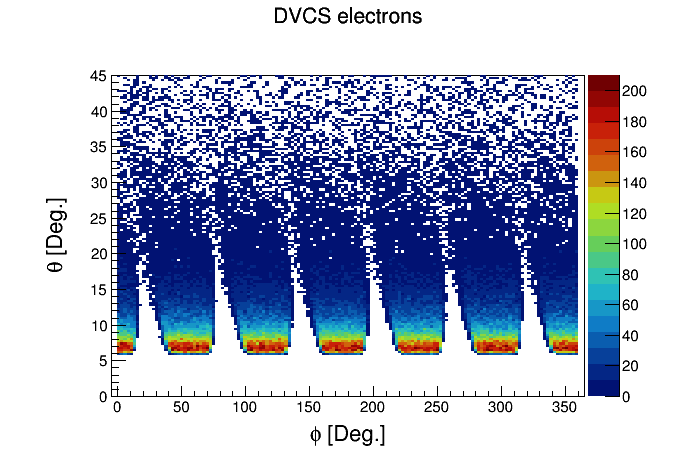}
  \includegraphics[width=0.49\textwidth,clip,trim=10mm 5mm 10mm 20mm]{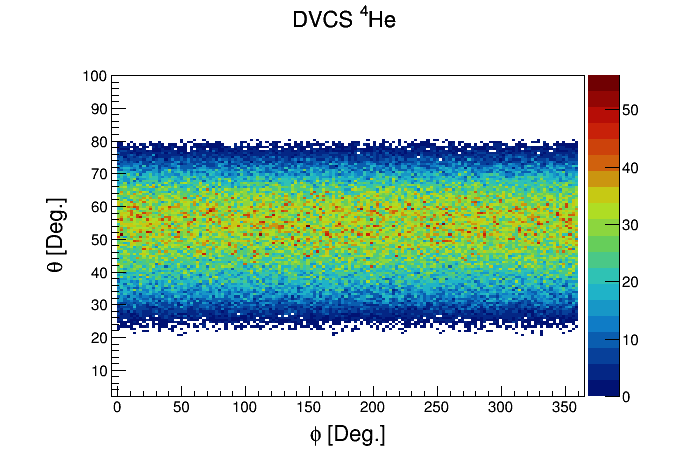}\\
  \includegraphics[width=0.49\textwidth,clip,trim=10mm 5mm 10mm 20mm]{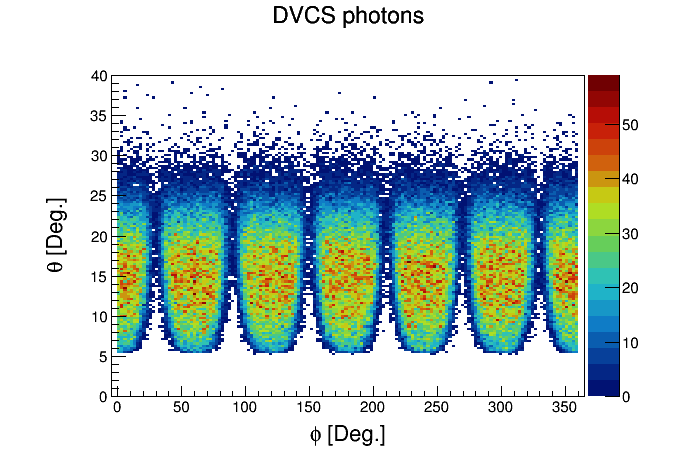}
  \caption{The angles $\theta$ and $\phi$ in the lab frame for all detected 
    particles in coherent DVCS off $^{4}$He. Top-left: the electron, top-right: the 
    $^{4}$He nucleus, and bottom: the real photon.\label{fig:phiversustheta}
  }
\end{figure}

\begin{figure}[htb]
  \includegraphics[width=0.49\textwidth,clip,trim=10mm 5mm 1mm 1mm]{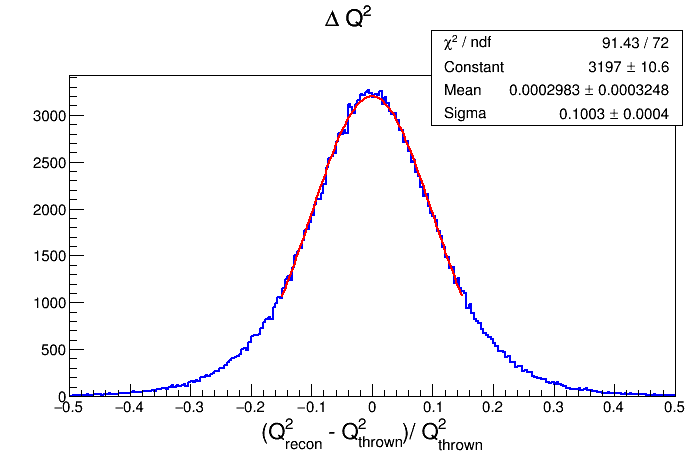}
  \includegraphics[width=0.49\textwidth,clip,trim=10mm 5mm 1mm 1mm]{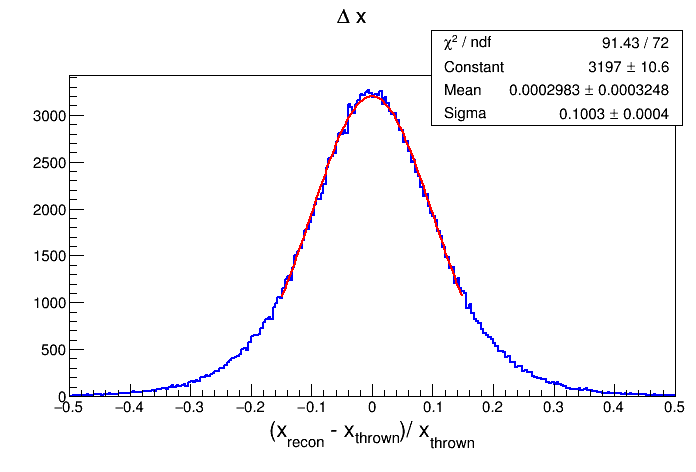}
  \includegraphics[width=0.49\textwidth,clip,trim=10mm 5mm 1mm 1mm]{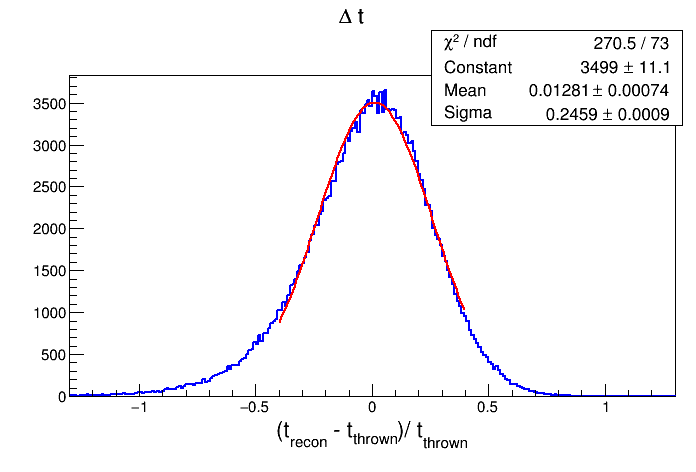}
  \includegraphics[width=0.49\textwidth,clip,trim=10mm 5mm 1mm 1mm]{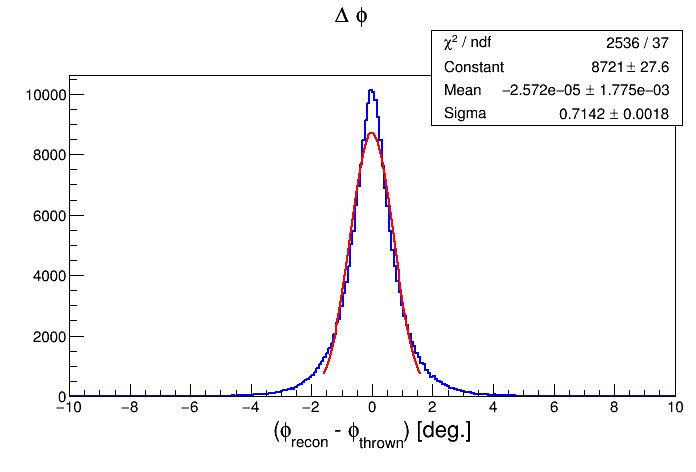}
  \caption{CLAS12 and ALERT resolutions for  $Q^{2}$ , $x_{B}$, $t$ and $\phi$.}
  \label{fig:clasresolutions}
\end{figure}

Figure \ref{fig:Q2-xB} shows the correlations between $Q^{2}$, $x_{B}$ and $-t$ 
variables which are determined by the acceptance of CLAS12 for electrons and 
ALERT for the recoiling $^{4}$He nuclei. Figure \ref{fig:phiversustheta} shows 
the correlation between the azimuthal angle $\phi$ and the polar angle $\theta$ 
in the laboratory frame for all detected particles from the coherent DVCS 
channel. The electron's $\phi$ versus $\theta$ distribution show the six CLAS12 
sectors. Figure \ref{fig:clasresolutions} presents the resolutions
for the kinematic variables $Q^{2}$, $x_{B}$, $t$ and $\phi$.

We define the additional four-vectors:
\begin{eqnarray}
\mathbf{P}^{e\textnormal{$^{4}$He}\gamma}_{X} &=& \mathbf{P}_{\gamma^{*}} 
    + \mathbf{P}_{\textnormal{$^{4}$He}} - (\mathbf{P}_{\gamma} + \mathbf{P'}_{\textnormal{$^{4}$He}})\\
\mathbf{P}^{e\textnormal{$^{4}$He}}_{X} &=& \mathbf{P}_{\gamma^{*}} + 
\mathbf{P}_{\textnormal{$^{4}$He}} -  \mathbf{P'}_{\textnormal{$^{4}$He}} \\
\mathbf{P}^{e\gamma}_{X} &=& \mathbf{P}_{\gamma^{*}} + 
    \mathbf{P}_{\textnormal{$^{4}$He}} - \mathbf{P}_{\gamma}
\end{eqnarray}
In order to access the beam spin asymmetry, one need to identify exclusive 
DVCS-BH events. To ensure exclusivity, only events with a good electron, one 
real photon and a recoiling $^4$He are selected as coherent events. To reduce 
even more the contribution of non-exclusive events, the following kinematical 
cuts have to be applied:
\begin{itemize}
  \item For exclusive coherent DVCS, the virtual photon, the emitted real 
    photon and the recoil helium have to be coplanar. The coplanarity angle 
    ($\Delta \phi$) defined as the difference in angle between these two 
    planes: the first defined by the virtual photon and the recoiling $^4$He 
    and the second defined by the real photon and the virtual one.
  \item Missing energy, mass and transverse momentum ($p^{T}_{X} = 
    \sqrt{(p^{x}_{X})^2 + (p^{y}_{X}})^2$) cuts on 
    $\mathbf{P}^{e\textnormal{$^{4}$He}\gamma}_{X}$.
  \item Missing mass cuts on the $e^{4}HeX$ and $e\gamma X$ systems, which are 
    defined as $(\mathbf{P}^{e\textnormal{$^{4}$He}}_{X})^{2}$ and 
    $(\mathbf{P}^{e\gamma}_{X})^{2}$ respectively.
  \item Cone angle cut between the measured real photon and the missing 
    particle in the $e\textnormal{$^{4}$He}X$ configuration. It is defined as:
    \begin{equation}
      \theta(\gamma, e\textnormal{$^{4}$He}X) = cos^{-1} \left(\frac{\overrightarrow{\mathbf{P}}_{\gamma} \cdot 
          \overrightarrow{\mathbf{P}}^{e\textnormal{$^{4}$He}}_{X}}{|\overrightarrow{\mathbf{P}}_{\gamma^{}}| 
      |\overrightarrow{\mathbf{P}}^{e\textnormal{$^{4}$He}}_{X}|}   \right).
    \end{equation}
\end{itemize}

Even with all the previously presented exclusive cuts, the selected events are 
not all true DVCS events. In our kinematic region, the main contamination comes 
from the exclusive electroproduction of $\pi^{0}$ ($e \textnormal{$^{4}$He} 
\rightarrow e ^{4}He \pi^{0} \rightarrow e \textnormal{$^{4}$He} \gamma \gamma 
$), in which one of the two photons from the $\pi^{0}$ decay passes the 
requirements for the DVCS events.  These events can however be subtracted to 
obtain the true number of DVCS events based on the experimentally measured 
number of $e\textnormal{$^{4}$He}\pi^0$ events.

\begin{figure}[phtb]
  \centering
  \includegraphics[width=0.99\textwidth,clip,trim=5mm 1mm 8mm 1mm]{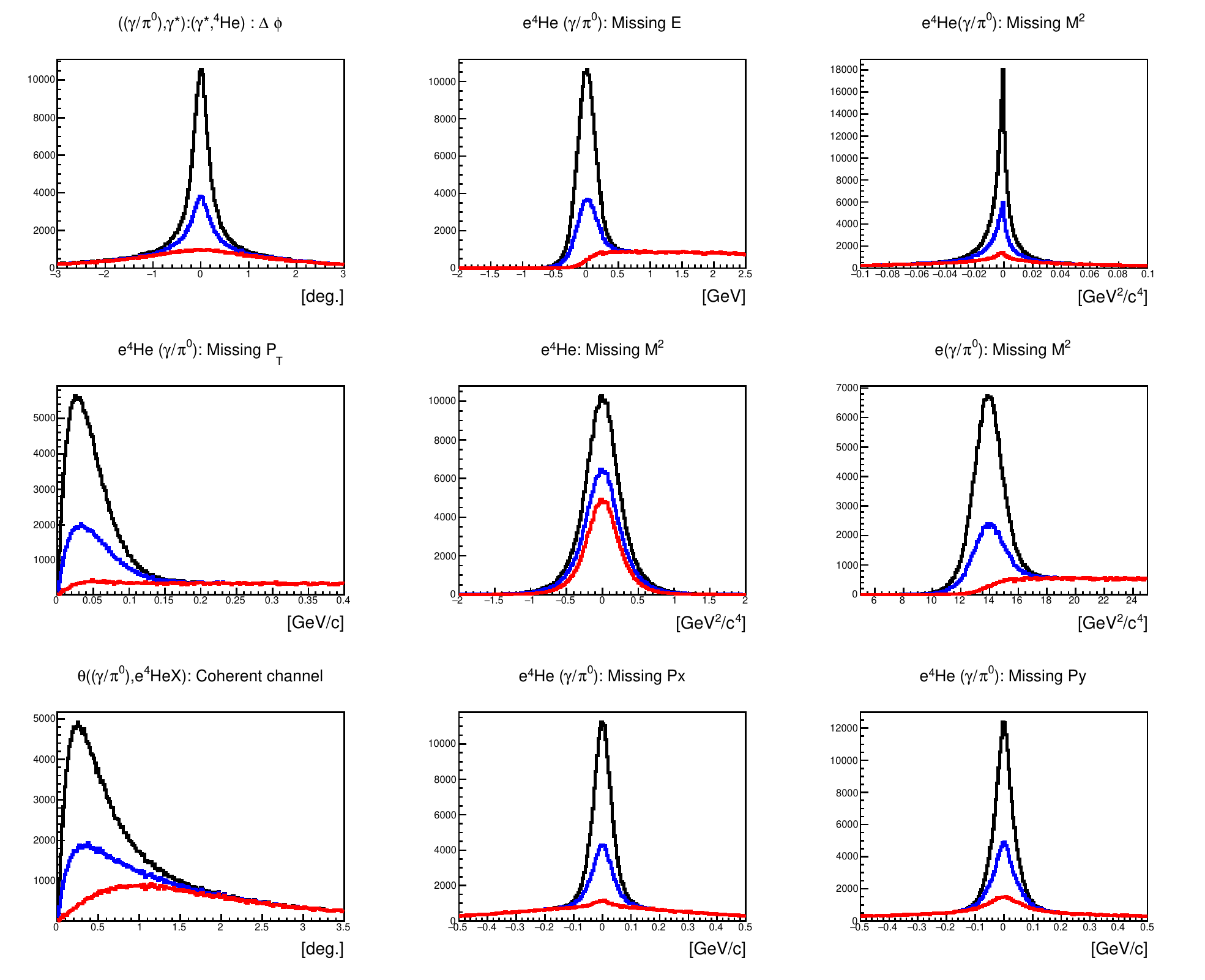}
  \caption{Contribution of non-exclusive ($\pi^{0}$) events to coherent DVCS 
    sample. The blue curves represent the DVCS events. The black curves are the 
    sum of the contributions from exclusive and non-exclusive events. The red 
    curves show the $\pi^{0}$ events where only one photon of the $\pi^{0}$ 
  two-photons decay is detected.\label{fig:before_exclusivity}}
\end{figure}

\begin{figure}[phtb]
  \centering
  \includegraphics[width=0.99\textwidth,clip,trim=5mm 1mm 8mm 1mm]{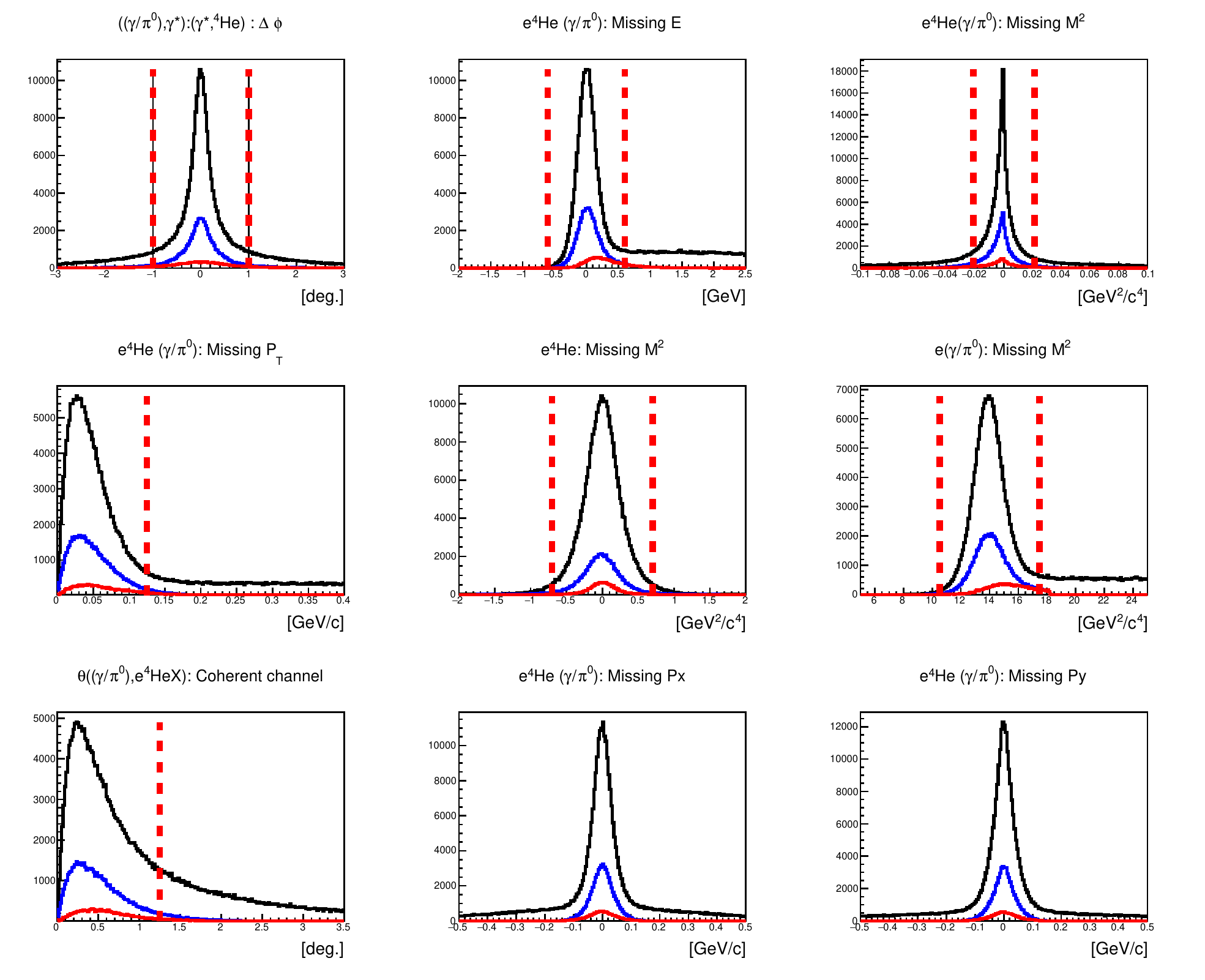}
  \caption{ The effectiveness of the exclusivity cuts on reducing the 
    background contamination to the coherent DVCS sample. The black curves are 
    for the exclusive and non-exclusive events before exclusivity cuts. The blue 
    curves are for the coherent DVCS events which satisfied the exclusivity 
    conditions except for the cuts on the variable being displayed. The red 
    histograms are for the $\pi^{0}$ background contamination that satisfied 
    the exclusivity conditions. The red vertical lines represent $3\sigma$ cuts 
    on each exclusive variable. The missing momentum in $x$  and $y$ directions 
    in the configuration $e\textnormal{$^{4}$He}\gamma X$, are shown for 
    information. See the text for the definition of the shown exclusive 
    variables.  }
  \label{fig:after_exclusivity}
\end{figure}

Figure \ref{fig:before_exclusivity} illustrates the contributions of exclusive 
$\pi^{0}$, wit only one photon of the decay is detected, events to the coherent 
DVCS data sample.  The number of simulated $\pi^{0}$ events is three times the 
number of the simulated single photon production events. The dependencies of 
the contamination from $\pi^{0}$ versus the exclusive distributions are shown 
in different panels of figure \ref{fig:before_exclusivity}. The black 
histograms are the total simulated events (DVCS and $\pi^{0}$ events), while 
the blue histograms are for the simulated DVCS events only, and the red 
histograms are for the simulated $\pi^{0}$ events where only one photon of the 
$\pi^{0}$ two-photons decay is detected due to CLAS acceptance and may 
contaminate the DVCS sample. Figure \ref{fig:after_exclusivity} illustrates the 
effectiveness of the exclusivity cuts on reducing the background contamination 
to the coherent DVCS sample.

\begin{figure}[phtb]
  \includegraphics[width=0.49\textwidth,clip,trim=20mm 1mm 25mm 1mm]{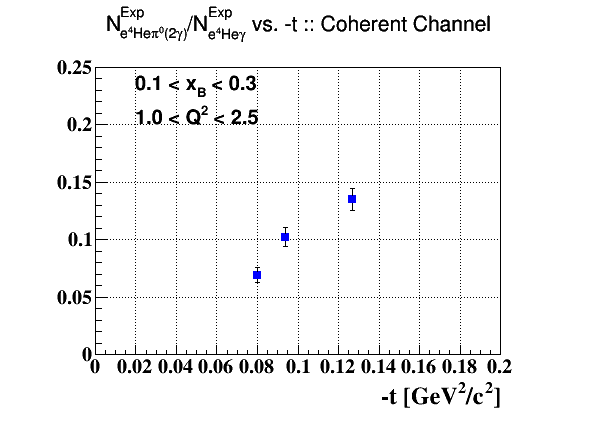}
  \includegraphics[width=0.49\textwidth,clip,trim=20mm 1mm 25mm 1mm]{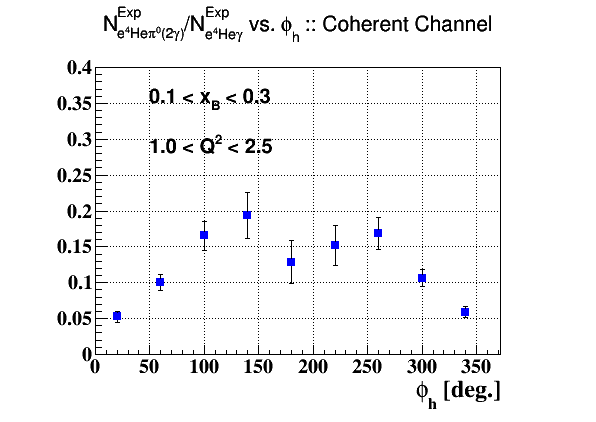}
  \caption{The $\pi^{0}$ to coherent DVCS production yield versus 
    four-momentum transfer (-t) integrated over the photon angle $\phi$ 
    (left), and versus $\phi$ (right) integrated over t as determined from 
    CLAS-eg6 analysis \cite{eg6_note}. Both plots are for 
  1.0~GeV$^{2}$~$<$~Q$^{2}$~$<$~2.5 GeV$^{2}$ and 0.1~$<~x_{B}~<$~0.3.}
  \label{fig:pi0yield}
\end{figure}

As mentioned previously, and as seen from the ratio of black to blue curves in 
figure \ref{fig:before_exclusivity}, the simulated $\pi^{0}$ events were three 
times the number of the DVCS events.  The true $\pi^{0}$ to coherent DVCS 
production ratio has been measured in eg6-experiment \cite{eg6_note} to be 
5$\%$ to 20$\%$ as can be seen in figure \ref{fig:pi0yield}.

Figure \ref{fig:after_exclusivity} shows the remaining $\pi^{0}$ events after 
the exclusivity cuts that contaminate the DVCS sample. The true number of the 
coherent DVCS events can be formulated as:
\begin{equation}
  N^{True}_{e\textnormal{$^{4}$He}\gamma} = N^{Exp.}_{e\textnormal{$^{4}$He}\gamma} -  N^{Exp.}_{e\textnormal{$^{4}$He}\pi^{0}(\gamma)},
  \label{equ_back_1}
\end{equation}
where $N^{True}_{e\textnormal{$^{4}$He}\gamma}$, 
$N^{Exp.}_{e\textnormal{$^{4}$He}\gamma}$ and $ 
N^{Exp.}_{e\textnormal{$^{4}$He}\pi^{0}(\gamma)}$ are the true number of 
coherent DVCS events, the experimentally measured number of 
$e\textnormal{$^{4}$He}\gamma$ events and the contamination number, 
respectively. The contamination can be calculated by using real data and 
simulation. We define, for each kinematic bin and for each beam helicity state
\begin{equation}
  N^{Exp.}_{e\textnormal{$^{4}$He}\pi^{0}(\gamma)} = 
  \frac{N^{Sim.}_{e\textnormal{$^{4}$He}\pi^{0}(\gamma)}}{N^{Sim.}_{e\textnormal{$^{4}$He}\pi^{0}(\gamma 
  \gamma)}} * N^{Exp.}_{e\textnormal{$^{4}$He}\pi^{0}(\gamma \gamma)},
  \label{equation: background_equ}
\end{equation}
where $N^{Exp.}_{e\textnormal{$^{4}$He}\pi^{0}(\gamma \gamma)}$ is the number 
of measured $e\textnormal{$^{4}$He}\pi^{0}$ events, for which both photons of 
the $\pi^{0}$ have been detected. The quantity 
$\frac{N^{Sim.}_{e\textnormal{$^{4}$He}\pi^{0}(\gamma)}}{N^{Sim.}_{e\textnormal{$^{4}$He}\pi^{0}(\gamma 
\gamma)}} $ is the acceptance ratio for detecting an 
$e\textnormal{$^{4}$He}\gamma$ event that originates from an 
$e\textnormal{$^{4}$He}\pi^{0}$ event. It can be derived from Monte-Carlo 
simulations by generating and simulating $e\textnormal{$^{4}$He}\pi^{0}$.  
$N^{Sim.}_{e\textnormal{$^{4}$He}\pi^{0}(\gamma)}$ is the number of such events 
passing the DVCS requirements, while 
$N^{Sim.}_{e\textnormal{$^{4}$He}\pi^{0}(\gamma \gamma)}$ is the number of 
simulated $e\textnormal{$^{4}$He}\pi^{0}$ events passing the exclusivity cuts 
for $e^{4}He\pi^{0}$ events.

\begin{figure}[htb]
  \centering
      \includegraphics[width=0.49\textwidth,clip,trim=5mm 1mm 10mm 15mm]{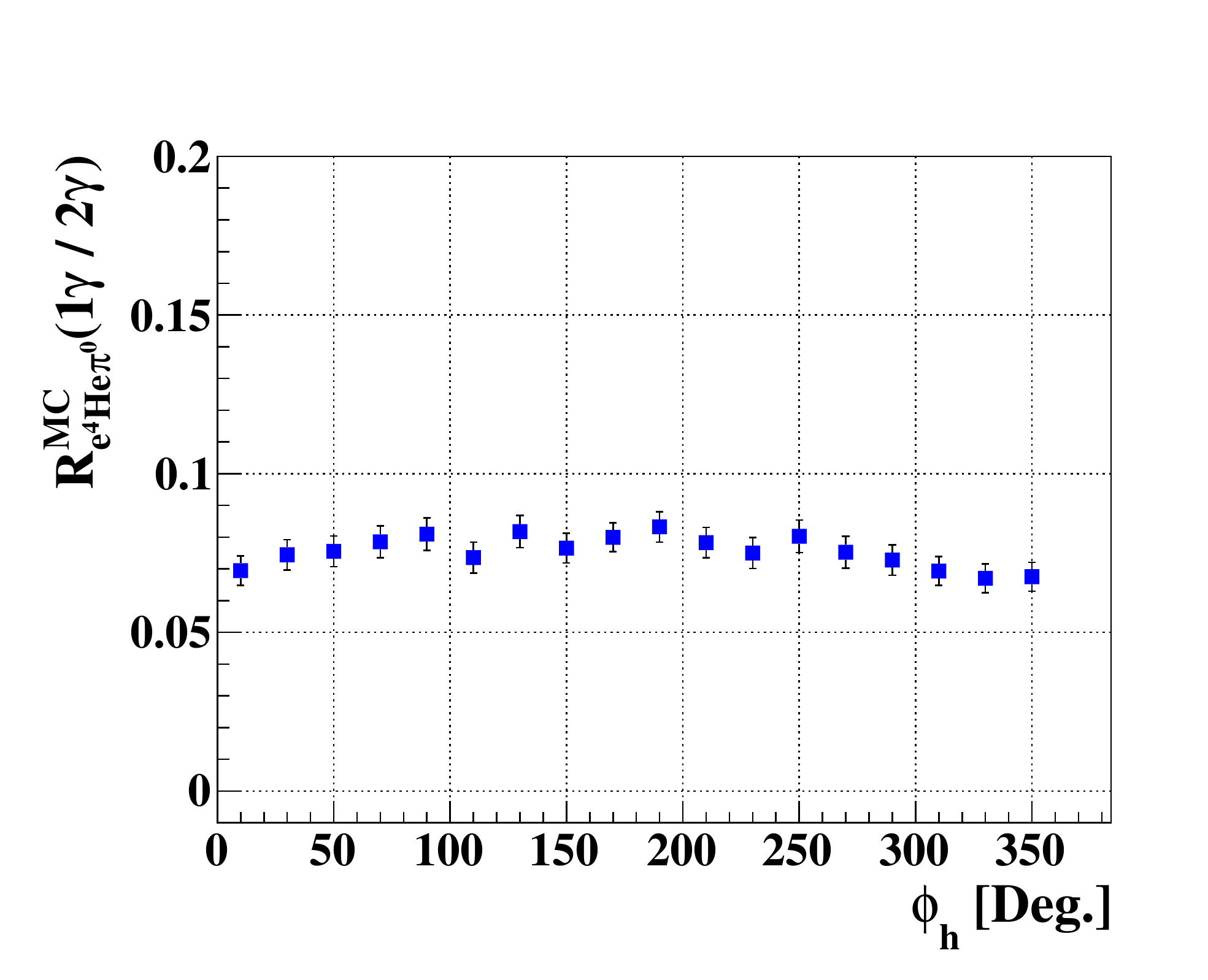}
  \caption{The coherent DVCS accpetance ratio as a function of $\phi$ 
    integrated over $t=$~\SIrange{0.06}{0.2}{\GeV^2}, $Q^{2}=$~\SIrange{1.0}{2.5}{\GeV^2}, and 
  $x_{B}=$~\numrange{0.1}{0.3} using both CLAS12 and ALERT detectors.}
  \label{fig:backgroundratio}
\end{figure}

Figure \ref{fig:backgroundratio} shows the coherent acceptance ratio as a 
function of $\phi_{h}$ using the CLAS12-ALERT setup. The mean value of the 
acceptance ratio for the coherent channel is around 8$\%$.

The polarized beam of CEBAF and the large acceptance of CLAS will allow us to 
extract the beam spin asymmetry $A_{LU}$ for various bins in Q$^2$, $x_B$, and 
$t$ and $\phi$ for both the coherent DVCS and the $\pi^{0}$ electroproduction
processes. The beam spin asymmetry in each bin is defined as:
\begin{equation}
  A_{LU} = \frac{1}{P_{B}} \frac{N^{+} - N^{-}}{N^{+} + N^{-} }.
\end{equation}
where $P_{B}$ is the beam polarization, and $N^{+}$ and $N^{-}$ are the number 
of events detected with positive and negative electron helicity, respectively.  
The statistical uncertainty of $A_{LU}$ is
\begin{equation}
  \sigma_{A_{LU}} = \frac{1}{P_{B}} \sqrt{ \frac{1 - (P_{B}A_{LU})^{2}}{N}}
\end{equation}
where $N (= N^{+} + N^{-}) $ is the total number of measured events.

%
\subsection{Event Generator}
\begin{figure}[htb]
  \centering
\includegraphics[width=0.48\textwidth,trim=25mm 1mm 10mm 20mm]{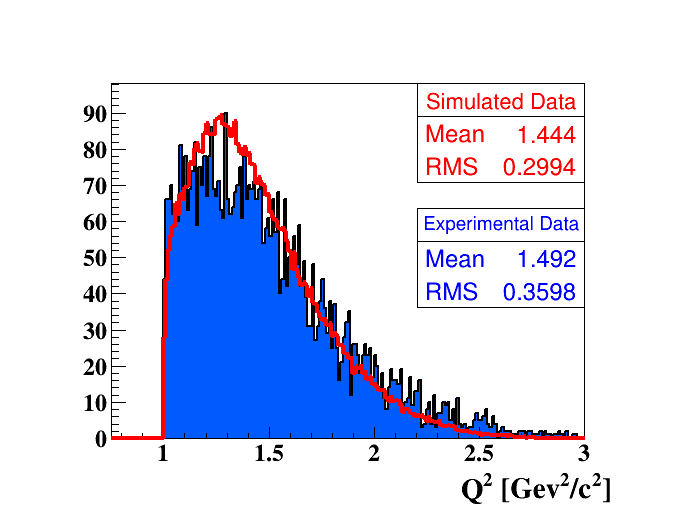}
\includegraphics[width=0.48\textwidth,trim=25mm 1mm 10mm 20mm]{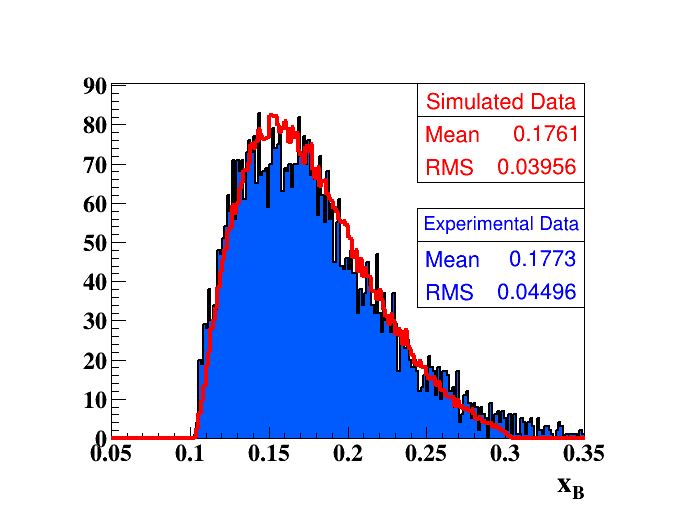}
\caption{Comparison between the simulated $e^{4}He\gamma$ DVCS events (in red 
lines) and the experimental DVCS events from eg6 (in shaded blue) with 6 GeV beam
as a function of the kinematic variables: $Q^{2}$ and $x_{B}$ \cite{eg6_note}. }
\label{fig:coh_comparison_with_simulation_1}
\end{figure}

In order to make projections of our results, we have used the following 
parametrization of the cross section which parameters were calibrated to 
reproduce the DVCS and exclusive $\pi^{0}$ electroproduction data from CLAS at 
6~GeV~\cite{FX_thesis}:
\begin{equation}
  \frac{d^{4}\sigma}{dQ^{2}dx_{B}dtd\phi} \propto 
  \left(\frac{Q^{2}_{0}}{Q^{2}}\right)^{\alpha} ~\frac{1}{1 + (\frac{x_{B} - 
  x_{c}}{c})^{2}} ~\frac{1}{(1+bt)^{\beta}} ~(1-d(1-\cos(\phi))).
  \label{equ:event_generator}
\end{equation}
This parametrization is the product of four factors which reproduce the DVCS 
and $\pi^{0}$, characteristics as follows:
\begin{itemize}
  \item the $Q^{2}$-dependent term with: $Q^{2}_{0}$ the minimum allowed value 
    and $\alpha$ a parameter which controls the shape of the distribution.   \item 
    the $x_{B}$ term accounts for the dependence of the cross section on the 
    parton distribution functions, with $x_{c}$ the mean value of the Bjorken 
    variable $x_{B}$.
  \item the $t$ term accounts for the t-dependence of the elastic form factors of 
    the helium and of the proton, via the parameters $b$ and $\beta$.
  \item the $\phi$ term accounts for the cross section dependence on this angle, 
    via the parameter $d$.
\end{itemize}
To reflect the change in the center of mass energy due to the higher beam 
energy of this proposal compared to E08-024 experiment, the parameter $x_{c}$ 
(the mean value of $x_{B}$) is calculated from the DIS mean kinematic values, 
while the parameters $b$ and $\beta$ were scaled with respect to the center of 
mass energy change from 6 GeV to 11 GeV. Table 
\ref{Table:event_generator_values} shows the values of the parameters used for 
the cross section parametrization of the four channels of interest: 
$e^{4}He\gamma$, $e^{4}He\pi^{0}$, $ep\gamma$, and $ep\pi^{0}$.
Figure \ref{fig:coh_comparison_with_simulation_1} shows a comparison between 
the experimentally identified coherent DVCS events from eg6 dataset and the 
simulated DVCS events as a function of the kinematic variables $Q^{2}$ and 
$x_{B}$.
\begin{table}[htb]
\begin{center}
\bgroup
\tabulinesep=1.3mm
\begin{tabu}{lccccc}
\tabucline[2pt]{-}                                                   
Parameter & Units & $e^{4}$He$\gamma$  & $e^{4}$He$\pi^{0}$ & $ep\gamma$  & $ep\pi^{0}$ \\
\tabucline[1pt]{-}                                                   
$Q^{2}_{0}$ &\si{\GeV^2/c} & 1.0 & 1.0 & 1.0 & 1.0 \\
$\alpha$ & & 2.5 & 3.0 & 1.5 & 1.5 \\
b &\si{\GeV^2/c} & -6.0  & -8.8  & -1.408 & -1.408 \\
$\beta$ & & 6.5 & 7.3 & 4.0 & 1.5\\
$x_{c}$ & & 0.27 & 0.3 & 0.2 & 0.5\\
c & & 0.2 & 0.3 & 0.2 & 0.5 \\
d & & 0.4 & 0 & 0.4 & 0 \\
\tabucline[2pt]{-}                                                   
\end{tabu}
\egroup
\caption{Values of the parameters used in DVCS event generator.}
\label{Table:event_generator_values}
\end{center}
\end{table}
%
\subsection{Projections}
\label{sec:DVCS-projs}
\begin{figure}[htb]
  \centering
  {%
    \setlength{\fboxsep}{0pt}%
    \setlength{\fboxrule}{1pt} \fbox{
  \includegraphics[width=0.9\textwidth,clip,trim=10mm 5mm 10mm 0mm]{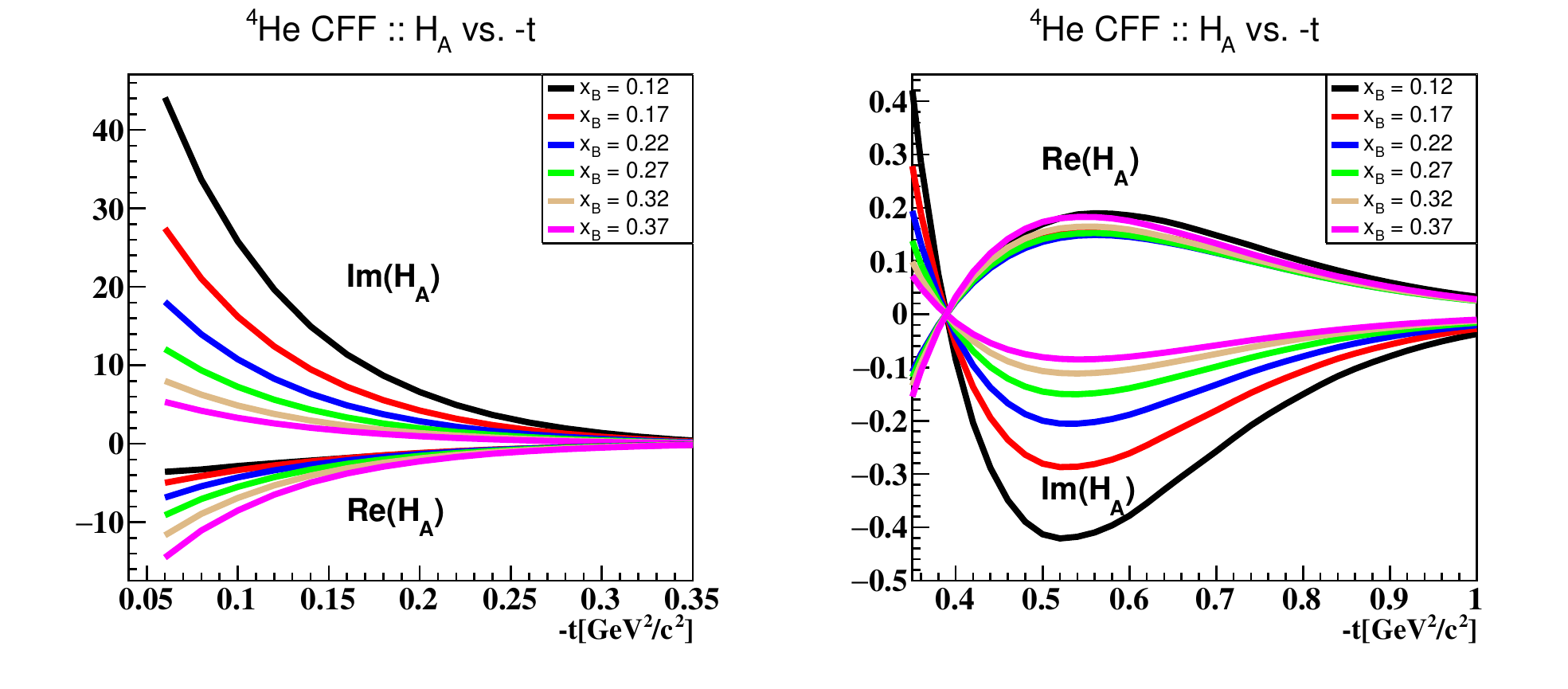}
}}
  \caption{From the impulse approximation model, the imaginary and the real part 
    of the $^{4}$He CFF as a function of $-t$, low $-t$ on the left and high $-t$ 
  on the right, at fixed $x_{B}$ values.}
  \label{fig:Impulse_approx_CFF}
\end{figure}

The projected precisions of the beam-spin asymmetries and Compton form factor 
for DVCS on $^{4}$He are presented in this section. Based on the Impulse 
Approximation model \cite{Guzey:2003jh}, the real and the imaginary parts of 
the $^{4}$He CFF were calculated and fed into our event generator to produce 
coherent DVCS events with beam-spin asymmetries following the formalism 
presented in chapter 2. Figure \ref{fig:Impulse_approx_CFF} presents the $t$ 
dependence of the real and the imaginary parts of the $^{4}$He CFF at different 
values of $x_{B}$.

\begin{figure}[htb]
  \centering
    \includegraphics[width=0.55\textwidth,clip,trim=5mm 1mm 10mm 23mm]{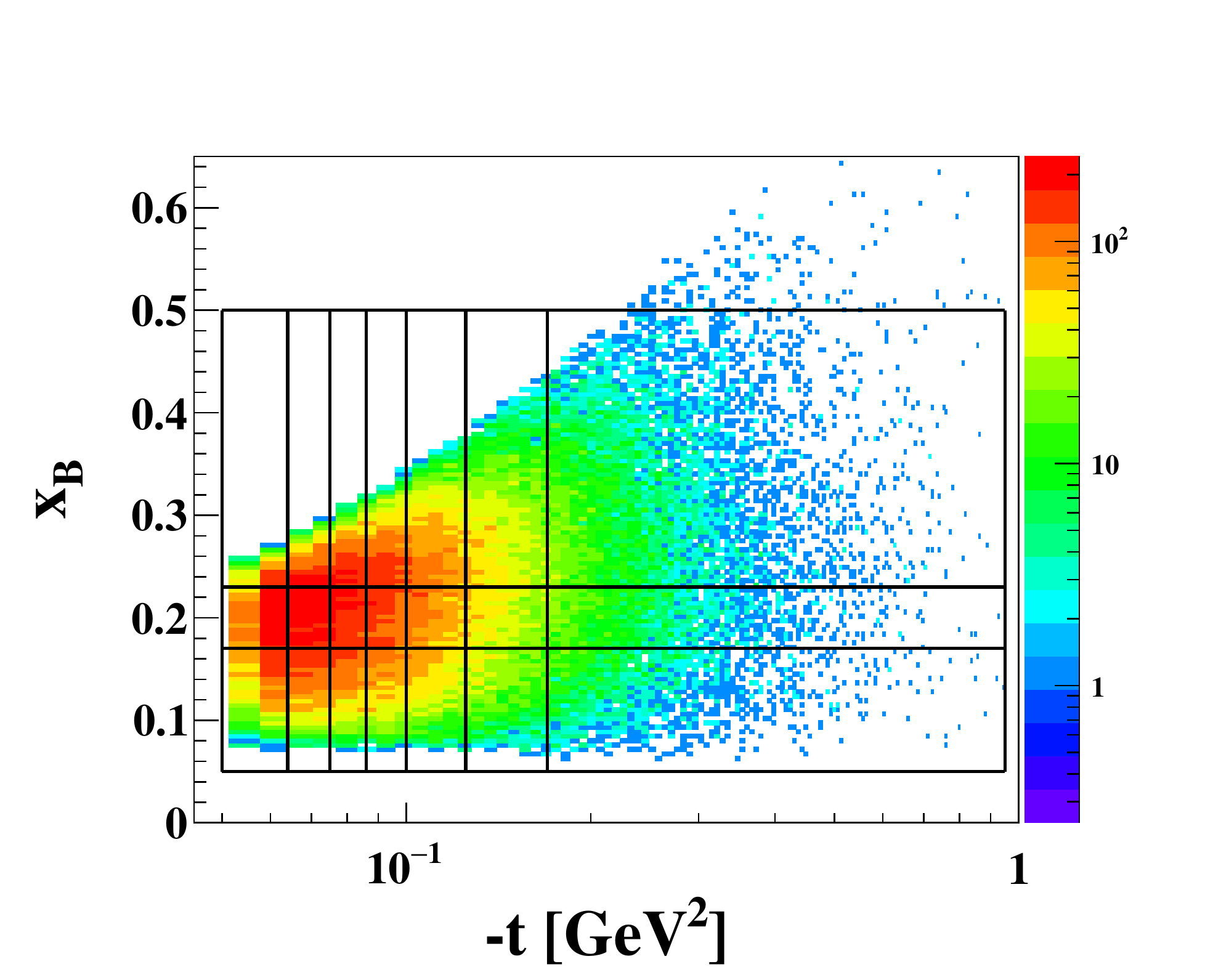}
  \caption{Data binning in $x_{B}$ vs $-t$ space.\label{fig:binning_x_t}}
\end{figure}

\begin{figure}[htb]
  \centering
  \includegraphics[width=0.48\textwidth,clip,trim=5mm 1mm 3mm 6mm]{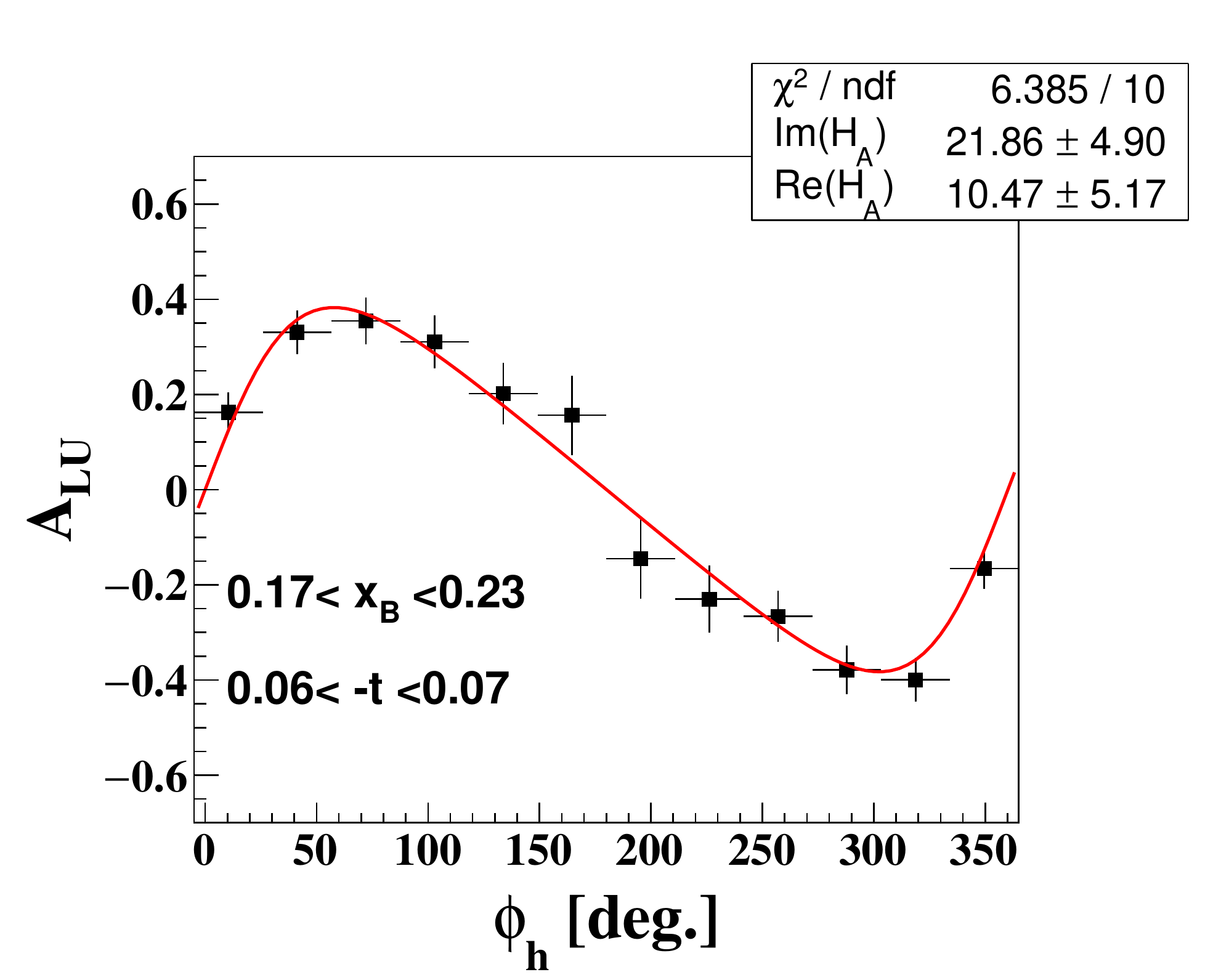}
  \includegraphics[width=0.48\textwidth,clip,trim=5mm 1mm 3mm 6mm]{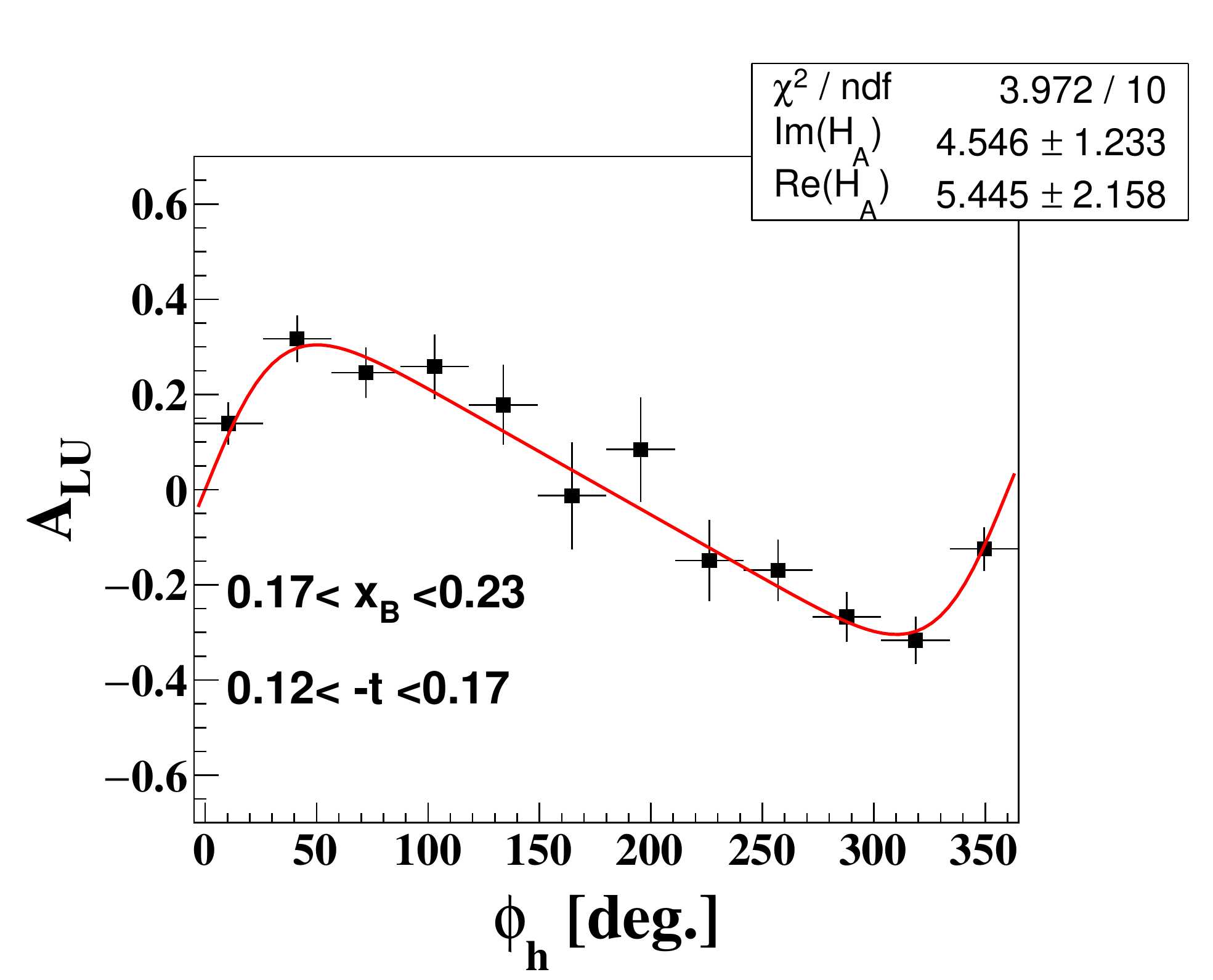}
  \caption{The coherent beam-spin asymmetry projections as a function of the 
    angle $\phi$ between the leptonic and the hadronic planes, for two different 
    bins $-t$ at the same $x_{B}$ range, integrated over $Q^{2}$ range. The red 
    solid curves represent a fit to the data in the full form of the asymmetry, 
    equation \ref{eq:A_LU-coh}, with the real and the imaginary parts of the CFF as 
  the free parameters of the fit. \label{fig:ALU-projections} }
  
\end{figure}

\begin{figure}[htb]
  \centering
  \includegraphics[width=0.6\textwidth]{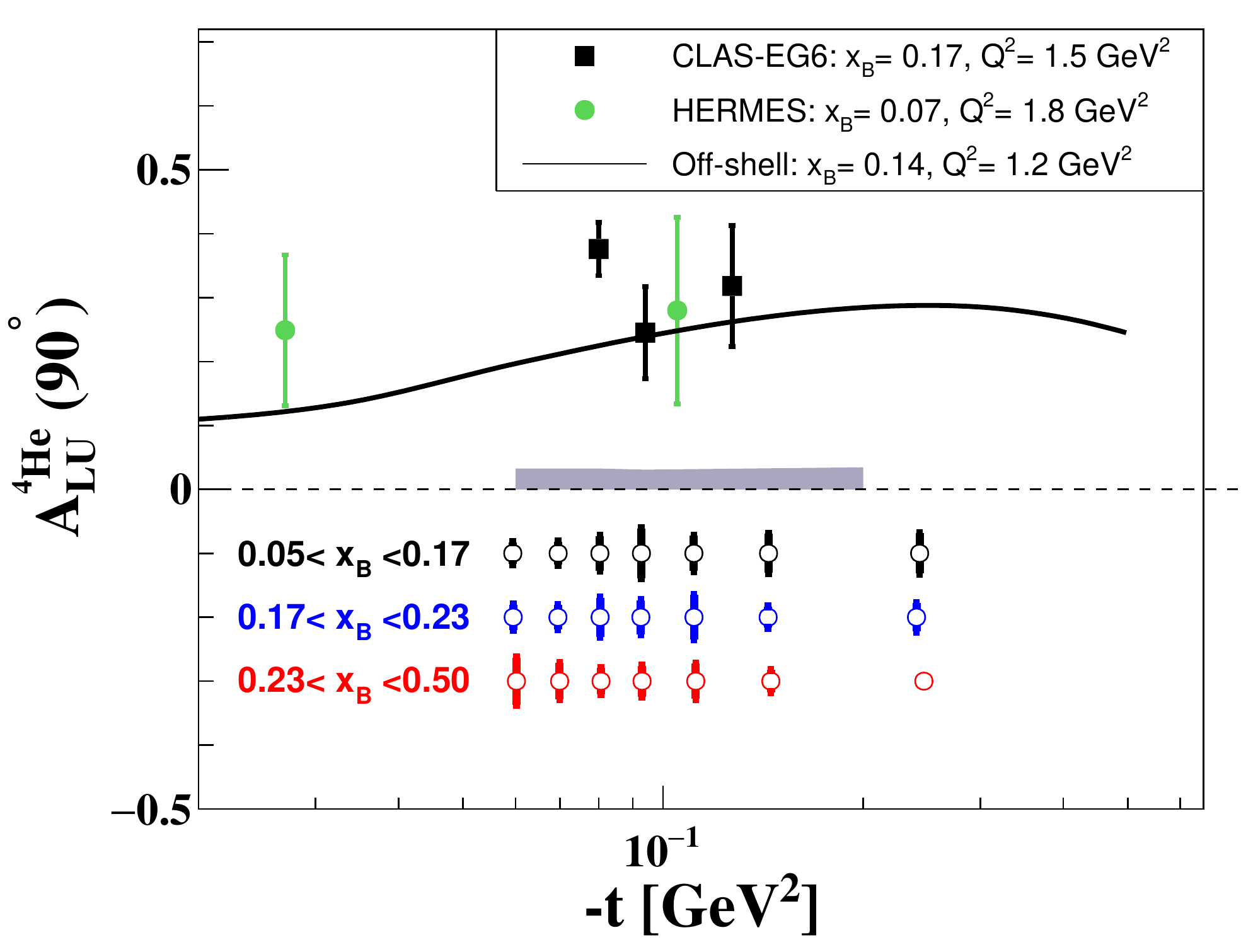}
  \caption{Projected precision for the $A_{LU}$ (90$^{\circ}$), from the fit, for 
    coherent DVCS on $^{4}$He versus $-t$ compared to the previous measurements 
    from CLAS-eg6 (black squares), HERMES (green circles) and spectral function 
  calculations (LT curves).}
  \label{fig:ALU-projections-90}
\end{figure}

\begin{figure}[htb]
  \centering
  \includegraphics[width=0.48\textwidth,clip,trim=4mm 1mm 15mm 0mm]{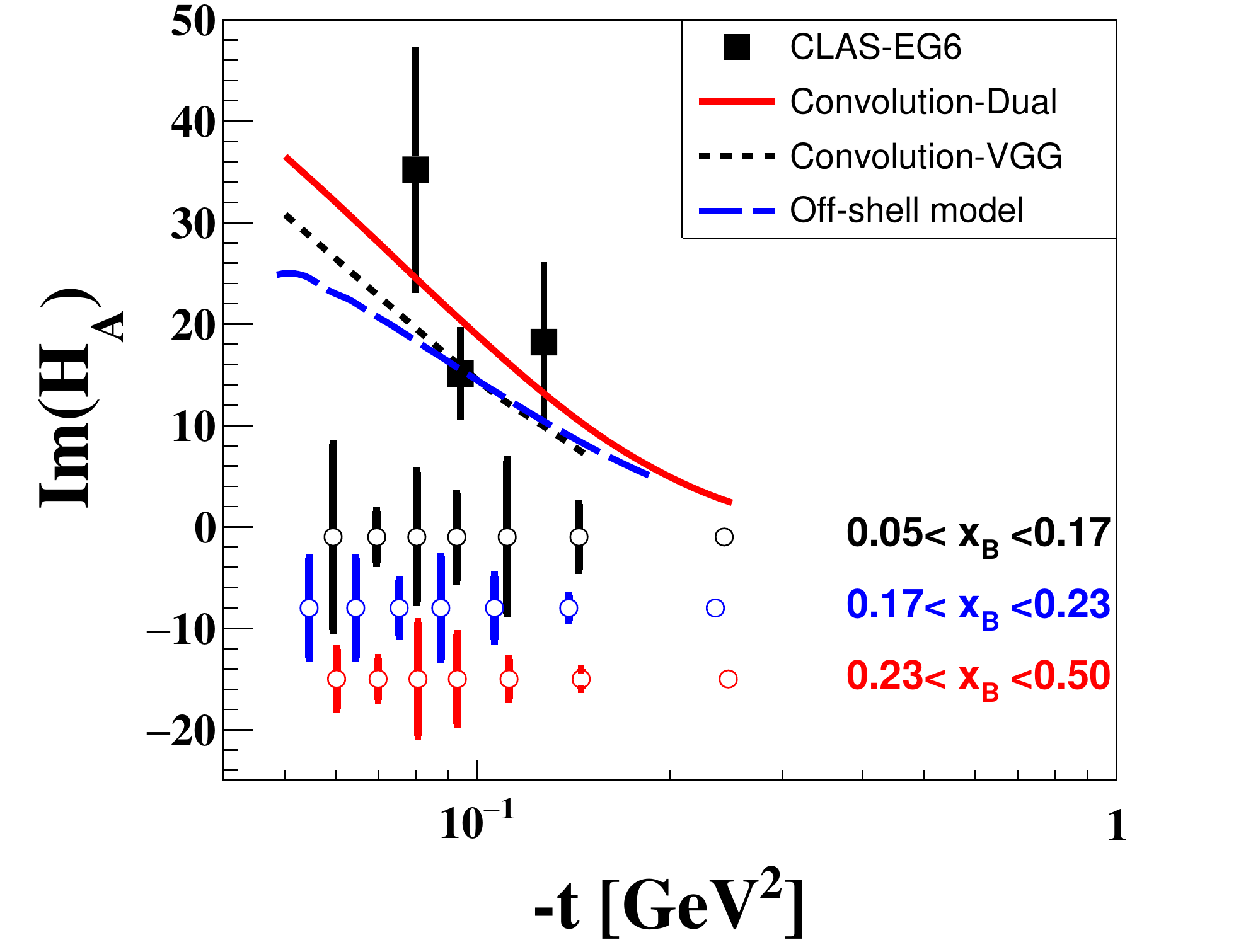}
  \includegraphics[width=0.48\textwidth,clip,trim=4mm 1mm 15mm 0mm]{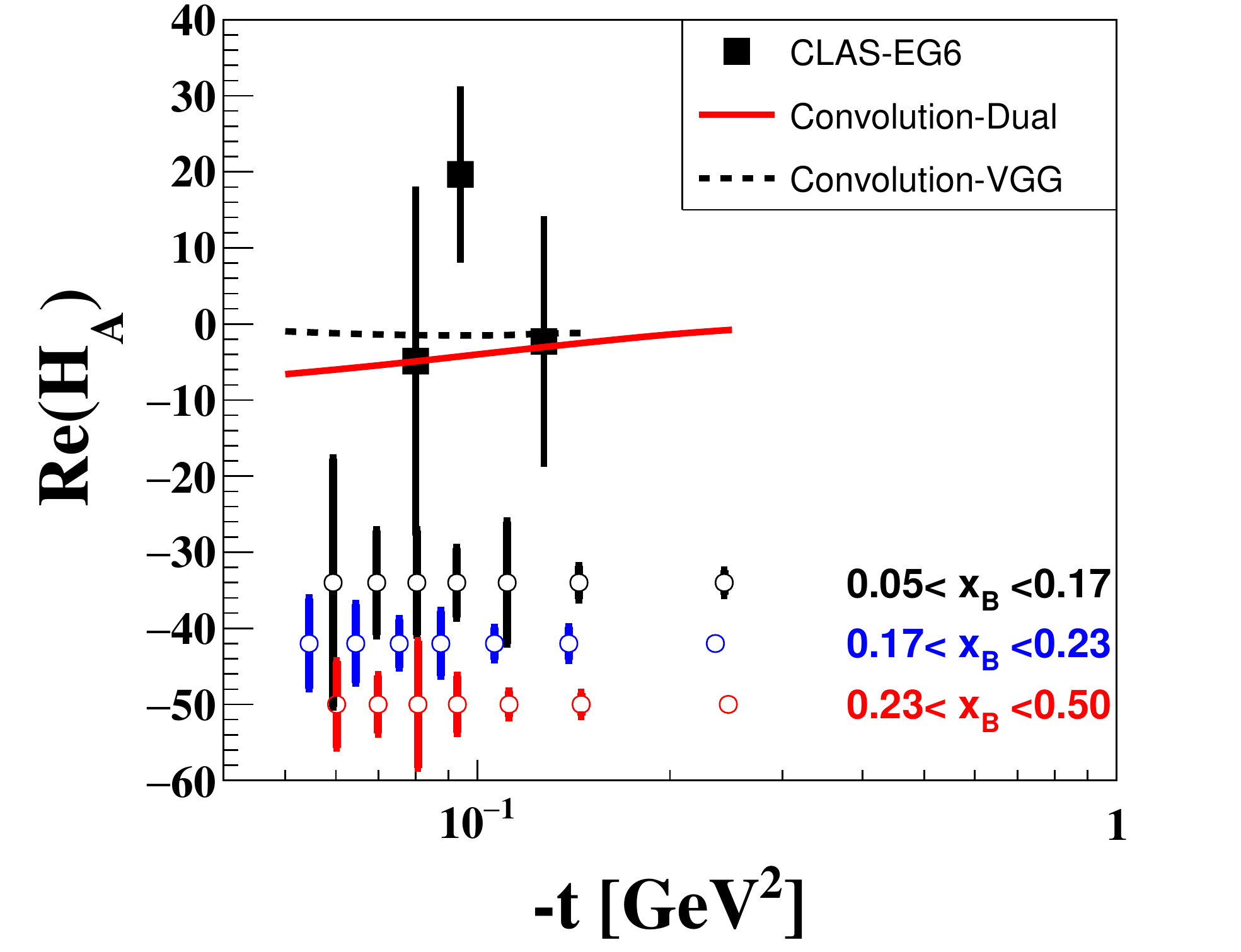}
  \caption{The projected statistical uncertainties for the  imaginary (top) and 
    real (bottom) parts of the CFF $H_{A}$, from the fits, as a function of $-t$ at 
  fixed ranges in $x_{B}$}
  \label{fig:CFF_projections}
\end{figure}

Figure \ref{fig:binning_x_t} shows the proposed binning in $x_B$ versus $-t$ 
space. The simulated data is integrated over the full $Q^{2}$ range. For the 
BSA $A_{LU}$ dependence on $-t$, the data has been binned into three bins in 
$x_{B}$, 7 bins in $-t$ and 12 bins in $\phi$.  The statistical error bars are 
calculated for 20 days at a luminosity of $0.75 \times 
10^{34}$~cm$^{-2}$s$^{-1}$ per nucleus (jointly with Tagged EMC proposal 
request) and 10 days at a luminosity of $1.5 \times 10^{34}$~cm$^{-2}$s$^{-1}$ 
per nucleus specifically dedicated to this proposal. The assumed beam 
polarization is 80$\%$. Figure \ref{fig:ALU-projections} shows the 
reconstructed beam-spin asymmetries as a function of the angle $\phi$ for two 
bins in $-t$ at a fixed $x_{B}$ value presenting a high and a low statistic 
bins. The projected precision of $A_{LU}$ at $\phi$ equal to 90$^{\circ}$ for 
the different bins is presented in figure \ref{fig:ALU-projections-90}. The 
projected uncertainties on the reconstructed real and imaginary parts of the 
CFF are shown in figure \ref{fig:CFF_projections}.


For the purpose of validating our projection results, that are extracted based 
on the parametrized cross section from CLAS-EG6 experiment, we performed an 
additional exercise. In Appendix \ref{app:BH_cross_section}, we extracted our
projections based on generating and simulating pure coherent BH events, which 
dominates most of the accessible phase-space of JLAB. In conslusion, our 
projections are well reproduced, within the statistical error bars, by the 
well-known pure BH process.

\subsection{Systematic uncertainties}

It is particularly convenient to use the BSA $A_{LU}$ as a DVCS observable, 
because most of the experimental systematic uncertainties, such as 
normalization and efficiencies that appear in the cross sections cancel out in 
the asymmetry ratio. However, some systematic uncertainties remain and they 
still contribute to the measured $A_{LU}$. The main known sources of systematic 
uncertainties are: the DVCS selection cuts, the fitting sensitivity to our 
binning, the beam polarization and the background (non exclusive $\pi^0$) 
acceptance ratio. In the following, we present estimates of the contribution 
from each source based on our prior knowledge during CLAS-eg6 DVCS analysis 
\cite{eg6_note} and our simulation studies of the proposed ALERT detector.

In order to evaluate the systematic uncertainties stemming from the DVCS 
selection cuts, the eg6-analysis was repeated with changing the width of the 
exclusive cuts. The resulting systematic uncertainty to the $A_{LU}$ asymmetry
was around 8$\%$ for the coherent DVCS channel. Because of the important 
improvement we expect with ALERT in terms of resolutions, we expect this 
uncertainty to be reduced to 5$\%$.

Regarding the sensitivity of the fit results to our binning, the eg6 data were 
binned into two different bins in $\phi$ and the reconstructed asymmetries were 
compared. The associated systematic uncertainty for $A_{LU}$ at $\phi = 90 
^{\circ}$ was found to be of 5.1$\%$. For the proposed measurements, we expect 
to achieve higher statistics and therefore we reduced the expected systematics 
to 3$\%$.
   
The beam polarization will be measured during the experiment by the Hall B 
M\o{}ller polarimeter. This polarimeter measures the angular distribution of 
the M\o{}ller electrons to obtain the beam polarization. The precision of the 
Hall B M\o{}ller polarimeter was measured to be around 3.5$\%$ 
\cite{PhysRevSTAB.7.042802}, which is expected to be improved with the upgrade.  
We assume therefore a 3.5$\%$ systematic uncertainty on the measured 
asymmetries similar to what was achieved during 6 GeV run.

To estimate the systematic uncertainty associated with the calculated 
acceptance ratio (R), two techniques can be used. The first is via repeating 
the analysis by implementing R differently, while the second technique is by 
using two generating models to calculate R. In CLAS-eg6 analysis both methods 
were investigated. A maximum variation of 0.6$\%$ has been observed on the 
coherent $A_{LU}$ at $\phi = 90^{\circ}$. An upper limit of 1$\%$ is assumed 
for the proposed measurements. 


The total systematic uncertainty on the measured $A_{LU}$ at $\phi = 90^{\circ}$ is the quadratic sum of the previously 
described individual uncertainties. Table \ref{Table:systematic_uncertainties} 
summarizes the systematic uncertainties for both CLAS-eg6 and the proposed 
measurements. 

\begin{table}[htb]
\begin{center}
\bgroup
\tabulinesep=1.3mm
  \begin{tabu}{lccc}
\tabucline[2pt]{-}                                                   
    Systematic source &  CLAS-EG6  & Proposed experiment &Systematic type\\
\tabucline[1pt]{-}                                                   
    DVCS cuts         & \SI{8}{\percent}          & 5$\%$   & bin to bin\\
    Data binning      & 5.1$\%$         & 3$\%$   & bin to bin\\
    Beam polarization & 3.5$\%$         & 3.5$\%$ & Normalization\\
    Acceptance ratio  & 0.6$\%$         & 1$\%$   & bin to bin\\
    \textbf{Total}    & \textbf{10}$\%$ & 7$\%$   & bin to bin\\
\tabucline[2pt]{-}                                                   
  \end{tabu}
  \egroup
  \caption{ The systematic uncertainties on the measured coherent beam-spin 
  asymmetries at $\phi = 90^{\circ}$ from CLAS-eg6 and the proposed experiment.}
  \label{Table:systematic_uncertainties}
\end{center}
\end{table}

\FloatBarrier

\section{Exclusive \texorpdfstring{$\phi$}{Phi} Electroproduction}

\begin{figure}[htb]
  \centering
      \includegraphics[width=0.99\textwidth,clip,trim=6mm 5mm 6mm 5mm]{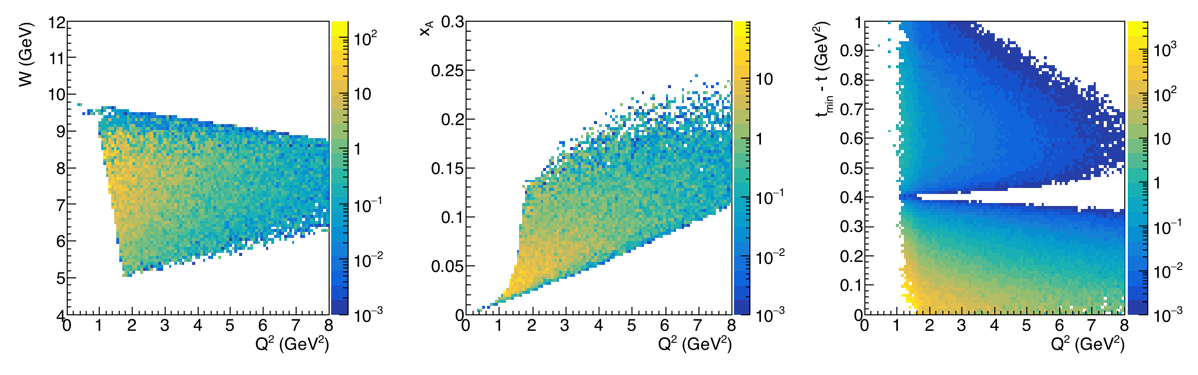}
  \caption{Total cross-section weighted kinematic distributions for electron, 
    kaon, and $^4$He coincidence events from $\phi$ production simulations.  Z-axis 
  units are in nb.} \label{fig:phiAcc}
\end{figure}

The CLAS12 detector will be used to detect one electron along with either a 
single kaon or a kaon pair, while the ALERT detector will detect the recoiling 
$^4$He.  The detected electron is constrained to the forward CLAS12 acceptance, 
and Figure~\ref{fig:phiAcc} shows the expected kinematic coverage after 
acceptance.  The CLAS12 detector acceptance was simulated using the java-based 
fastMC, with resolution smearing taken from the older fortran fastMC. ALERT 
detector acceptance was taken from the above sections and  351 simulation 
results.

\begin{figure}[htb]
  \centering
  \includegraphics[width=0.5\textwidth,clip,trim=0mm 0mm 0mm 20mm]{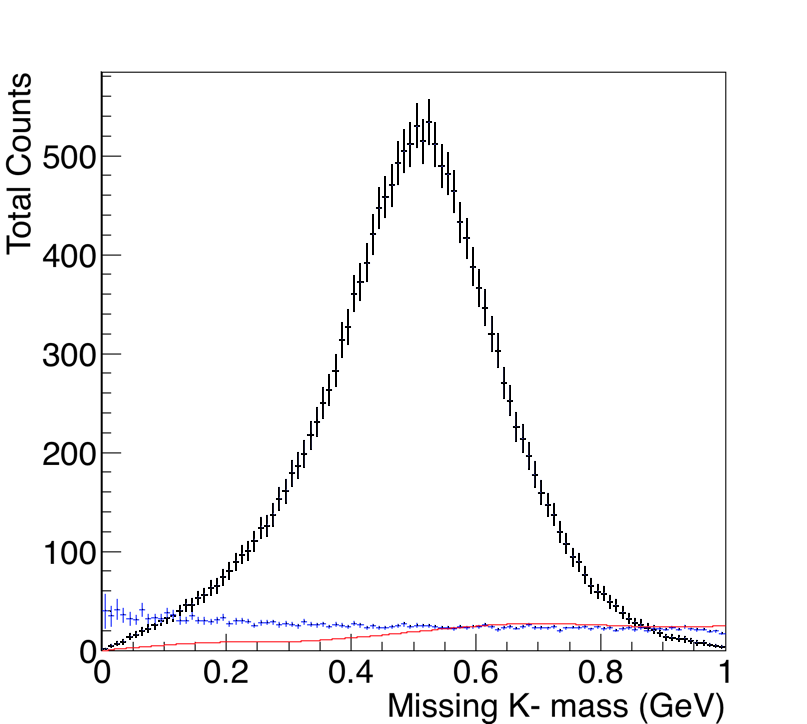}
  \caption{The black histogram shows the expected K$^-$ mass resolution from reconstruction using the electron and K$^+$ detected in CLAS12, and $^4$He detected in ALERT with 
    counts for proposed running time.  The light-blue histogram presents the 
    estimated background from coherent $\omega \rightarrow \pi^+\pi^-\pi^0$ and 
    $\rho \rightarrow \pi^+\pi^-$, with a misidentified K$^+$ and a production 
    cross-section 1000 times the $\phi$ cross-section.  The red histogram 
    estimates a misidentified $^3$He or $^2$H incoherent background using 
    PYTHIA.  See text for details.} \label{fig:phi_Kaon_mass}
\end{figure}

\begin{figure}[htb]
  \centering
  \includegraphics[width=0.45\textwidth,clip,trim=5mm 5mm 23cm 5mm]{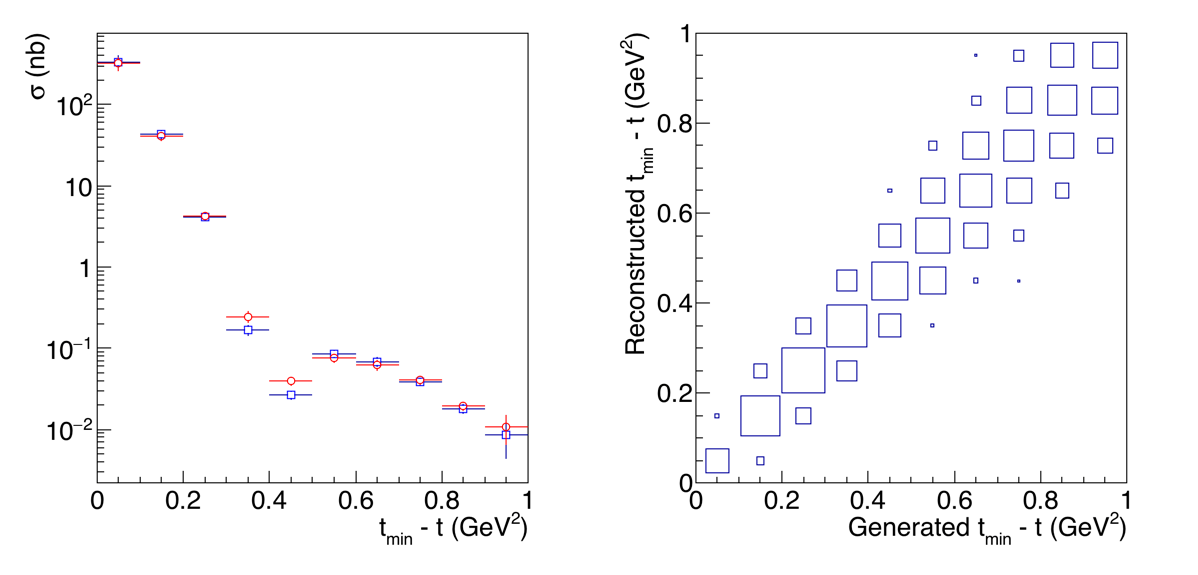}
  \includegraphics[width=0.45\textwidth,clip,trim=21.5cm 5mm 20mm 5mm]{fig_NuclGPD/phi_tbin_migr.png}
  \caption{Quantification of $t$ bin migration effects are shown in these two 
    plots.  The left plot shows cross-section versus reconstructed $t_{min} - 
    t$ with red-circles, and the input generated $t_{min} - t$ with blue 
    squares.  On the right, reconstructed versus generated $t_{min} - t$ is 
  plotted, which illustrates the expected bin-migration.} \label{fig:phi_tmig}
\end{figure}

In the case of a single kaon, the missing kaon will be reconstructed through 
missing momentum and energy.  The missing 4-vector will be constrained to have 
a kaon mass, and the reconstructed 4-vector of the kaon and the missing kaon 
will be constrained to have an invariant mass of the $\phi$ mass.  
Figure~\ref{fig:phi_Kaon_mass} shows the reconstructed mass of a missing K$^-$ 
after smearing the electron and K$^+$ with fastMC resolution smearing, and the 
detected $^4$He with a momentum resolution in the ALERT detector of $\Delta p < 
10\%$.  A cut can then be made on the missing kaon mass from 0.2 to 0.8 GeV$^2$ 
to help eliminate background.   The ALERT detector resolution drives both the 
missing kaon mass resolution and the resolution of the $t$-variable 
calculation.  Figure~\ref{fig:phi_tmig} shows the expected ($t$ - $t_{min}$)-bin 
migration after resolution effects.

\subsection{Production and Background Rates}

\begin{figure}[htb]
  \centering
      \includegraphics[width=0.45\textwidth,clip,trim=5mm 7mm 15mm 5mm]{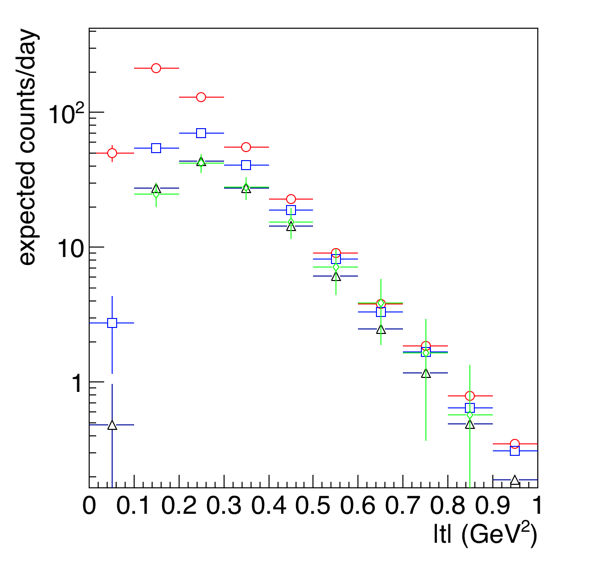}
  \caption{Expected counts per day for coherent $\phi$ production.  Three 
    analysis channels that will be investigated include the fully exclusive $K^+ 
    \, K^-$ (black-triangles), missing K$^-$ (red-circles), missing K$^+$ 
    (blue-squares), and missing $K_L^0$ with $K_S^0 \rightarrow \pi^+ \, \pi^-$  
  (green-diamonds).} \label{fig:phi_rates_t}
\end{figure}

In Figure~\ref{fig:phi_rates_t}, the expected counts per day for the primary 
decay channel is shown versus $t$.  In this calculation, the total luminosity 
is assumed to be 7.5$\times\,10^{33}$ nuclei/cm$^2$/s, and the kaon and alpha 
detection efficiency is set at 50\% in addition to the simulated acceptance as 
an additional safety margin on the expected counts.  In addition to the $\phi 
\rightarrow K^+\, K^-$ channel, we can gain additional statistics from the 
$\phi \rightarrow K^0_L K^0_S$ channel, with $K_S^0 \rightarrow \pi^+\, \pi^-$. 
In this case, the $\pi^+\,\pi^-$ pair detection is assumed to have an 
efficiency of 70\% in addition to the simulated acceptance.

Kaon PID will be performed through a combination of the TOF and veto from the 
LTCC, as proposed in PR12-12-007\cite{Girod:2012PR}.  At kaon momenta $<$ 5 
GeV, the TOF will provide a 1-$\sigma$ or better separation of kaons and pions.  
In combination with the charged pion momentum threshold of 2.5 GeV/c for the 
LTCC, kaon PID is not expected to be an issue.  An additional cross-check of 
kaon identification can be performed with the CLAS12 RICH detector.

Background pion rates are expected to be small when the ALERT detector is 
required to tag a recoiling $^4$He in the event; in this case, most pion 
background will come from non-$\phi$ meson production which can be cleaned up 
through missing mass cuts on the missing Kaon and reconstructed $\phi$.  For 
estimation of this background, phase-space for $\rho \rightarrow \pi^+ \pi-$ 
and $\omega \rightarrow \pi^+ \pi^- \pi^0$ was generated.  Since a 
comprehensive cross-section calculation and parameterization for $\rho / 
\omega$ electroproduction off $^4$He is non-trivial and somewhat outside the 
scope of the analysis, the cross-section for both $\rho$ and $\omega$ is 
estimated to be 1000 times the $\phi$ cross-section. This factor of 1000 is a 
conservative estimate based on the ratio of rho/omega to phi production 
cross-sections off the proton at JLab6 energies, which are on the order of a 
few hundred times larger.  Misidentification of the pions or protons as kaons 
is simulated according to the expected TOF separation at low-momentum, and with 
a conservative $95\%$ rejection in the LTCC above 3 GeV momentum for pions.  
The results are shown in Figure~\ref{fig:phi_Kaon_mass}. Another possible source 
of contamination will come from misidentification of $^4$He in the ALERT 
detector.  
described above, should still provide a very clean separation of background.

A PYTHIA simulation for $\gamma + N \rightarrow X$ was performed, forcing a 
re-scattering between the recoil nucleon and the residual nucleus in $^4$He.  
Re-scattering is required since a residual $^3$He nucleus, with a residual 
momentum equal to its fermi-momentum from being bound in a $^4$He, is almost 
completely outside ALERT's momentum acceptance.  If the re-scattered $^3$He or 
$^2$H enters ALERT acceptance, it is assumed to have a $^4$He misidentification 
probability of $10\%$ (upper limit of expected misidentification).   True kaons 
and misidentified protons or pions are also accepted in CLAS12 as described 
above and the total rate is calculated assuming the experiment's production 
luminosity.  The total rate per day using worst-case estimates is calculated to 
be approximately 25 (or less than one-tenth the production rate), averaged over 
the entire accepted phase-space.  This should not greatly impact the $\phi$ 
identification. The estimated counts from this background are also shown in 
Figure~\ref{fig:phi_Kaon_mass} for comparison to the signal peak.


\subsection{Event Generator}\label{sec:phiEG}
Event generation for $\phi$ production off $^4$He is done in two steps.  First, 
the cross-section for $\phi$ production off a proton target is generated, and 
then the charge form-factor of $^4$He is folded in, while the corresponding 
charge form-factor for the proton is divided out.  The phase-space for 
generation is created by sampling uniformly in Q$^2$, $x_B$, and $t$.

Since the cross-section must first be calculated off the proton, the relevant 
value for $t$ must be recalculated. For this we define a new variable $t_p$ 
which is calculated assuming a target proton with a momentum uniformly 
distributed up to the $^4$He fermi-momentum in the initial state, and a 
scattered proton with 1/4 the momentum of the recoiling $^4$He with a uniformly 
distributed fermi-momentum in the final state.  The cross-section is then given 
by:
\begin{eqnarray}
  \frac{d\sigma_{^4He}}{dt}(t) = 
  \frac{d\sigma_p}{dt}(t_p)\left(\frac{A\,F_{C,^4He}(t)}{F_{C,p}(t_p)}\right)^2
  \label{eqn:xsec_ttp}
\end{eqnarray}
where $A$ is the nucleon number of $^4$He, and $F_C$ the charge form factor of 
$^4$He is parametrized using the world data through its first minimum in $t$ 
following: $F_{C,^4He} = (1 - (2.5t)^6)e^{11.7t}$. The calculation of the 
$\phi$ production cross-section off the proton follows the exact formalism as 
put forth by the accepted CLAS12 proposal PR12-12-007 \cite{Girod:2012PR} and 
discussed in section \ref{sec:DVMPFormalism}. The differential cross-section 
for the case of an unpolarized electron is
\begin{eqnarray}
 \frac{d^3\sigma}{dx_B\,dQ^2\,dt} = 
 \Gamma(x_B,Q^2,E)\left(\frac{d\sigma_T}{dt}(W,Q^2,t) + 
 \epsilon\frac{d\sigma_L}{dt}(W,Q^2,t)\right),
\end{eqnarray}
where the virtual photon flux is defined using the Hand~\cite{Hand:1963bb} 
convention
\begin{equation}
  \Gamma = 
  \frac{\alpha}{2\pi}\frac{E^{\prime}}{E}\frac{K}{Q^2}\frac{1}{1-\epsilon}
\end{equation}
and $K = \nu - Q^2/2M$.
The transverse cross-section is parameterized in $W$ and Q$^2$ to fit world 
data as shown in Figure~\ref{fig:PR1207_verify2}. The ratio of longitudinal to 
transverse cross-section is also fit to world data as a function of Q$^2$ and 
is shown in Figure~\ref{fig:PR1207_verify}. The $t$-dependence is incorporated 
as an exponential with a slope that depends on $W$. The exact functional forms 
for each of these is omitted here for brevity, but can be found in PR12-12-007 
\cite{Girod:2012PR}.  The total calculated cross-section for $\phi$ 
electroproduction off $^4$He follows that formalism, except everywhere 
$d\sigma/dt$ enters the calculation, the calculation of Eqn.~\ref{eqn:xsec_ttp} 
is used instead.  The plots of the cross-section versus world data from 
PR12-12-007 are shown in Figures~\ref{fig:PR1207_verify} 
and~\ref{fig:PR1207_verify2}.

\subsection{Projections}

\begin{figure}[htb]
  \centering
  \includegraphics[width=0.45\textwidth,clip,trim=5mm 5mm 0mm 5mm]{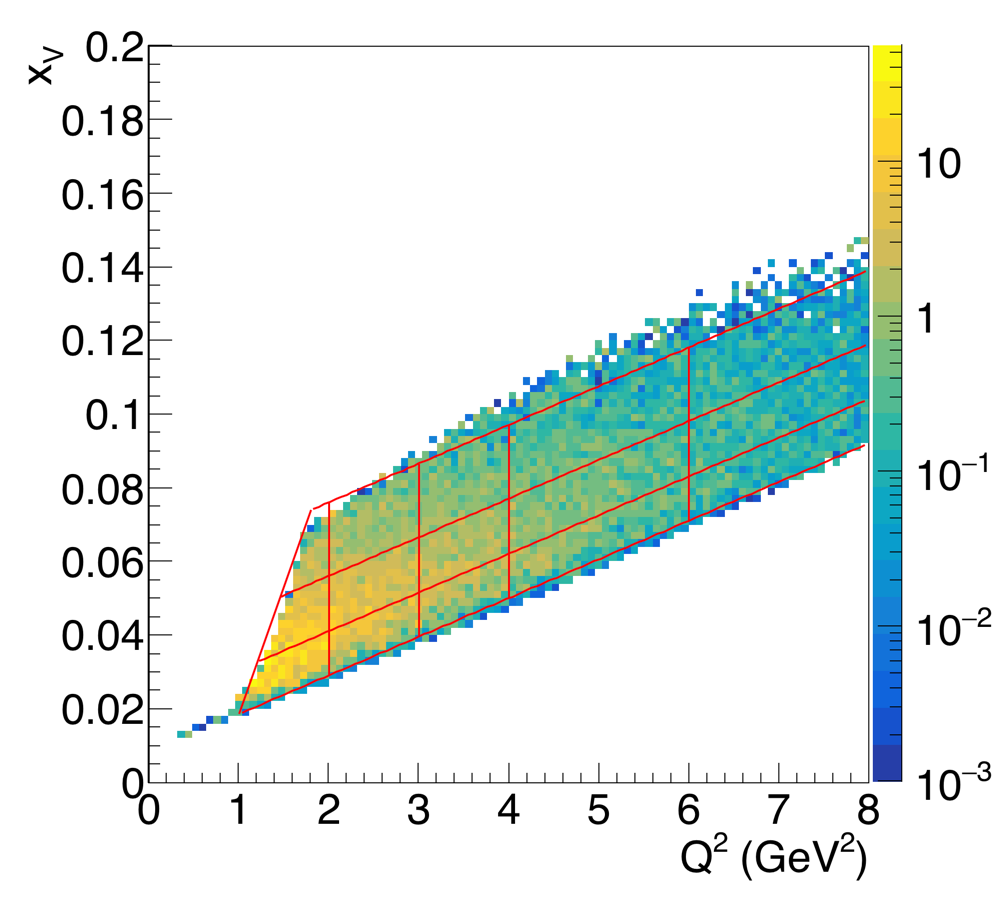}
  \includegraphics[width=0.45\textwidth,clip,trim=5mm 5mm 0mm 5mm]{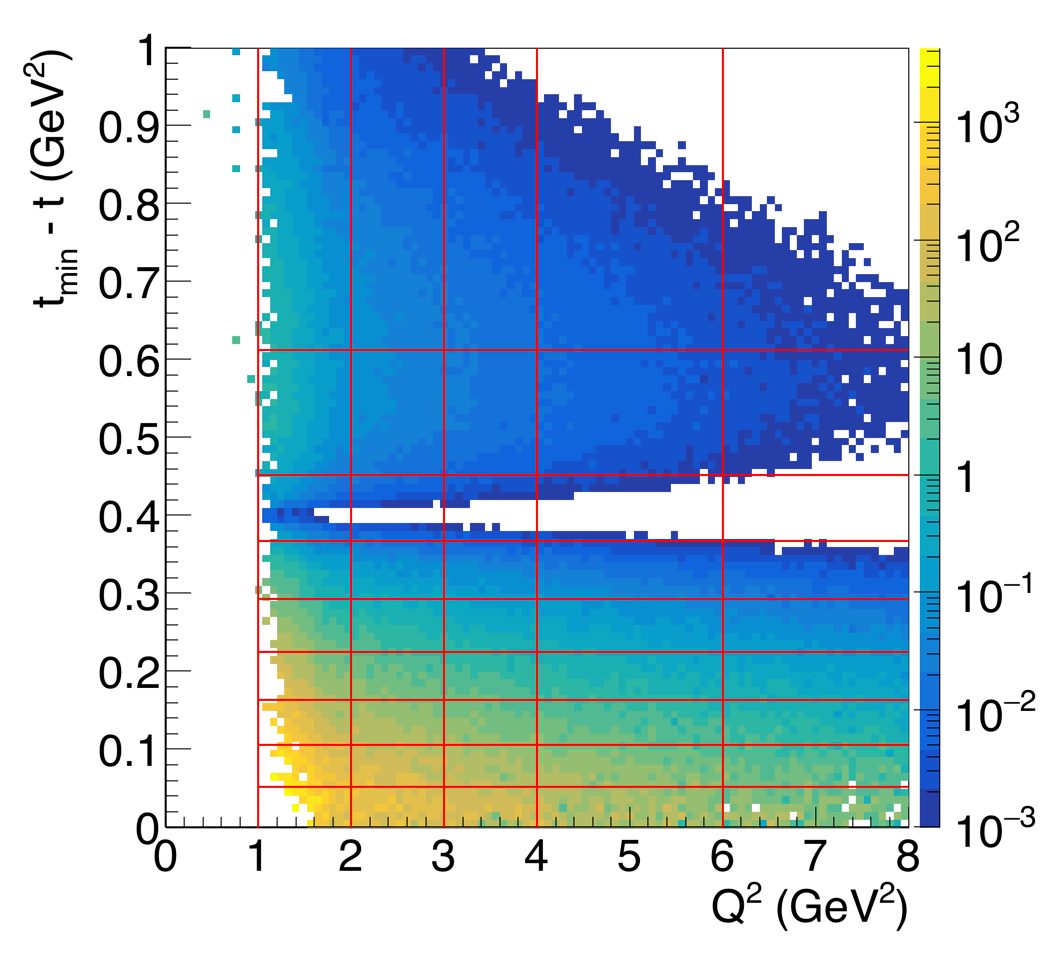}
  \caption{A possible binning over the accepted phase-space for gluon GPD extraction with $\phi$ production.} 
  \label{fig:phi_binning}
\end{figure}

The extraction of gluon GPDs will need the cross-section calculated in bins of 
$Q^2$, $x_V$ and $t_{min} - t$.  The variable $x_V$ is similar to $x_A$ but 
takes into account the production of a vector meson with mass greater than zero, and is 
useful for direct comparison between DVMP and DVCS.  A summary of the many 
different notations for $x$ calculation using different target masses and 
vector masses is shown below:
\begin{eqnarray}
  x_V &=& \frac{Q^2 + M_V^2}{W^2 + Q^2 + M_{^4He}^2} = \left(\frac{Q^2 + M_V^2}{Q^2}\right)x_A \\
  x_{Vp} &=& \frac{Q^2 + M_V^2}{W^2 + Q^2 + M_{p}^2} = \left(\frac{Q^2 + M_V^2}{Q^2}\right)x_B
\end{eqnarray}
The exact binning will depend on total run-time, but a feasible binning for the 
requested beam-time is shown in Figure~\ref{fig:phi_binning}.  In this binning 
configuration, the largest occupancy bins will have greater than 1000 signal 
events.  The smaller occupancy bins can have 100 or less events, and may be 
folded together where necessary.  Additionally, an identical binning to the 
DVCS analysis can be performed for a more direct comparison of results; the 
data are expected to overlap in much of the t, Q$^2$ and x$_{V,A}$ phase-space.

\begin{figure}[htb]
  \centering
      \includegraphics[width=0.45\textwidth,clip,trim=5mm 5mm 0mm 5mm]{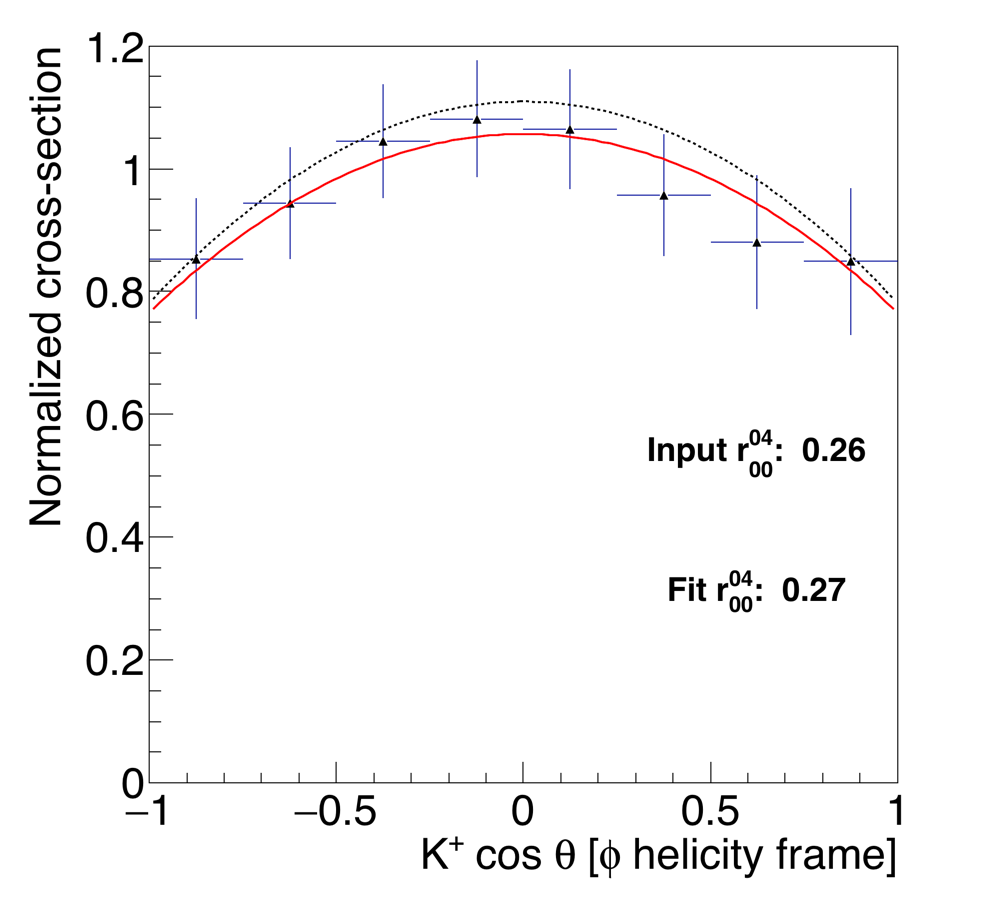}
  \includegraphics[width=0.45\textwidth,clip,trim=5mm 5mm 0mm 5mm]{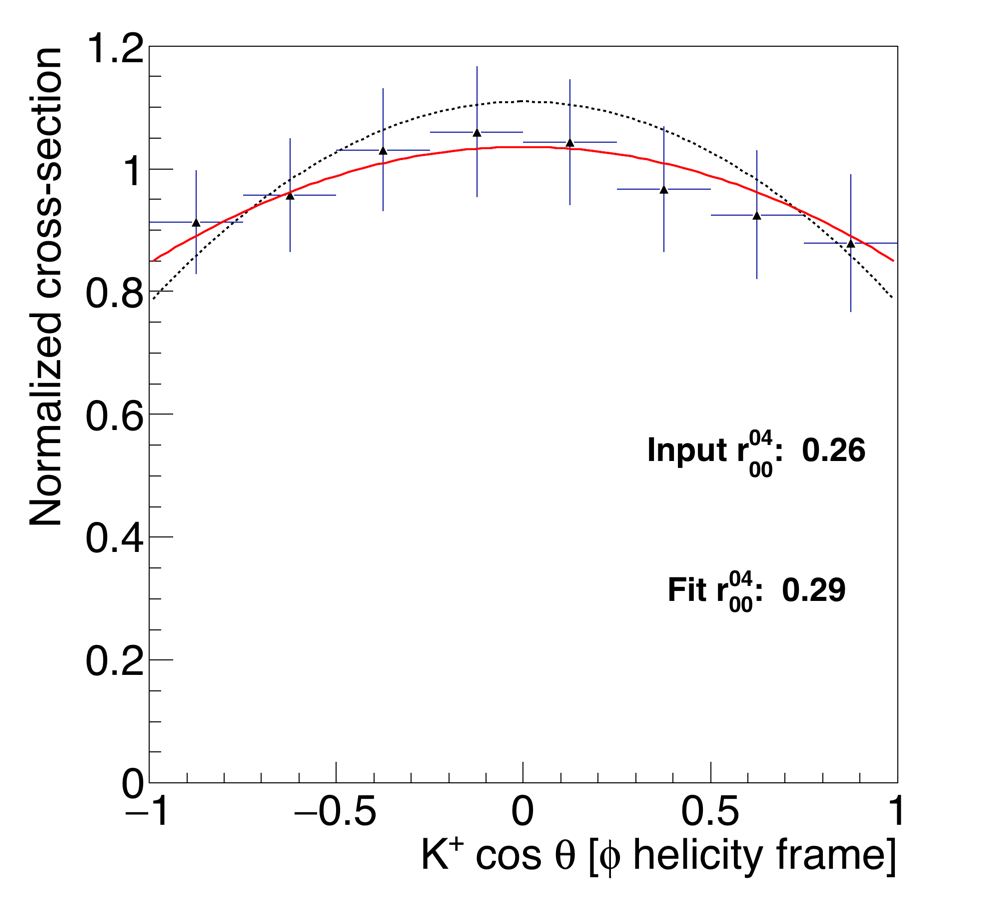}
  \caption{A fit to the cos($\theta$) distribution of the K$^{+}$ in the 
    reconstructed $\phi$-helicity frame within a bin of values $[0.02 < t_{min} 
    - t < 0.04\,GeV^2]$, $[0.025 < x_V < 0.05]$, and $[1.5 < Q^2 < 2.0\, 
    GeV^2]$.  This is an example of an extraction for an intermediate occupancy 
    bin which is calculated to have a few hundred events during the run period.  
    The dashed line shows the distribution that was generated.  The data are 
    then fitted after acceptance and resolution smearing for comparison to the 
    generated values.  Uncertainty on the parameter $r_{00}^{04}$ for the fit 
    on both pannels is 0.03 (just over 10\%).  The left plot assumes a momentum 
    resolution of 5\% for the ALERT detection of $^4$He, and the right plot 
    assumes a momentum resolution of 10\% .}
  \label{fig:phi_r0400}
\end{figure}

\begin{figure}[htb]
  \centering
  \includegraphics[scale=0.3]{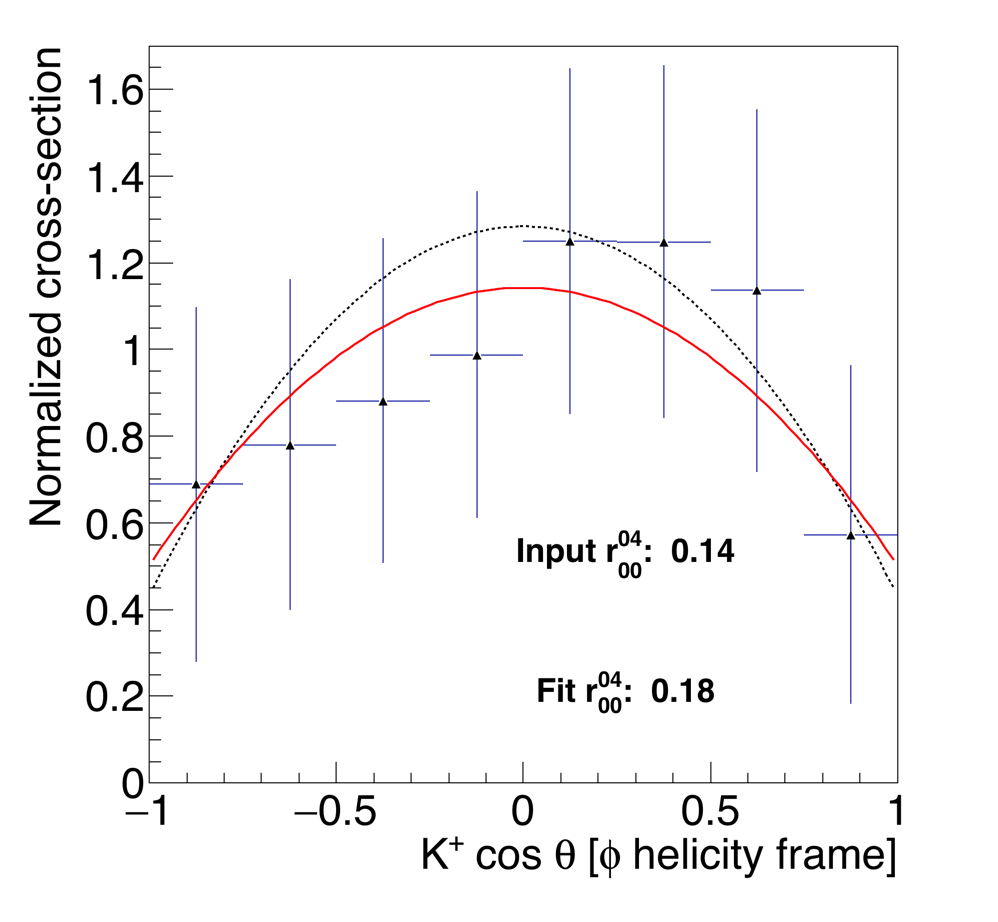}
  \caption{A fit to the cos($\theta$) distribution of the K$^{+}$ in the 
    reconstructed $\phi$-helicity frame within a bin of values $[0.1 < t_{min} 
    - t < 0.15\,GeV^2]$, $[0.04 < x_V < 0.07]$, and $[3.0 < Q^2 < 4.0\, 
    GeV^2]$. This is a low occupancy bin, expected to have around 100 events 
    over the run period, and is shown for comparison to the high statistics fit 
    from Figure~\ref{fig:phi_r0400}.  The dashed line shows the distribution that 
    was generated.  The data are then fitted after acceptance and resolution 
    smearing for comparison to the generated values. Uncertainty on the parameter 
    $r_{00}^{04}$ for the fit shown above is 0.08 (just under 50\%). An ALERT 
  detector momentum resolution of 10\% is assumed.}
  \label{fig:phi_r0400_low}
\end{figure}

As discussed in section \ref{sec:sigLFormalism}, to extract the ratio $R$, it 
is necessary to boost to the $\phi$-helicity frame and fit the cos($\theta$) 
distribution of one of the decay kaons, as described in 
Eqn.~\ref{eqn:phi_r0400}.  The resolution of this $\theta$ distribution is 
highly dependent on the ALERT momentum resolution of the $^4$He.  Shown in 
Figure~\ref{fig:phi_r0400} is the extraction of $r_{00}^{04}$ for a bin with 
$[0.02 < t - t_{min} < 0.04\,GeV^2]$, $[0.025 < x_V < 0.05]$, and $[1.5 < Q^2 < 
2.0\, GeV^2]$.  The two panels show the effect of resolution on the extraction; 
the left plot has standard CLAS12 resolutions plus a 5\% momentum resolution 
for the $^4$He detected by ALERT, and the right plot shows the same except a 
resolution of 10\% in ALERT.  The general characteristic of increasing 
resolution, is a flattening of the $\cos\theta$ distribution.  Additional 
constraints may be able to improve the momentum resolution of the ALERT 
detector and even correct the $\theta$ distributions.  For comparison, an R 
extraction in a less populated bin is shown in Figure~\ref{fig:phi_r0400_low}.

Once $R$ is extracted, the gluon GPD is defined as the square-root of the 
normalized longitudinal differential cross-section:
\begin{eqnarray}
  |\left<H_g\right>|(t) \propto \sqrt{\frac{d\sigma_L}{dt}(t_{min} - t) \Big/ 
  \frac{d\sigma_L}{dt}(0)}
\end{eqnarray}
This normalization of the cross-section to the $t = t_{min}$ point simplifies 
the analysis and cancels some of the systematic effects that would otherwise 
increase the uncertainty.  As an example, consider the gluon transverse density 
profile is shown in Figure~\ref{fig:phi_gdens} which is a result of the Hankel 
transform (\ref{eq:HankelTransform}). The extraction is performed on the 
simulated events with all acceptance, smearing, and background effects 
included.  The binning choice reflects that of the example extraction performed 
above for DVCS: $x_{Vp}$ between 0.18 and 0.25, with an additional $Q^2$ cut 
between 2.0 and 3.0 $GeV^2$. A second transformation is performed on the 
generated cross-section before any acceptance or resolution effects.  The 
difference between this pre-acceptance/smearing transformation and the post 
acceptance/smearing transformation is used to estimate the total systematic 
uncertainty on the transverse profile calculation. The combined systematic and 
statistical uncertainty is shown in the width of the band for the gluon density 
calculation in Figure~\ref{fig:phi_gdens}.


\section{Parton Distributions in the Transverse plane} 
\label{sec:impactParDists}

\begin{figure}[htb]
   \centering
      \includegraphics[width=0.65\textwidth,clip,trim=5mm 5mm 5mm 5mm]{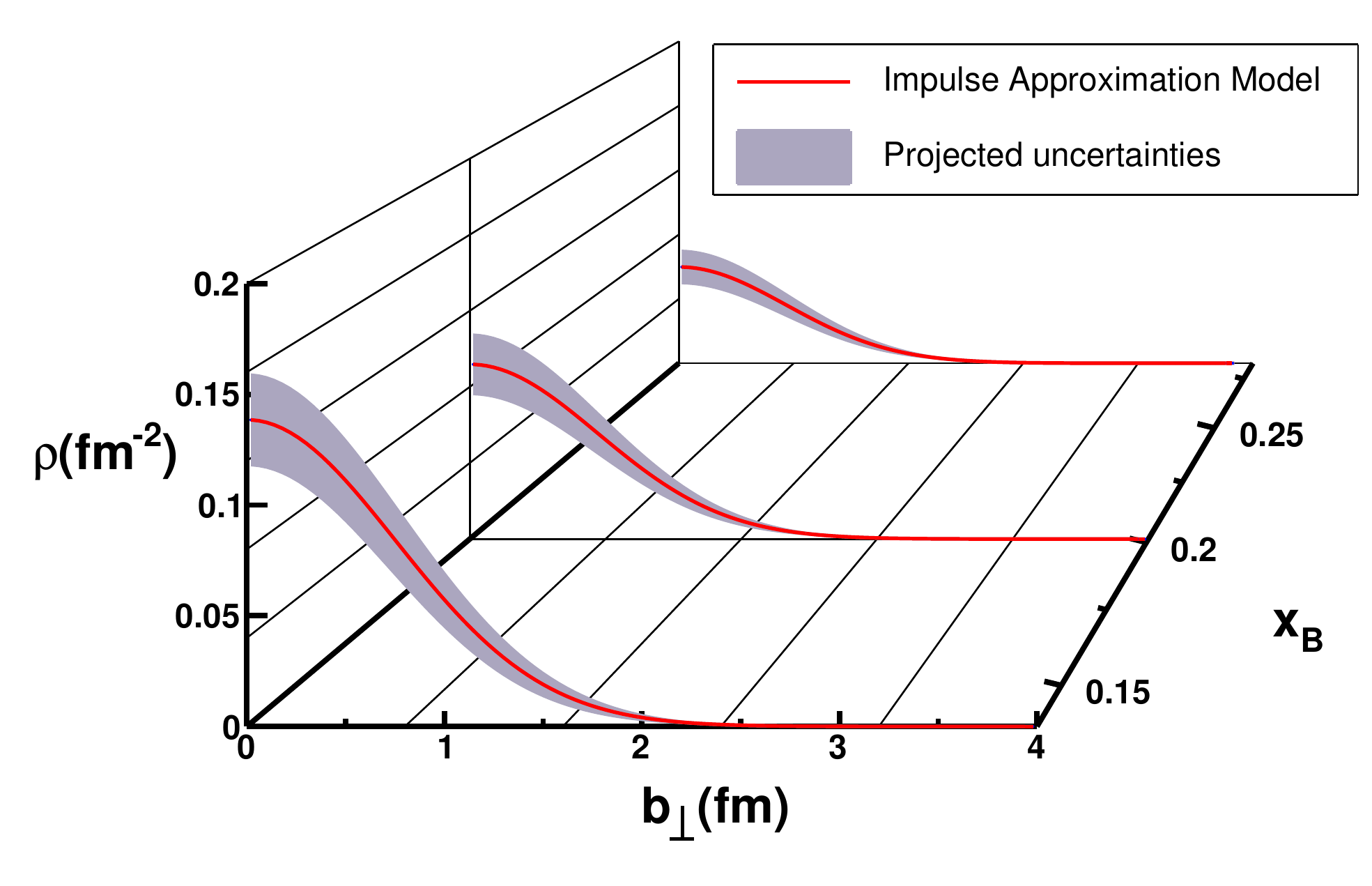}
   \caption{The statistical uncertainties of the  parton density profiles as a 
     function of the impact parameter, $b_{\perp}$, based on the 
     $\mathcal{H}^{A}$ CFF extracted from the Impulse Approximation (IA) at the mean 
   $x_{B}$ values in the different bins. }
   \label{fig:density_profile}
 \end{figure}

\begin{figure}[htb]
  \centering
  \includegraphics[width=0.5\textwidth]{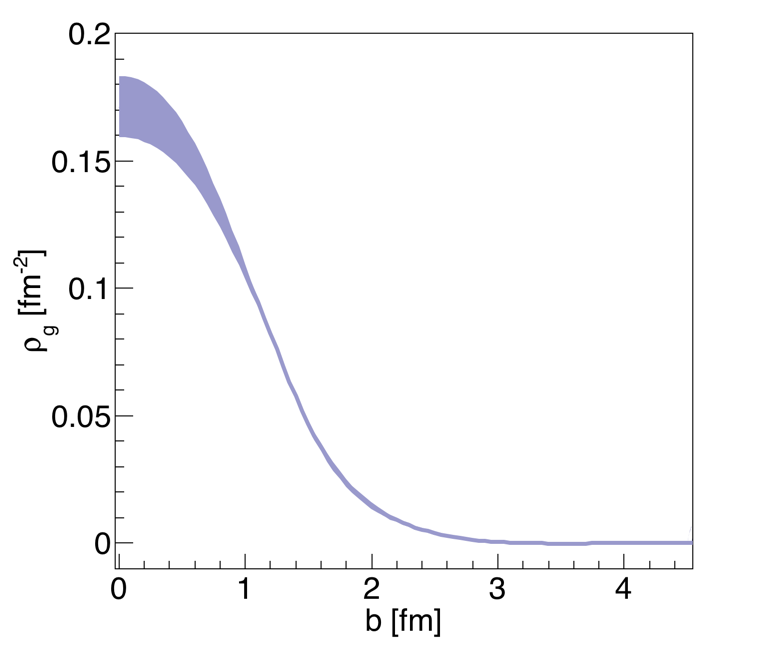}
  \caption{An example of the calculated average gluon density for an $x_{Vp}$ 
    bin between 0.18 and 0.25, and a $Q^2$ bin between 2.0 and 3.0 $GeV^2$.  
  The band represents the combined statistical and systematic uncertainties.}
  \label{fig:phi_gdens}
 \end{figure}

Using equation (\ref{eq:HankelTransform}) the quark and gluon transverse 
density profiles of $^{4}$He are extracted from the GPDs $H_A$ and $H_g$, 
respectively. Projections are shown in Figure~\ref{fig:density_profile} and 
Figure~\ref{fig:phi_gdens}.

The projected uncertainties on the quark density profiles shown in 
Figure~\ref{fig:density_profile} were determined using the statistical 
uncertainties on $\mathcal{H}_{A}$ (see Figure~\ref{fig:CFF_projections}). 
The uncertainties on the density profiles are calculated by 
varying the extracted t-dependent GPD within the projected uncertainties.
The red lines on Figure~\ref{fig:density_profile} represent the density 
profiles extracted from 
the model.

\begin{figure}[htb]
  \centering
  \includegraphics[width=0.5\textwidth]{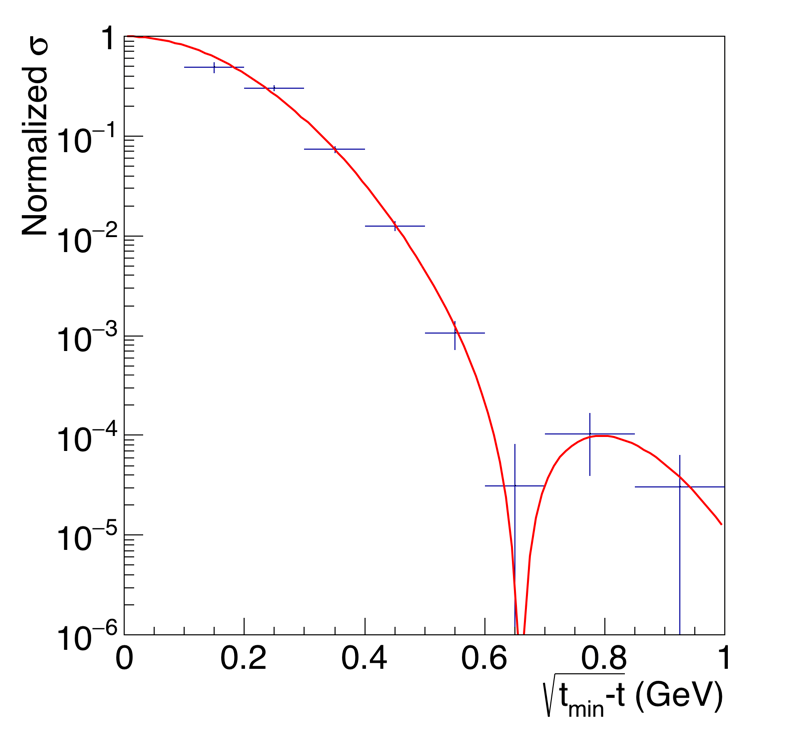}
  \caption{The normalized cross-section as a function of $\sqrt{t_{min} - t}$ for an 
    $x_{Vp}$ bin between 0.18 and 0.25, and a $Q^2$ bin between 2.0 and 3.0 
    $GeV^2$.  Total running time for this experiment provides adequate 
    statistics to locate the first diffractive minimum if it exists from $\phi$ 
    production within ALERT $t$ acceptance. The graph points have 
    uncertainties that include all statistical and resolution/bin-smearing 
    effects.  The red line is a fit to the graph with a similar form of 
  $|F_C|^2$ for the $^4$He nucleus.}
  \label{fig:phi_roott}
 \end{figure}

An noteworthy feature of the 2-dimensional Fourier transform is the 
low-$b$ sensitivity to  the high-$t$  part of the GPDs. That is, higher 
momentum transfers are more sensitive to smaller distances.
Investigating the location of the first diffractive minimum in $F_C$ for $^4$He 
is important for the discussion of the comparison between charge and gluon 
densities. The requested beam-time for this experiment provides just the 
required statistics to quantitatively discern the location of the first 
diffractive minimum, if it exists, within the $t$ range available.   
Figure~\ref{fig:phi_roott} shows the expected $\sqrt{t_{min} - t}$ spectrum for 
the gluon profile extraction, with uncertainties that reflect statistics and 
systematics after all acceptance and resolution smearing and the same binning 
used to extract the gluon density in Figure~\ref{fig:phi_gdens} above: $0.18 < 
x_{Vp} < 0.25$, and $2.0 < Q^2 < 3.0\,GeV^2$. The red line shows a fit used to 
find the diffractive minimum. Fewer statistics would still allow a high 
precision gluonic RMS radius calculation, where low-$t$ events are most 
important. Higher statistics would allow us to better locate a diffractive 
minimum if additional nuclear effects reduce the sharpness of the minimum.

\chapter*{Summary and Answers to PAC44\markboth{\bf Summary and Answers to PAC44}{}}
\label{chap:conclusion}
\addcontentsline{toc}{chapter}{Summary and Answers to PAC44}

\setlength\parskip{\baselineskip}%
\section*{Answers to PAC44 issues\markboth{\bf Answer to PAC44 issues}{}}
\addcontentsline{toc}{section}{Answer to PAC44 issues}

{ \it \textbf{Issues:}}

{ \it 
The Drift Chamber/scintillator technology needs to be demonstrated. We observe 
that a strong program of prototype studies is already underway. }

{\bf Answer:} We feel the technology has no major unknowns, wire chambers and
scintillators have been used for decades as detectors of low energy nuclei
and their properties have been well established. We present in the
proposal a conceptual design demonstrating the feasibility of the detector,
it is common practice to work on the optimization of a certain number of 
parameters after the proposal is approved. In particular, because it is easier to
fund and man a project that has an approved status than a future proposal.
Nevertheless, we remain open to discuss 
the topic in more depth if the committee has any concerns.

{ \it 
The TAC report voiced concerns about the length of the straw cell target and 
the substantial effort needed to integrate the DAQ for this detector into the 
CLAS12 DAQ. }

{\bf Answer:} The TAC and PAC44 raised concerns about the target cell. We have 
added extra discussion  in section \ref{sec:targetCell}, which includes a table 
of existing or planned targets that are similar to the one we proposed. In summary,
our proposed target is twice as wide as the ones used in the 6 
GeV era for the BONuS and eg6 run and should therefore cause no issues. Note 
that the experiment 12-06-113 (BONuS12) is approved with a longer and thinner 
target. Their design will be reviewed by JLab for their experiment readiness 
review (ERR) before the PAC45 meeting. The result of this review should settle 
the question, but in any case, we propose a safer solution based on the 
successful experiments of the 6 GeV era. 

The TAC and PAC44 raised issues regarding integration of ALERT into the CLAS12 
DAQ. First, they raised a concern that the resources necessary for this integration 
are not clearly identified. We have added text in section 
\ref{sec:jlabContributions} outlining the resources provided by each group 
and the technical support they are expected to provide. Secondly, they mentioned 
a concern about the ``substantial effort needed to integrate the DAQ for this 
detector into the CLAS12 DAQ''. We want to emphasize that the read-out systems for 
ALERT are already being used in the CLAS12 DAQ to readout Micromegas detectors. 
Therefore, we will use and build on the experience gained from these systems.

{ \it 
The proposal does not clearly identify the resources (beyond generic 
JLAB/CLAS12 effort) necessary for DAQ integration which may be a substantial 
project. }

{\bf Answer:} As mentioned above, we do not feel this contribution is major,
nevertheless we made this part clearer in the proposal.

{ \it 
During review the collaboration discovered an error in converting the 
luminosity to beam current. This resulted  in a revision that will either 
require doubling the current or the target density. The beam current change 
would require changes to the Hall B beam dump, while raising the target density 
could impact the physics reach of the experiment by raising the minimum 
momentum threshold. }

{\bf Answer:} During the PAC44 proposal submission process the wrong beam 
current was requested. It was a factor of 2 too low. This increased beam 
current brought into contention the issue of possible Hall B beam current 
limits. We chose to use the higher beam current in this new version. 
Based on discussions with the Hall-B and accelerator staff, the only
necessary upgrade necessary to run at \SI{1}{\uA} is with the Hall-B beam blocker.

{ \it 
The precise interplay between final state interactions (FSI) and the tails of 
the initial state momentum distribution in DVCS on 4He was a topic of some 
debate. The collaboration makes an argument that the excellent acceptance of 
the apparatus allows novel constraints that allow selection of kinematic ranges 
where FSI is suppressed. While the originally suggested method to unambiguously 
identify areas of FSI was revised during the review, the committee remains 
unconvinced that the new kinematic selections suggested do not also cut into 
interesting regimes for the initial state kinematics. The committee believes 
that this is model dependent and would like to see more quantitative arguments 
than were provided in this version of the proposal. }

{\bf Answer:} We acknowledge there was an overstatement of the possibilities 
of the Tagged-DVCS proposal on this topic, this has been corrected. We now show 
a reduction, in opposition to the complete suppression previously claimed, in 
events that differ from the PWIA result. This finding is based on a simulation 
using a simple model of FSIs together with a Monte-Carlo event generator.

{ \it 
\textbf{Summary:}

The committee was generally enthusiastic about the diverse science program 
presented in this proposal; in particular the tagged EMC studies and the unique 
study of coherent GPD's on the 4He nucleus. However, the substantial 
modifications made in the proposal during review indicate that it could be 
substantially improved on a reasonably short time scale. We would welcome a new 
proposal that addresses the issues identified by the committee and by the 
collaboration. }

{\bf Answer:} We hope that the new proposals will answer all the questions 
raised by the PAC44 and will make the physics case even more compelling.

{ \it 
We also note that there are multiple experiments, proposed and 
approved, to study the EMC effect, including several with novel methods of 
studying the recoil system. We appreciate the comparisons of recoil 
technologies in this proposal and would welcome a broader physics discussion of 
how the proposed measurements contribute to a lab-wide strategy for exploring 
the EMC effect. }

{\bf Answer:} While no strategy document has been drafted after them, we want to point out
to the PAC that the community of physicist interested by the partonic 
structure of nuclei meets regularly, with often a large focus on what can be done at JLab
(see workshops at Trento\footnote{New Directions in Nuclear Deep Inelastic Scattering \url{http://www.ectstar.eu/node/1221}}, 
Miami\footnote{Next generation nuclear physics with JLab12 and EIC \url{https://www.jlab.org/indico/event/121/}}, 
MIT\footnote{Quantitative challenges in EMC and SRC Research and Data-Mining \url{http://web.mit.edu/schmidta/www/src_workshop/}}, 
and Orsay\footnote{Partons and Nuclei \url{https://indico.in2p3.fr/event/14438/}} for example). 
Nonetheless, we added
in the tagged EMC proposal summary an extension about the 12~GeV approved experiments  
related to the EMC effect. This short annex will hopefully clarify the context 
and the uniqueness of the present experiments.

\section*{Summary and Beam Time Request\markboth{\bf Summary and Beam Time Request}{}}
\addcontentsline{toc}{section}{Summary and Beam Time Request}
We are proposing an experimental  program that will provide data, for the 
first time, required for a global analysis that will extract partonic GPDs 
in a dense nucleus, in this case GPD 
$H_{^4{\rm He}}$ for both quarks and gluons together in $^4$He.  
The DVCS process has been the hallmark of the 3-D investigation of the nucleon 
structure at large $x$ at Jefferson Lab. This investigation  was extended 
successfully to the $^4$He nucleus in the 6 GeV era albeit with limited 
kinematic leverage and statistics. This proposal will not only provide Compton 
form factors for the quarks contributions but also uses the electroproduction 
of $\phi$ to explore the contribution of gluons at large $x$,  in tandem with 
that of the quarks. \\

This program will be a precursor of physics to be explored at a future EIC, 
namely 3-D imaging of spin zero nuclei. Other mesons will be accessible in the 
same data stream of this proposed experiment extending the physics reach of 
this proposal. For example, a flavor decomposition of these form factors will 
be possible by investigating the deep exclusive production of pseudoscalar and 
vector mesons with masses below the $\phi$ meson, like $\pi^0$, $\rho$, 
$\omega$. Given the limited energy reach of Jefferson Lab, the most promising 
access to the gluonic structure of nucleon and nuclei is by using the $\phi$ 
meson production as a probe as was done in proposal E12-12-007. Of course a 
future EIC will allow the use of heavier mesons  such as  the $J/\Psi$ and the 
$\Upsilon$ for a "cleaner" gluonic probe, nevertheless we believe that at large $x$ 
the strange quark contribution could be separated from that of the gluons in a 
global analysis where DVCS and DVMP data from different pseudoscalar and vector mesons, thus this coherent proposal.\\ 

In order to achieve the statistical uncertainties presented in this proposal, 
we request 20 days of running with 11~GeV electron beam at a luminosity of $3 
\times 10^{34}$~cm$^{-2}$s$^{-1}$ per nucleon (same beam time request as the 
tagged EMC proposal) and 10 days at $6 \times 10^{34}$~cm$^{-2}$s$^{-1}$ per 
nucleon with helium target (specific to this proposal), both with 80\% 
longitudinally polarized beam. We will also need 5 days of commissioning of the 
ALERT detector at 2.2~GeV with helium and hydrogen targets.
\section*{Relation to other proposals\markboth{\bf Relation to other 
proposals}{}}
\addcontentsline{toc}{section}{Relation to other proposals}
Our proposal has no direct relation to other approved proposals, except that 
similar measurements have been proposed and approved for the proton. These are 
the first coherent DVCS and DVMP measurements that are proposed for a $^4$He 
target.

\appendix
\chead[]{\let\uppercase\relax\leftmark}

\chapter{Twist-2 $e ~^4He \rightarrow e ~^4He ~\gamma$ cross section} 
\label{app:Helium_cross_section}
Following the definitions of reference \cite{Belitsky:2008bz}, the variables 
that appear in equations \ref{TTBH} to \ref{TTinter} are defined as:

 $\mathcal{P}_{1}(\phi)$ and $\mathcal{P}_{2}(\phi)$ are BH propagators and 
 defined as:
\small
\begin{align}
&\mathcal{P}_{1}(\phi) = \frac{(k - q')^{2}}{Q^{2}} = - \frac{1}{y (1 + \epsilon^{2})} 
\big[ J + 2 K \cos(\phi) \big] \\
&\mathcal{P}_{2}(\phi) = \frac{(k - \Delta)^{2}}{Q^{2}} = 1 + \frac{t}{Q^{2}} + 
\frac{1}{y (1 + \epsilon^{2})} \big[ J + 2 K \cos(\phi) \big]
\end{align}
~~~~~~~~~~~with,
\begin{align}
& J = \bigg( 1 - y - \frac{y \epsilon^{2}}{2} \bigg) \bigg(1 + \frac{t}{Q^{2}} \bigg) - 
(1 - x_{A})(2 - y) \frac{t}{Q^{2}} \\
& K^{2} = - \delta t \, (1 - x_{A}) \bigg( 1 - y - \frac{y^{2} \epsilon^{2}}{4} \bigg) 
\bigg\{ \sqrt{1 + \epsilon^{2}} + \frac{4 x_{A} (1-x_{A}) + \epsilon^{2}}{4 (1 - x_{A})}
\delta t \bigg\} \\
& \delta t = \frac{t - t_{min}}{Q^{2}} = \frac{t}{Q^{2}} + \frac{2(1-x_{A}) \left(1- \sqrt{1 + 
\epsilon^{2}} \right) + \epsilon^{2}}{4 x_{A} (1- x_{A}) + \epsilon^{2}}
\end{align}
\normalsize
where $t_{min}$ represents the kinematic boundary of the process and defined as:
\small
\begin{equation}
t_{min} = -Q^2 \frac{2(1-x_A)(1 - \sqrt{1+\epsilon^2}) + \epsilon^2}{4 
x_A(1-x_A) + \epsilon^2}
\end{equation}
\normalsize
The Fourier coefficients, in equations \ref{TTBH}, \ref{TTDVCS} and \ref{TTinter}, of a spin-0 target are defined as:
\small
\begin{eqnarray}
c_0^{BH} = & \bigg[ & \left\{ {(2-y)}^2 + y^2{(1+\epsilon^2)}^2 \right\} 
\left\{ \frac{\epsilon^2 Q^2}{t} + 4 (1-x_A) + (4x_A+\epsilon^2) \frac{t}{Q^2} 
\right\} \nonumber \\
& \phantom{\bigg[} & + 2 \epsilon^2 \left\{ 4(1-y)(3+2\epsilon^2) + y^2(2-\epsilon^4) 
\right\} - 4 x_A^2{(2-y)}^2 (2+\epsilon^2) \frac{t}{Q^2} \nonumber \\
& \phantom{\bigg[} & + 8 K^2 \frac{\epsilon^2 Q^2}{t} \,\,\,\,\,\,\, \bigg] F_A^2(t)  \\
c_1^{BH} = & \phantom{\bigg[} & -8 (2-y) K \left\{ 2 x_A + \epsilon^2 - 
\frac{\epsilon^2 Q^2}{t} \right\} F_A^2(t)  \\
c_2^{BH} = & \phantom{\bigg[} & 8 K^2 \frac{\epsilon^2 Q^2}{t} F_A^2(t) 
\end{eqnarray} 
\normalsize
where $F_A(t)$ is the electromagnetic form factor of the $^4$He. At leading twist, the $|\mathcal{T}_{DVCS}|^{2}$ writes as a function of only one CFF according to
\small
\begin{equation}
   c_0^{DVCS}= 2 \frac{2-2y+y^2 + \frac{\epsilon^2}{2}y^2}{1 + \epsilon^2} \, 
   {\mathcal H}_A {\mathcal H}^{\star}_A \label{eq:c0DVCS}
\end{equation}
\normalsize
and the interference amplitude coefficients are written as:
\small
\begin{equation}
s_{1}^{INT} = F_{A}(t) \Im m(\mathcal{H}_{A}) S_{++}(1),
\end{equation}
with
\begin{eqnarray}
   S_{++}(1) &=& \frac{-8K(2-y)y}{1+\epsilon^2} \left( 1 + 
\frac{1-xA+\frac{\sqrt{1+\epsilon^2}-1}{2}}{1+\epsilon^2} 
\frac{t-t_{min}}{Q^{2}} \right) \cdot F_{A}(t) \label{eq:s1I}
\end{eqnarray}

\begin{eqnarray}
\small
c_0^{INT} &=& F_A(t) \Re e(\mathcal{H}_{A}) C_{++}(0),
\end{eqnarray}
with \begin{eqnarray}  C_{++}(0) &=&
\frac{-4(2-y)(1+\sqrt{1+\epsilon^{2}})}{(1+\epsilon^{2})^2}  \bigg\{ 
   \frac{\widetilde{K}^2}{Q^2}  \frac{(2-y)^2}{\sqrt{1+\epsilon^{2}}} \, \\
   &+& \frac{t}{Q^2}  \left( 1 - y - \frac{\epsilon^2}{4} y^2 \right)  
(2-x_{A}) \left(  1 + \frac{2x_A(2-x_A + \frac{\sqrt{1+\epsilon^{2}}-1}{2} + 
\frac{\epsilon^{2}}{2x_A})\frac{t}{Q^2} + \epsilon^{2}}{(2-x_A) 
(1+\sqrt{1+\epsilon^{2}})}  \right)  \bigg\} \nonumber
 \label{eq:c0I} 
 \end{eqnarray}

\begin{eqnarray}
   c_1^{INT} &=&  F_A(t) \Re e(\mathcal{H}_{A}) C_{++}(1),
\end{eqnarray}
with  

\begin{eqnarray}
  C_{++}(1) &=&
  \frac{-16K(1-y+\frac{\epsilon^{2}}{4}y^2)}{(1+\epsilon^{2})^{5/2}}\bigg\{\left(1+(1-x_A)\frac{\sqrt{1+\epsilon^{2}}-1}{2x_A} 
    + \frac{\epsilon^{2}}{4x_A}\right) 
    \frac{x_At}{Q^2}-\frac{3\epsilon^{2}}{4.0} \bigg\} \nonumber \\&-& 4K \left( 
    2-2y+y^2+\frac{\epsilon^{2}}{2}y^2\right)\frac{1+\sqrt{1+\epsilon^{2}}-\epsilon^{2}}{(1+e2)^{5/2}}\bigg\{1-(1-3x_A)\frac{t}{Q^2}\nonumber\\&\,\,\,\,&\,\,\,\,\,\,\,\,\,\,\,\,\,\,\,\,\,\,\,\,\,+\frac{1-\sqrt{1+\epsilon^{2}}+3\epsilon^{2}}{1+\sqrt{1+\epsilon^{2}}-\epsilon^{2}} 
  \frac{x_A*t}{Q^2}\bigg\} \label{eq:c1I}
\end{eqnarray}

\chapter{Projections using Bethe-Heitler cross sections} \label{app:BH_cross_section}

As it has been shown in equation \ref{eq:sigdiff}, the amplitude of the 
photon-leptoproduction cross section is decomposed into three terms: the pure 
DVCS scattering amplitude (${\mathcal{T}}_{DVCS}$), the pure BH amplitude ($ 
{\mathcal{T}}_{BH}$), and the interference amplitude 
(${\mathcal{I}}^{\lambda}_{BH*DVCS}$). Within the accessible phase-space of 
JLAB, the measured cross section is mostly dominant by the well known BH 
reaction. Herein, we perform an exercise of extracting projections of our 
results based on generating and simulating pure BH events for the purpose of 
validating our projection results that were extracted based on the parametrized 
cross section from CLAS-EG6 experiment.

\begin{figure}[tbp]
 \includegraphics[scale=0.4]{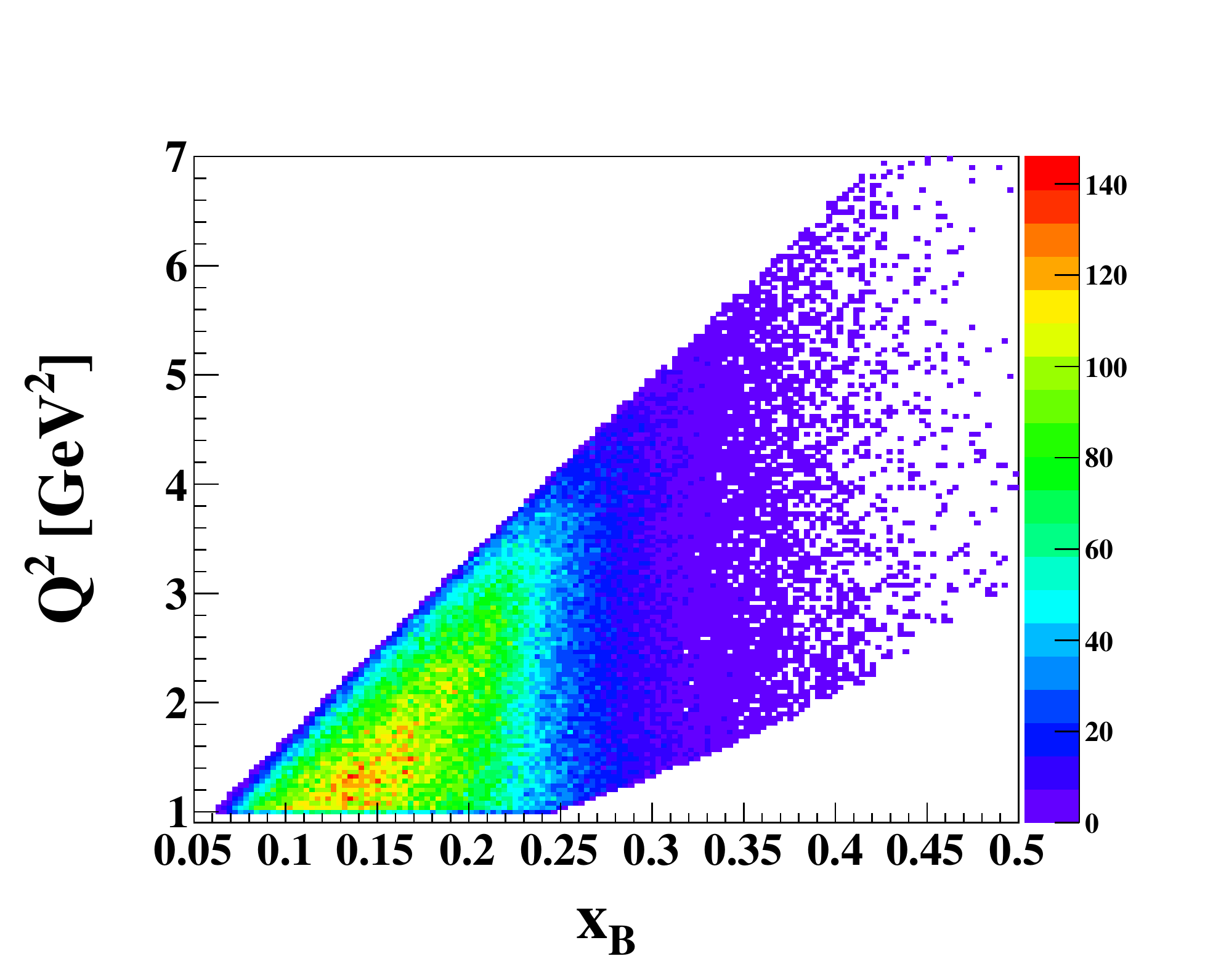}
 \includegraphics[scale=0.4]{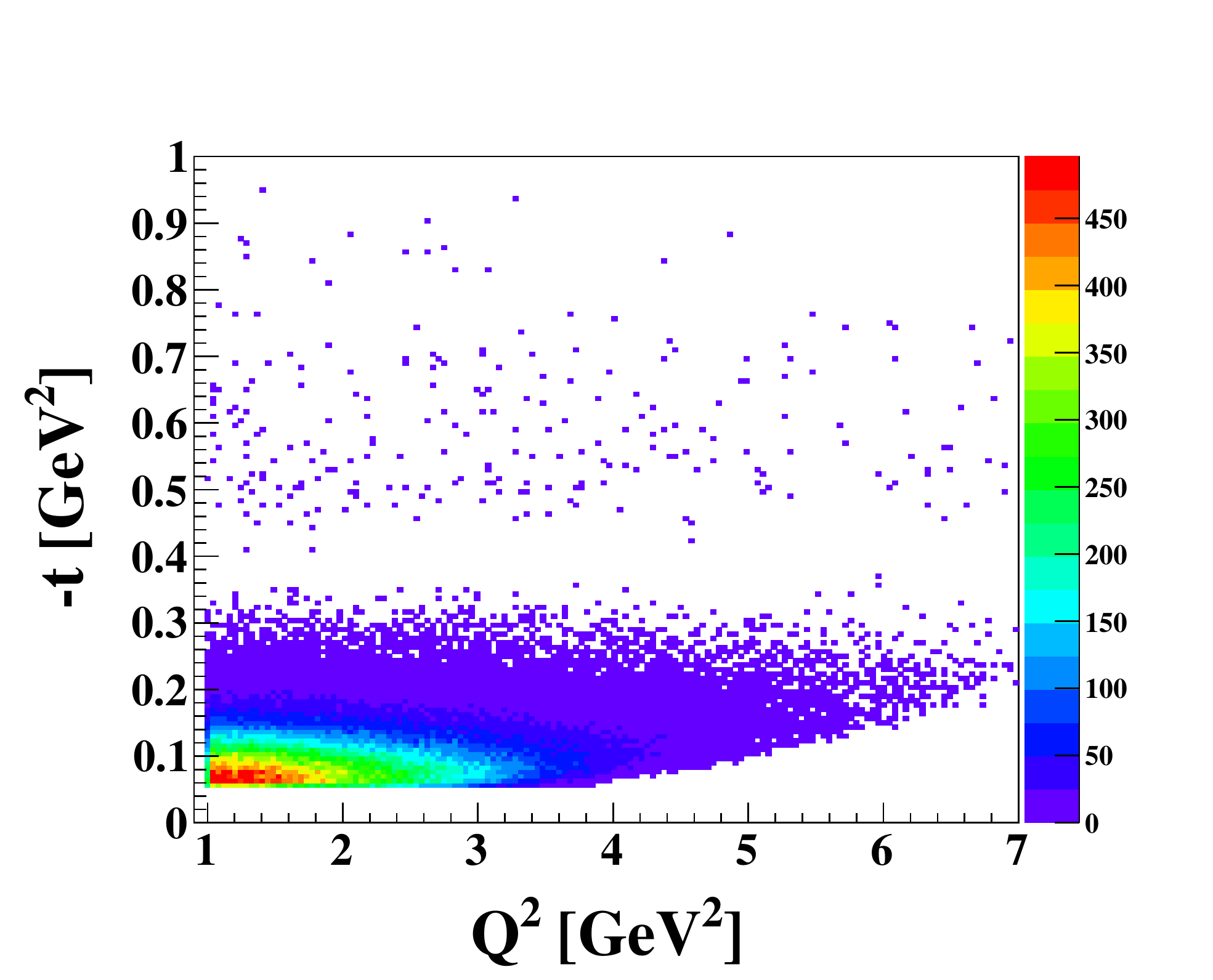}
 \caption{Correlation between $Q^{2}$ and $x_{B}$ (on the left) and between 
 $-t$ and $Q^{2}$ for pure coherent BH off $^{4}$He.} \label{fig:Q2-xB-BH}
 \end{figure}

\begin{figure}[!h]
   \centering
\includegraphics[scale=0.52]{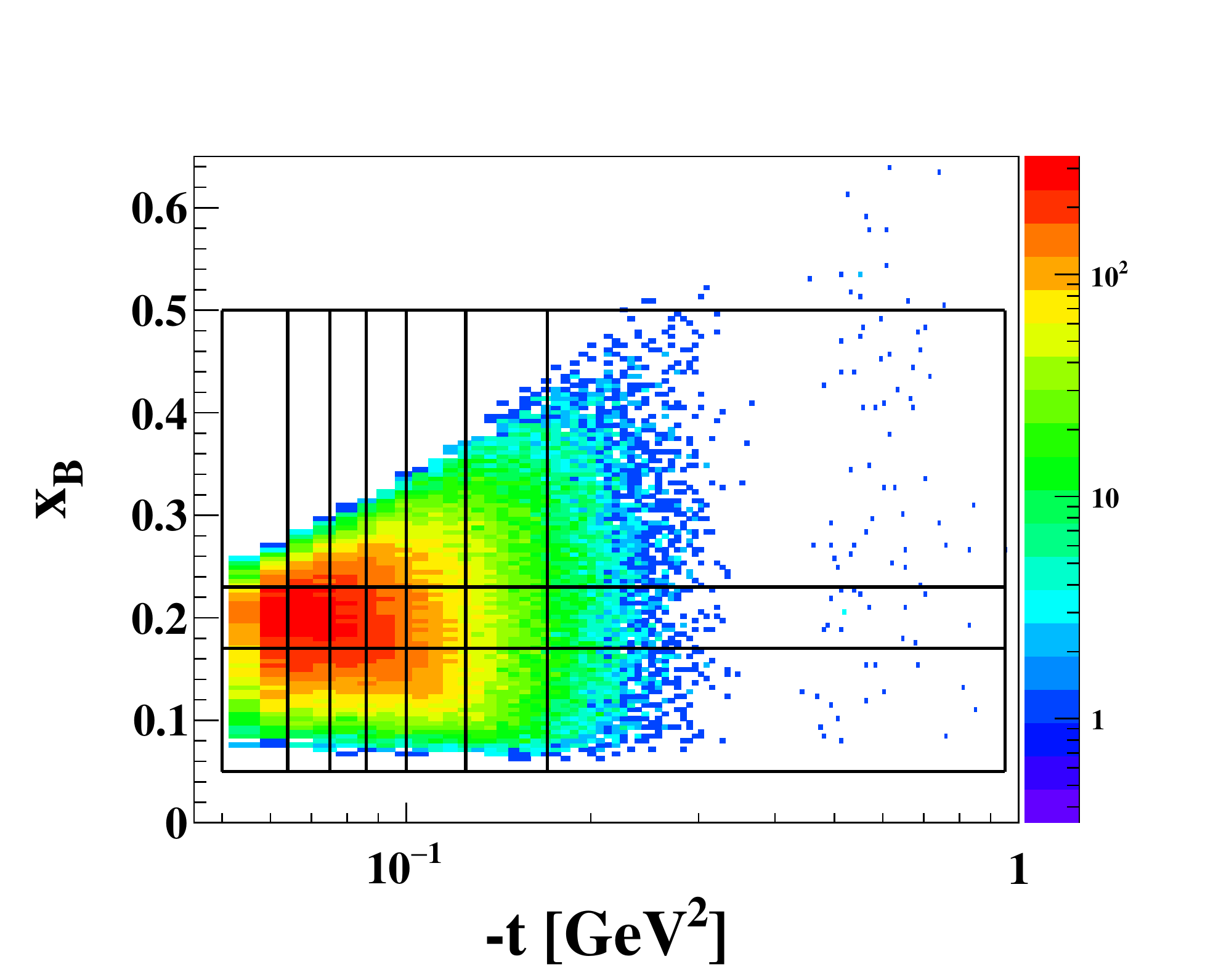}
\vspace{-0.5cm}
\caption{Data binning in $x_{B}$ vs $-t$ space for the reconstructed pure 
coherent BH off $^{4}$He. }
\label{fig:binning_x_t-BH}
\end{figure}

\begin{figure}[tbp]
   \centering
\includegraphics[scale=0.4]{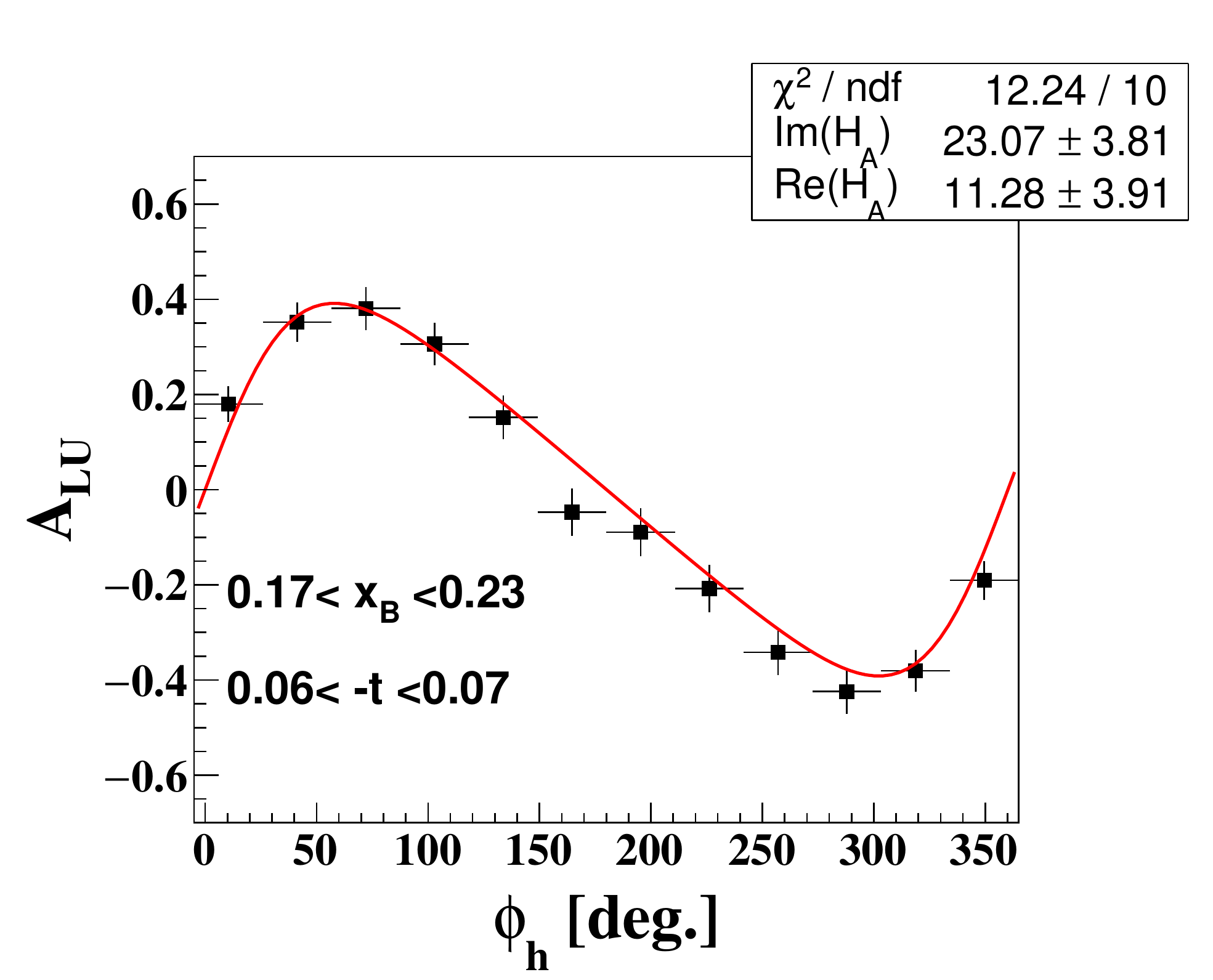}
\includegraphics[scale=0.4]{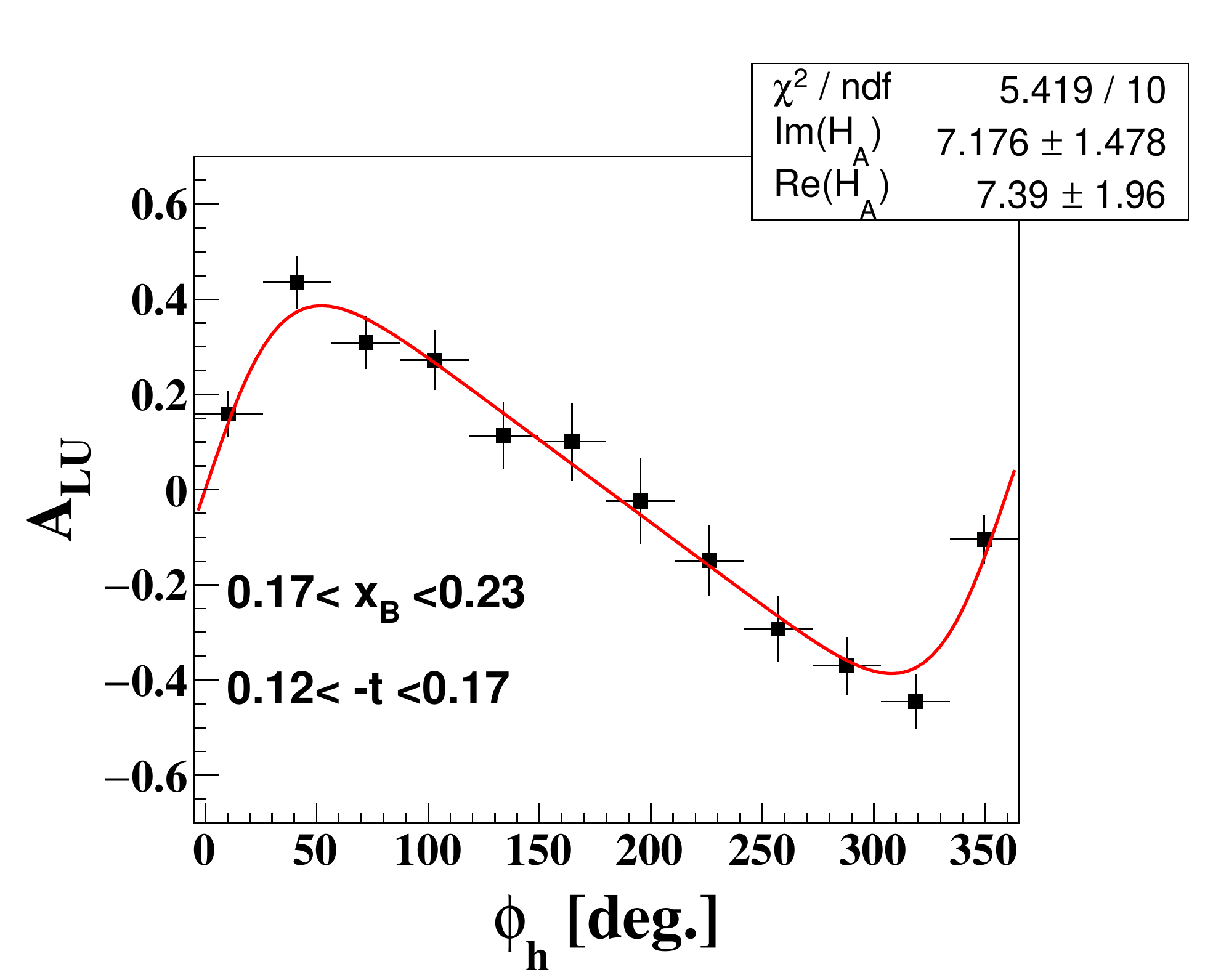}
\caption{The coherent beam-spin asymmetry projections as a function of the 
   angle $\phi$ between the leptonic and the hadronic planes, for two different 
   bins $-t$ at the same $x_{B}$ range, integrated over $Q^{2}$ range. The red 
   solid curves represent a fit to the data in the full form of the asymmetry, 
equation \ref{eq:A_LU-coh}, with the real and the imaginary parts of the CFF as 
the free parameters of the fit.  }
\label{fig:ALU-projections-BH}
\end{figure}

\begin{figure}[!h]
\centering
\includegraphics[scale=0.45]{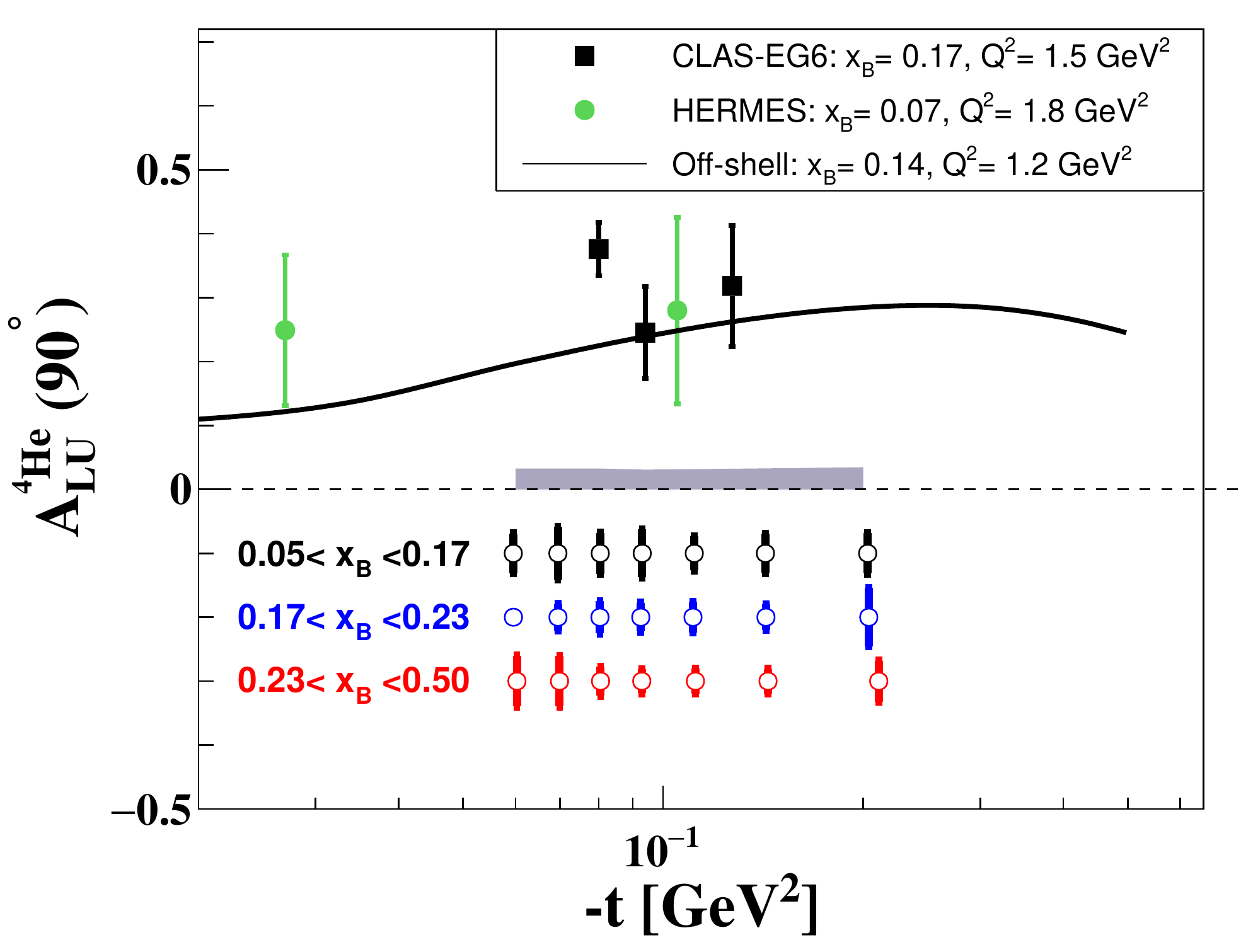}
\caption{Projected precision for the $A_{LU}$ (90$^{\circ}$), from the fit, for 
coherent DVCS on $^{4}$He versus $-t$ compared to the previous measurements 
from CLAS-eg6 (black squares), HERMES (green circles) and spectral function 
calculations (LT curves).}
\label{fig:ALU-projections-90-BH}
\end{figure}

\begin{figure}[!h]
\centering
\includegraphics[scale=0.45]{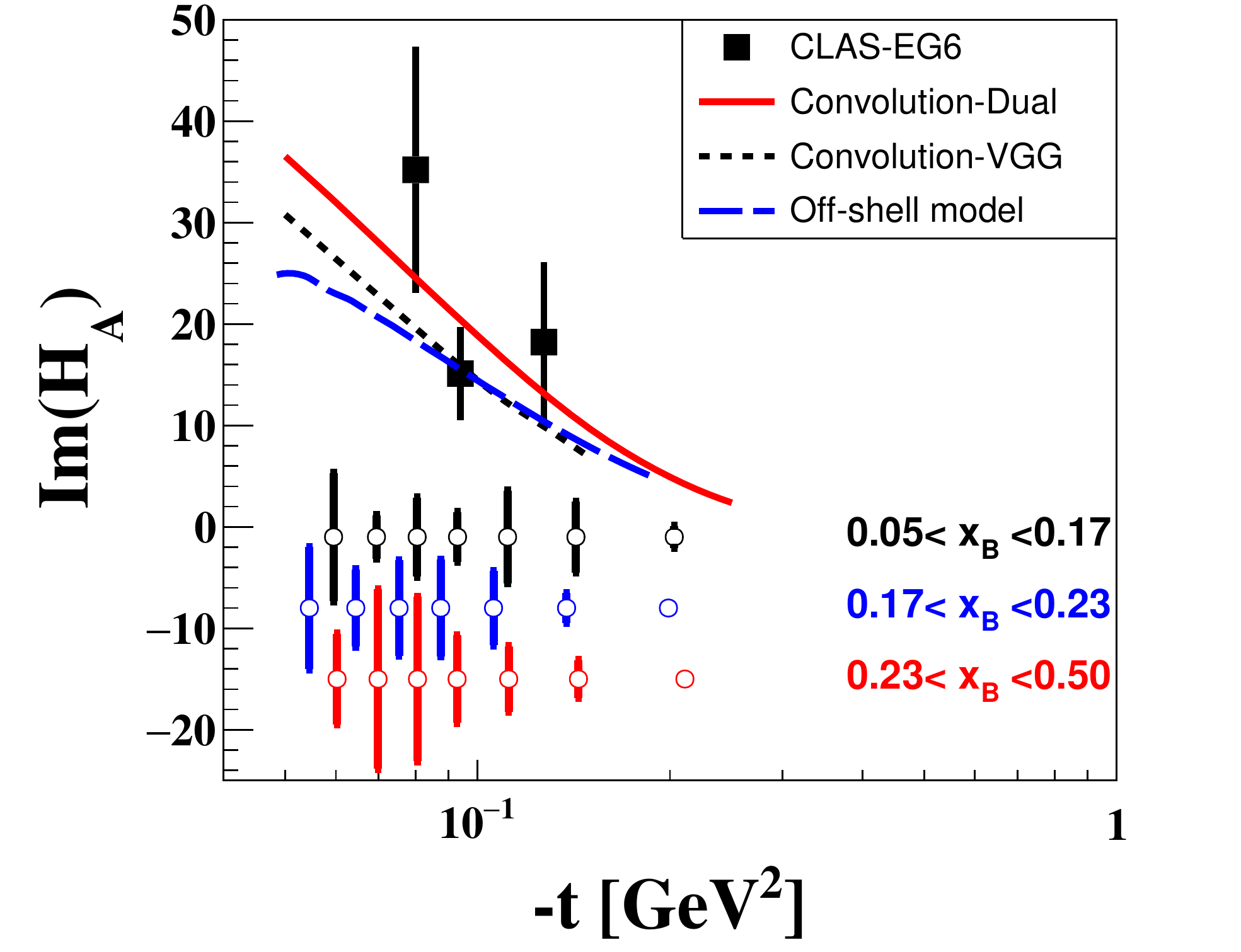}
\includegraphics[scale=0.45]{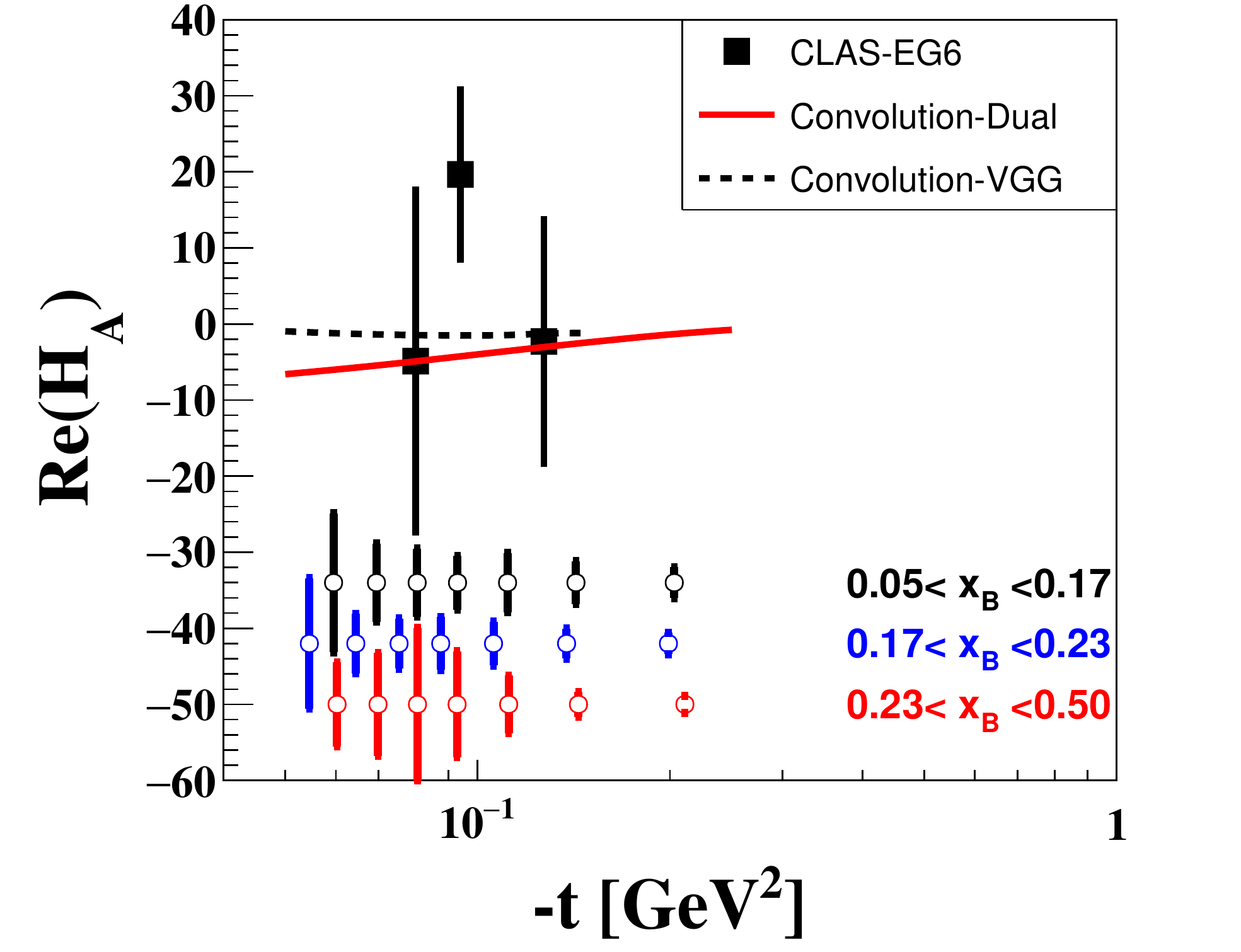}
\caption{The projected statistical uncertainties for the  imaginary (top) and 
real (bottom) parts of the CFF $H_{A}$, from the fits, as a function of $-t$ at 
fixed ranges in $x_{B}$}
\label{fig:CFF_projections-BH}
\end{figure}

\begin{figure}[!h]
\centering
\includegraphics[scale=0.55]{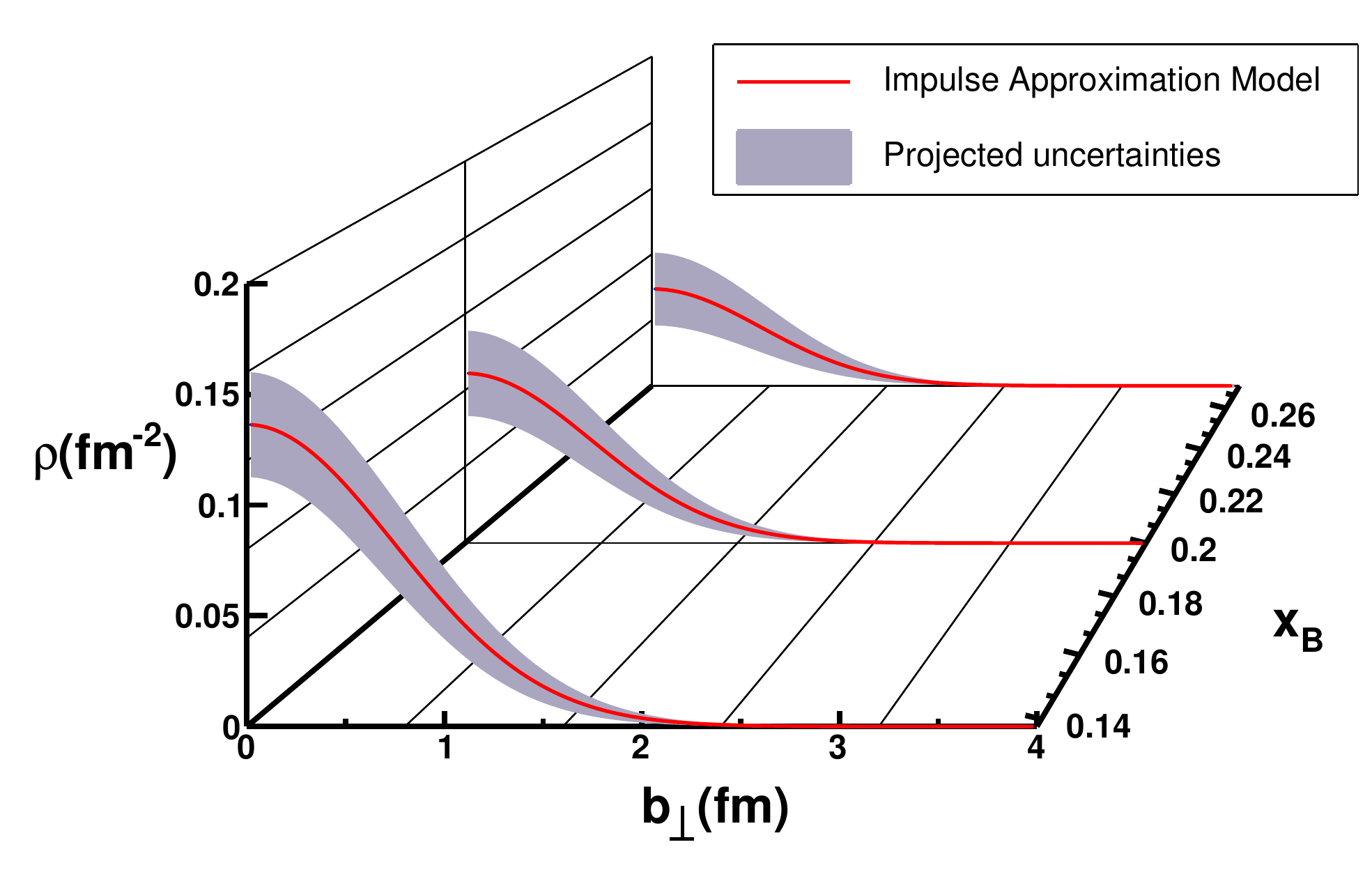}
\caption{The statistical uncertainties of the  parton density profiles as a 
   function of the impact parameter, $b_{\perp}$, based on the 
$\mathcal{H}^{A}$ CFF extracted from the Impulse Approximation (IA) at the mean 
$x_{B}$ values in the different bins. }
\label{fig:density_profile-BH}
\end{figure}

The statistical error bars that are shown in the following figures were 
calculated using the same beam time and luminosities listed in section 
\ref{sec:DVCS-projs}.  Figure \ref{fig:Q2-xB-BH} shows the kinematical 
correlations for the reconstructed BH events. For binning the data and 
extracting the projections, we have performed the same binning that was shown 
in figure \ref{fig:binning_x_t}. Figure \ref{fig:binning_x_t-BH} represents the 
same binning laid over the reconstructed BH events. Figure 
\ref{fig:ALU-projections-BH} shows the reconstructed beam-spin asymmetries as a 
function of the angle $\phi$ for two bins in $-t$ at a fixed $x_{B}$ value 
presenting a high and a low statistic bins. The projected precision of $A_{LU}$ 
at $\phi$ equal to 90$^{\circ}$ for the different bins is presented in figure 
\ref{fig:ALU-projections-90-BH}. The projected uncertainties on the 
reconstructed real and imaginary parts of the CFF are shown in figure 
\ref{fig:CFF_projections-BH}. Finally, figure \ref{fig:density_profile-BH} 
shows the projected uncertainties on the quark density profiles using the 
statistical error bars on the imaginary part of the $^{4}$He CFF presented in 
figure \ref{fig:CFF_projections-BH}. 

In summary, this exercise has proven that the parametrized cross section that 
has been used to extract our projections in section \ref{sec:DVCS-projs} is 
valid and our projections are well reproduced, within the statistical error 
bars, by the well-known pure BH process. The slight differences between the 
results presented in \ref{sec:DVCS-projs} and here are associated to the fact 
that the parametrized cross section was extracted based on real data from 
CLAS-EG6 that includes both processes, BH and DVCS.

\bibliographystyle{ieeetr}
\bibliography{biblio}

\end{document}